\documentclass[review,number,sort&compress]{elsarticle}
\usepackage{float}
\usepackage{svg}
\usepackage{lineno,}
\usepackage{caption}
\usepackage{subcaption}
\usepackage{multicol}
\usepackage{multirow}
\usepackage[version=4]{mhchem}
\usepackage{amsmath}
\usepackage[utf8]{inputenc}
\usepackage{xr}
\modulolinenumbers[5]
\journal{Materials Today Physics}

\begin{document}

\begin{frontmatter}

\title{Bridging the Gap Between Simulated and Experimental Ionic Conductivities in Lithium Superionic Conductors}
% \tnotetext[mytitlenote]{This is a footnote associated with the title.}

% \address{Radarweg 29, Amsterdam}
% \fntext[myfootnote]{Since 1880.}

\author[mymainaddress]{Ji Qi}
% \ead{j1qi@eng.ucsd.edu}
\author[mysecondaryaddress]{Swastika Banerjee}
\author[mysecondaryaddress]{Yunxing Zuo}
\author[mysecondaryaddress]{Chi Chen}
\author[mysecondaryaddress]{Zhuoying Zhu}
\author[mysecondaryaddress]{H.C. Manas Likhit}
\author[mysecondaryaddress]{Xiangguo Li}
\author[mysecondaryaddress]{Shyue Ping Ong\corref{mycorrespondingauthor}}
\cortext[mycorrespondingauthor]{Please direct correspondences to}
\ead{ongsp@eng.ucsd.edu}

\address[mymainaddress]{Materials Science and Engineering Program, University of California San Diego, 9500 Gilman Dr, Mail Code 0448, La Jolla, CA 92093-0448, United States}
\address[mysecondaryaddress]{Department of NanoEngineering, University of California San Diego, 9500 Gilman Dr, Mail Code 0448, La Jolla, CA 92093-0448, United States}

\begin{abstract}
Lithium superionic conductors (LSCs) are of major importance as solid electrolytes for next-generation all-solid-state lithium-ion batteries. While \textit{ab initio} molecular dynamics have been extensively applied to study these materials, there are often large discrepancies between predicted and experimentally measured ionic conductivities and activation energies due to the high temperatures and short time scales of such simulations. Here, we present a strategy to bridge this gap using moment tensor potentials (MTPs). We show that MTPs trained on energies and forces computed using the van der Waals optB88 functional yield much more accurate lattice parameters, which in turn leads to accurate prediction of ionic conductivities and activation energies for the \ce{Li_{0.33}La_{0.56}TiO3}, \ce{Li3YCl6} and \ce{Li7P3S11} LSCs. NPT MD simulations using the optB88 MTPs also reveal that all three LSCs undergo a transition between two quasi-linear Arrhenius regimes at relatively low temperatures. This transition can be traced to an expansion in the number and diversity of diffusion pathways, in some cases with a change in the dimensionality of diffusion. This work presents not only an approach to develop high accuracy MTPs, but also outlines the diffusion characteristics for LSCs which is otherwise inaccessible through \textit{ab initio} computation.
\end{abstract}

\begin{keyword}
% Immediately after the abstract, provide a maximum of 6 keywords, using American spelling and avoiding general and plural terms and multiple concepts (avoid, for example, 'and', 'of'). Be sparing with abbreviations: only abbreviations firmly established in the field may be eligible. These keywords will be used for indexing purposes.
Lithium superionic conductors; Machine learning interatomic potentials; PBE versus optB88 functionals; Non-Arrhenius behavior; Room temperature ionic conductivity.
\end{keyword}

\end{frontmatter}

% \linenumbers

\section{Introduction}

Lithium superionic conductors (LSCs) are the critical enabling solid electrolyte (SE) component in next-generation all-solid-state rechargeable lithium-ion batteries \cite{tarasconIssuesChallengesFacing2001a, armandBuildingBetterBatteries2008, xuElectrolytesInterphasesLiIon2014, wangDesignPrinciplesSolidstate2015}. Replacing the traditional flammable organic solvent electrolyte, ceramic LSCs exhibit superior safety and are also a potential pathway to higher energy density cell architectures and utilization of lithium metal anodes. As the name implies, a key property of LSCs is a high ionic conductivity, typically ranging from O(10$^{-1}$) mS cm$^{-1}$ to O(10) mS cm$^{-1}$ (rivaling that of liquid electrolytes) at room temperature. The anion chemistry of an LSC has a major influence on their properties. Sulfide LSCs, such as the \ce{Li10GeP2S12} (LGPS) family \cite{kamayaLithiumSuperionicConductor2011b, bronLi10SnP2013, whiteleyEmpoweringLithiumMetal2014, katoHighpowerAllsolidstateBatteries2016a}, \ce{Li7P3S11} \cite{yamaneCrystalStructureSuperionic2007, seinoSulphideLithiumSuper2014a, wenzelInterphaseFormationDegradation2016a, buscheSituMonitoringFast2016, chuInsightsPerformanceLimits2016a} and \ce{Li3PS4} \cite{tachezIonicConductivityPhase1984, liuAnomalousHighIonic2013a, yamadaAllSolidStateLithium2015}, tend to have very high ionic conductivities due to the large, polarizable \ce{S^{2-}}, but suffer from narrow electrochemical stability windows, air- and moisture-sensitivity. Oxide LSCs, such as the \ce{Li7La3Zr2O12} garnet family \cite{muruganFastLithiumIon2007a} and LISICONs \cite{huIonicConductivityLithium1977, kuwanoNewLiIon1980, bruceIonicConductivityLISICON1982}, typically have lower ionic conductivities compared to the sulfides, but are much more electrochemically and chemically stable. Recently, a promising new class of halide LSCs, \ce{Li3YCl6} and \ce{Li3YBr6}, has been discovered that exhibits a good compromise of ionic conductivities (0.51 mS cm$^{-1}$ for \ce{Li3YCl6} and 1.7 mS cm$^{-1}$ for \ce{Li3YBr6}) and electrochemical stabilities between those of the sulfides and oxides \cite{asanoSolidHalideElectrolytes2018}.

Molecular dynamics (MD) simulations have been extensively used in the study of ion conduction in LSCs. In particular, \textit{ab initio} MD (AIMD), i.e., simulations where the energies and forces are directly obtained by solving the Schr\"{o}dinger equation via density functional theory (DFT), have emerged as a powerful tool in recent years as they can be transferably and broadly applied to the entire range of LSC chemistries \cite{moFirstPrinciplesStudy2012a, ongPhaseStabilityElectrochemical2013, miaraEffectRbTa2013, dengRationalCompositionOptimization2015, chuInsightsPerformanceLimits2016a, heOriginFastIon2017, zhuLiPSLi2017, dengDataDrivenFirstPrinciplesMethods2017, wangLithiumChloridesBromides2019}. However, the high cost of AIMD simulations means that they are usually performed at elevated temperatures to obtain sufficient diffusion statistics, sometimes far in excess of the melting points of some LSCs, and for relatively short simulation time frames ($\sim$ 100 ps). As a consequence, extrapolated room-temperature ionic conductivity and diffusivity have large error bars \cite{heStatisticalVariancesDiffusional2018}. Further, there may be phase transitions or transitions in diffusion mechanisms occurring between room temperature and simulated high temperatures, invalidating the Arrhenius assumption used in extrapolation. Such non-Arrhenius behavior and phase transitions have been reported in many LSCs, including \ce{Li_{3x}La_{2/3-x}TiO_3} \cite{inagumaHighIonicConductivity1993a, salkusDeterminationNonArrheniusBehaviour2011}, \ce{Li3PS4} \cite{tachezIonicConductivityPhase1984}, and LGPS \cite{kwonSynthesisStructureConduction2015}. Another source of error arises from the fact that most AIMD simulations of LSCs are performed in the NVT ensemble using the equilibrium volume from a 0K density functional theory (DFT) relaxation calculation. The most common DFT functional used is the Perdew-Burke-Ernzerhof (PBE) generalized gradient approximation (GGA) \cite{perdewGeneralizedGradientApproximation1996a}, which tends to overestimate the lattice parameters of solids and differ from experimental values by up to 2-3\% \cite{heydEnergyBandGaps2005,klimesVanWaalsDensity2011}. These differences in lattice parameters can have a major effect on ion diffusion and activation barriers \cite{ongPhaseStabilityElectrochemical2013, moradabadiEffectLatticeDopant2020}.

The net result of the mismatch in working temperatures and lattice parameters between simulations and experiments is that room-temperature ionic diffusivity and conductivity of LSCs computed from AIMD simulations often disagree substantially with those measured experimentally, e.g., via electrochemical impedance spectroscopy (EIS). For example, Chu et al. \cite{chuInsightsPerformanceLimits2016a} previously predicted a room-temperature ionic conductivity of 57 mS cm$^{-1}$ for the \ce{Li7P3S11} LSC using AIMD simulations, far in excess of the highest experimentally measured room-temperature ionic conductivity of 17 mS cm$^{-1}$ \cite{seinoSulphideLithiumSuper2014a}. Similarly, Wang et al. \cite{wangLithiumChloridesBromides2019} predicted an ionic conductivity of 14 mS cm$^{-1}$ for the \ce{Li3YCl6} LSC, again far in excess of the experimentally reported 0.51 mS cm$^{-1}$ \cite{asanoSolidHalideElectrolytes2018}.

Classical MD simulations using an interatomic potential (IAP) to parameterize the potential energy surface (PES) are a potential solution to enable low-temperature and long-timescale studies. In recent years, machine learning (ML) the PES as a function of local environment descriptors has emerged as an especially promising, and reproducible approach to develop IAPs with near-DFT accuracy in energies and forces \cite{behlerGeneralizedNeuralNetworkRepresentation2007a, bartokGaussianApproximationPotentials2010, thompsonSpectralNeighborAnalysis2015, shapeevMomentTensorPotentials2016a, wangDeePMDkitDeepLearning2018, chenAccurateForceField2017, liStudyLiAtom2017a, liQuantumaccurateSpectralNeighbor2018, dengElectrostaticSpectralNeighbor2019a, liComplexStrengtheningMechanisms2020a, zuoPerformanceCostAssessment2020b, wangLithiumIonConduction2020, huangDeepPotentialGeneration2021}. However, most ML-IAPs that have been developed in the literature still rely on DFT calculations performed using the PBE functional; as such, their performance are still limited by the accuracy of the DFT training data.

\begin{figure}
    \centering
    \begin{subfigure}[b]{0.49\textwidth}
        \centering
        \graphicspath{ {./} }
	    \includegraphics[height=3.5cm]{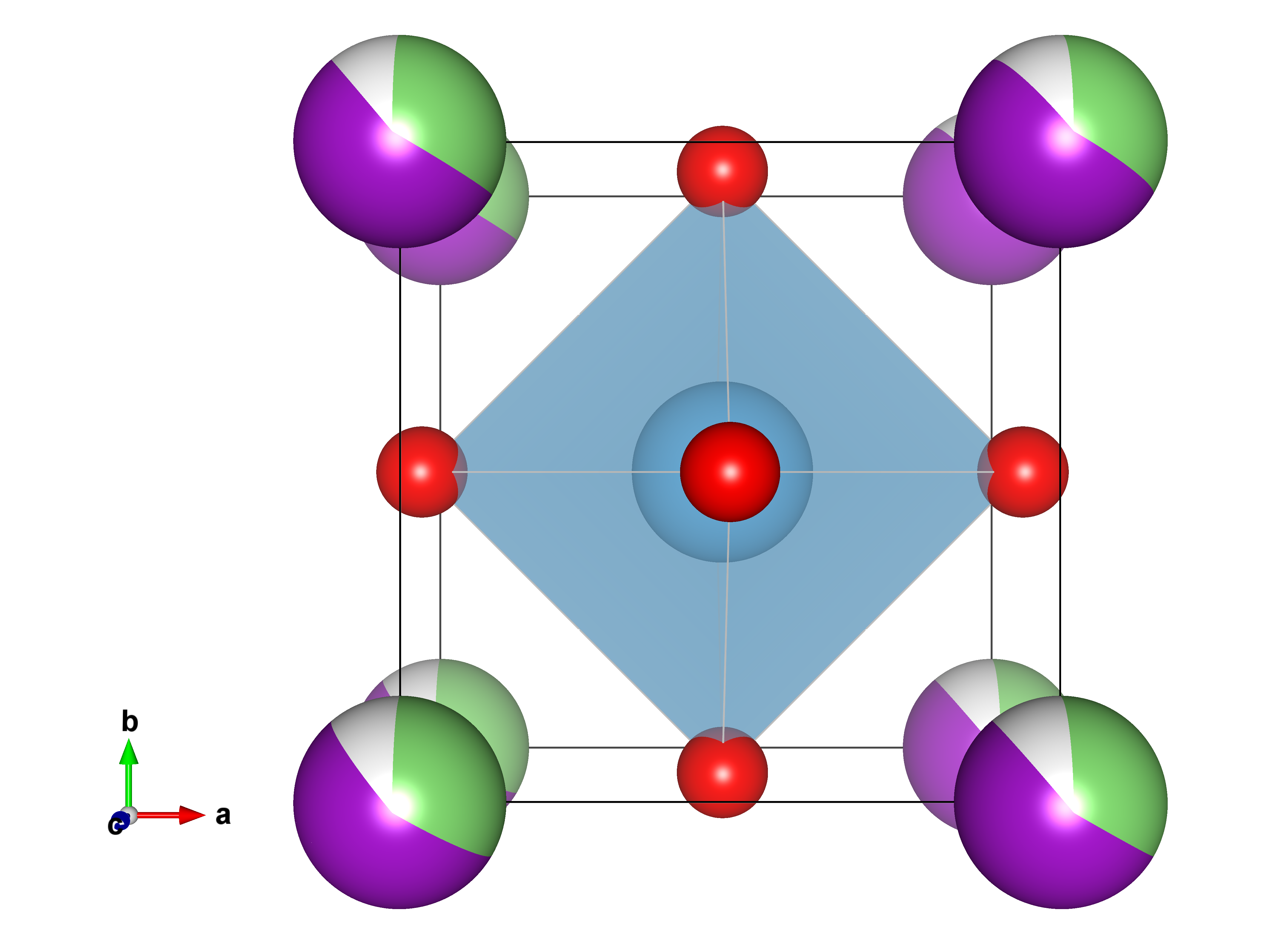}
    	\caption{\label{subfig:unitcell_LLTO}\ce{Li_{0.33}La_{0.56}TiO3}}
    \end{subfigure}
    \begin{subfigure}[b]{0.49\textwidth}
        \centering
        \graphicspath{ {./} }
	    \includegraphics[height=4cm]{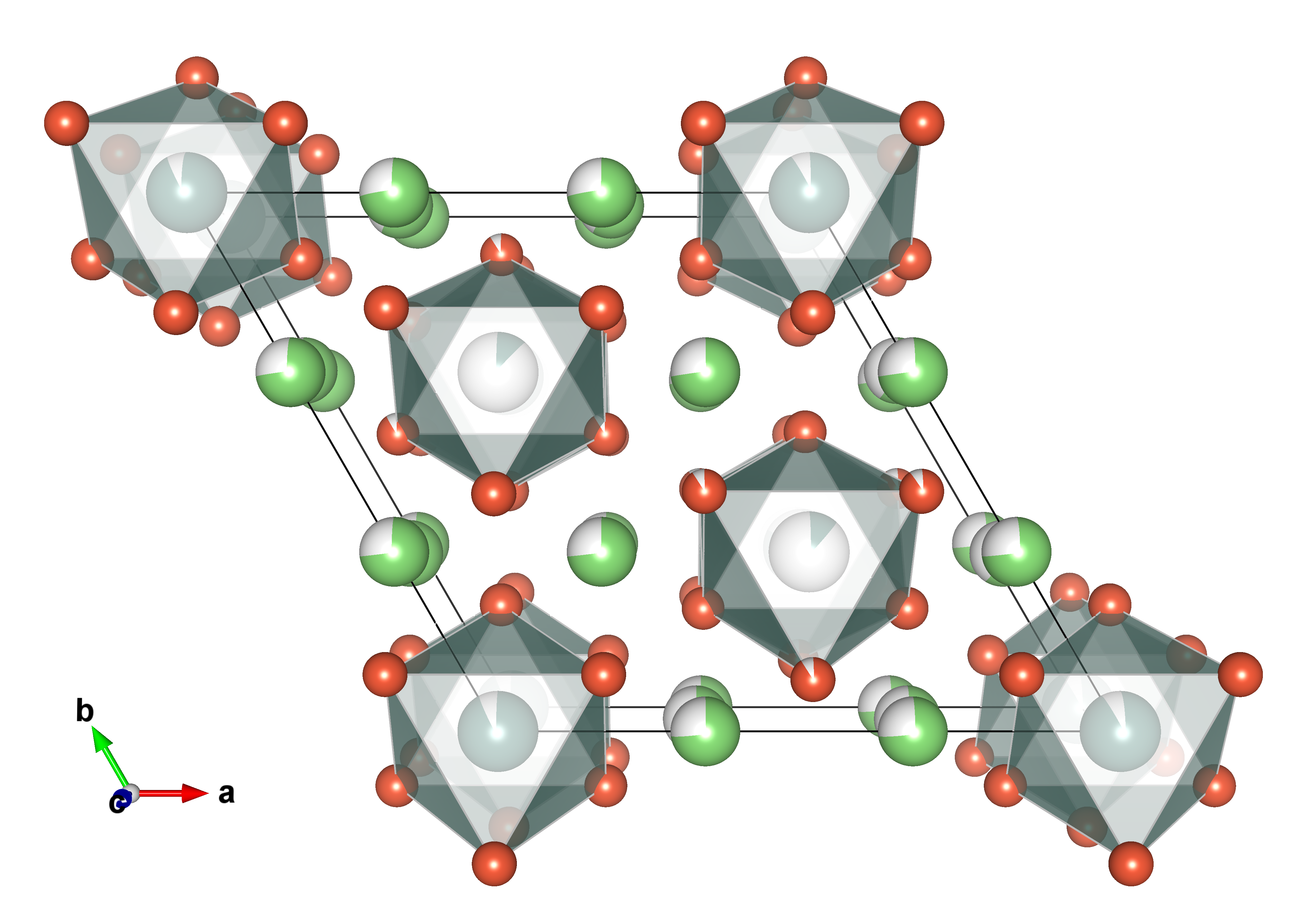}
    	\caption{\label{subfig:unitcell_LYC}\ce{Li3YCl6}}
    \end{subfigure}
    \begin{subfigure}[b]{0.49\textwidth}
        \centering
        \graphicspath{ {./} }
	    \includegraphics[height=4cm]{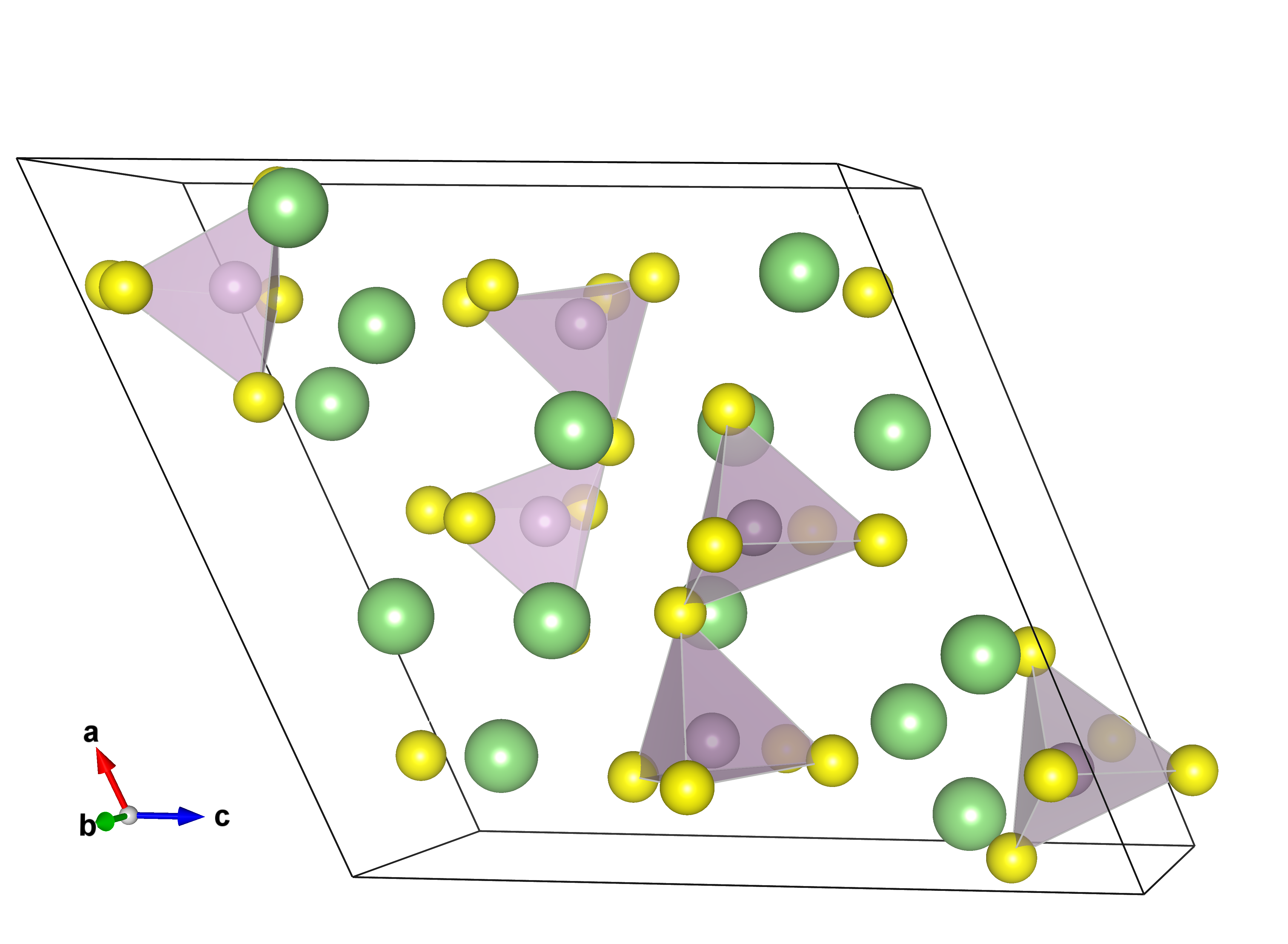}
    	\caption{\label{subfig:unitcell_Li7P3S11}\ce{Li7P3S11}}
    \end{subfigure}
    \begin{subfigure}[b]{0.4\textwidth}
        \centering
        \graphicspath{ {./} }
	    \includegraphics[height=4cm]{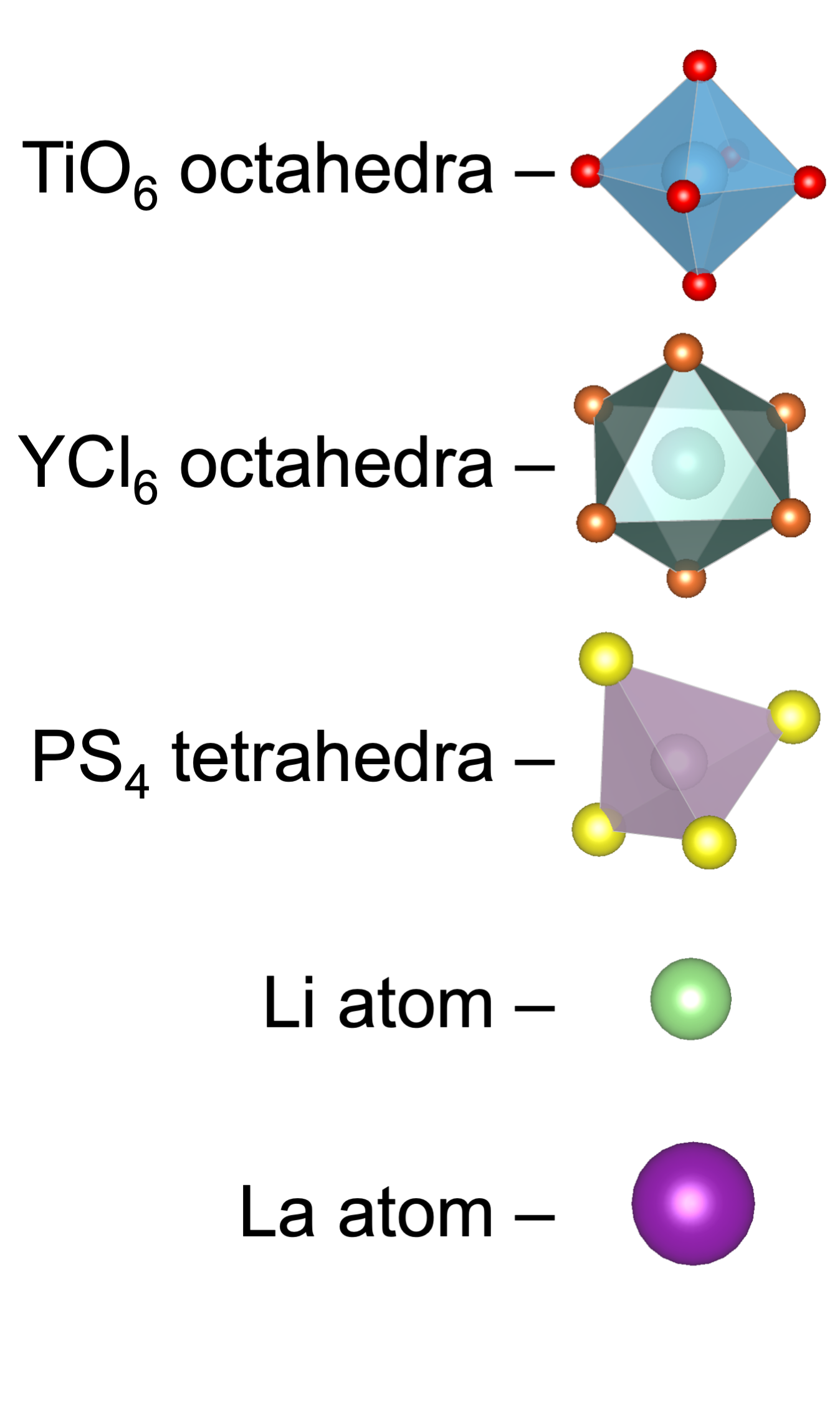}
    \end{subfigure}
    \caption{Crystal structures of (a) \ce{Li_{0.33}La_{0.56}TiO3} (Space group $Pm\bar{3}m$, No. 221), (b) \ce{Li3YCl6} (Space group $P\bar{3}m1$, No. 164), and (c) \ce{Li7P3S11}(Space group $P\bar{1}$, No. 2).}
\label{fig:lsc}
\end{figure}

In this work, we show that the gap between experimental and simulated ionic conductivities in LSCs can be bridged by developing ML-IAPs under the moment tensor potential (MTP) formalism \cite{shapeevMomentTensorPotentials2016a, gubaevAcceleratingHighthroughputSearches2019} using training data from the optB88 van der Waals (vdW) DFT functional \cite{klimesChemicalAccuracyVan2010, klimesVanWaalsDensity2011}. Three LSCs, \ce{Li_{0.33}La_{0.56}TiO3} (LLTO), \ce{Li3YCl6} and \ce{Li7P3S11} spanning a diversity of anion chemistries have been selected as the model systems for investigation, as shown in Figure \ref{fig:lsc}. These LSCs have been selected because of their major interest to the battery research community, as well as the fact that previous AIMD calculations have either yielded large disagreements with experimentally reported room temperature conductivity or else have not been performed as in the case of LLTO. We demonstrate that in all three cases, the discrepancy between computed and measured conductivities can be explained by a transition between quasi-linear Arrhenius regimes arising from the activation of additional diffusion pathways.

\section{Material and Methods}

\begin{figure}
    \centering
    \includegraphics[height=4cm]{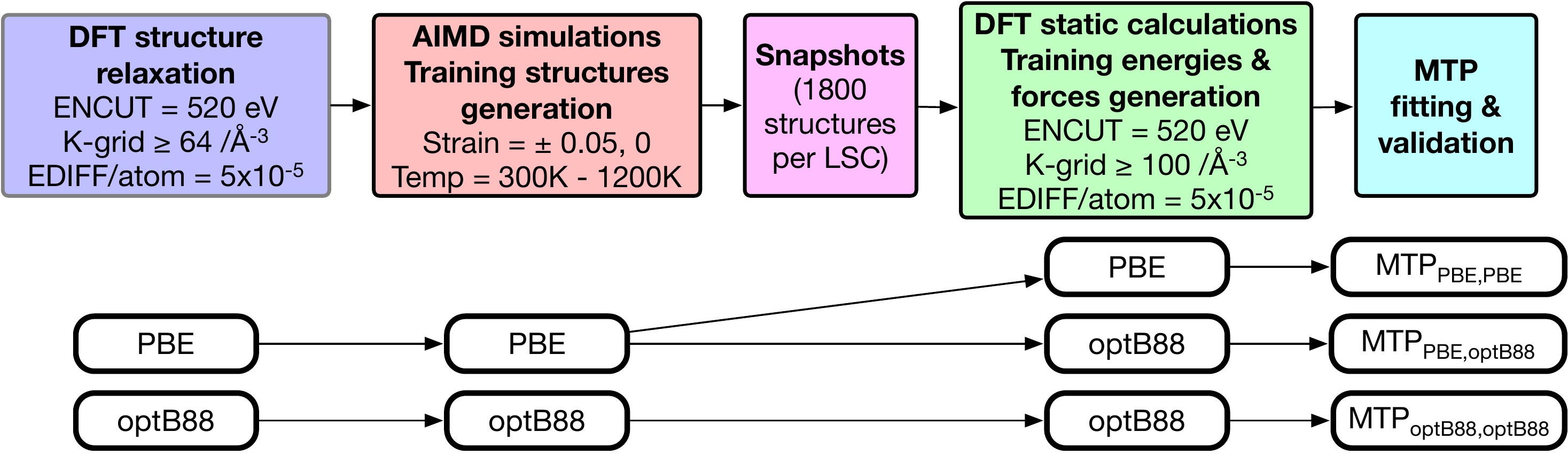}
\caption{Flowchart of the stepwise construction of MTPs for LSCs. DFT functionals utilized in each step and the names of the as-trained MTPs were listed.}
\label{fig:workflow}
\end{figure}

Figure \ref{fig:workflow} summarizes the overall workflow for the construction of MTPs for the LSCs investigated in this work as well as the DFT functional choices investigated. 

\subsection{Structure Construction}
\label{Section: construction of supercell}

Supercells of LLTO, \ce{Li3YCl6} and \ce{Li7P3S11} with lattice parameters greater than 10 \AA~were constructed to minimize interactions between periodic images. For LLTO, a $3\times3\times1$ supercell of \ce{Li_{0.33}La_{0.56}TiO3}, equivalent to $x = 0.11$ in the general formula of  \ce{Li_{3x}La_{2/3-x}TiO_3} \cite{inagumaHighIonicConductivity1993a, haradaLithiumIonConductivity1998, haradaOrderDisorderAsite1999}, was initially generated to enumerate symmetrically distinct orderings of Li/La/Vacancy on the perovskite A site. These orderings were fully relaxed by DFT. The lowest energy ordering  was then stacked along the $c$ direction to obtain a $3\times3\times3$ supercell. During relaxation, shifts of the lithium ion position from the A-sites of perovskite were observed (see Figure S1), which is consistent with previous theoretical studies on LLTO \cite{qianLithiumLanthanumTitanium2012,chengIntegratedApproachStructural2014,romeroExperimentalTheoreticalRaman2016}. \ce{Li3YCl6} has previously been identified to be an isomorph of \ce{Li3ErCl6} (ICSD No. 50151, space group $P\bar{3}m1$, No. 164) \cite{asanoSolidHalideElectrolytes2018}. Starting from the experimentally reported disordered \ce{Li3YCl6} structure, the site occupancies were rounded to the nearest rational numbers based on a total of 3 formula units per unit cell (see Table S1) and enumeration of distinct orderings was performed. The lowest energy relaxed structure was then selected to construct a $1\times1\times2$ supercell. A $1\times2\times1$ supercell of \ce{Li7P3S11} was constructed from the experimentally refined crystal structure \cite{yamaneCrystalStructureSuperionic2007}. The formation energies ($E_f$) and energy above the convex hull ($E_{hull}$) for all three LSCs calculated with the PBE and optB88 functionals are given in Table S2. All three LSCs are predicted to have $E_{hull} <$ 0.05 eV/atom. The optB88 $E_f$ are 5-10\% lower than the PBE values, which is consistent with the $\sim$5\% higher atomization energies predicted by optB88 for ionic solids \cite{klimesVanWaalsDensity2011}.

\subsection{DFT Calculations and AIMD Simulations}

All DFT calculations were performed using the Vienna \textit{ab initio} simulation package (VASP) with the projector augmented-wave (PAW) approach \cite{blochlProjectorAugmentedwaveMethod1994a, kresseEfficientIterativeSchemes1996a}. For initial structural relaxations (Step 1 in Figure \ref{fig:workflow}), spin-polarized calculations were performed with an energy cutoff of 520 eV and a k-point density of at least $64/\text{\AA}^{-3}$, similar to those used in the Materials Project (MP) \cite{jainCommentaryMaterialsProject2013a}. 

In Step 2, non-spin polarized \textit{ab initio} molecular dynamics (AIMD) simulations using NVT ensemble were carried out on the relaxed supercells with a plane-wave energy cutoff of 280 eV and a minimal $\Gamma$-centered $1\times1\times1$ k-mesh. A time step of 2 fs and the Nose–Hoover thermostat \cite{noseUnifiedFormulationConstant1984, hooverCanonicalDynamicsEquilibrium1985} were used. A similar protocol was followed as previous works \cite{liQuantumaccurateSpectralNeighbor2018, dengElectrostaticSpectralNeighbor2019a, zuoPerformanceCostAssessment2020b}, wherein simulations were performed at three strains (0, $\pm$ 0.05) and four temperatures (300K to 1200K with 300K intervals) to diversify the training structures. The initial structures were heated from 0K to the target temperatures with a temperature gradient of 0.25 K/fs and equilibrated for at least 30 ps. Snapshots were then extracted from a production run of 15 ps at 0.1 ps intervals, i.e., 150 structures for each temperature and strain. Hence, for each LSC, a total of 1800 training structures ($150\times4\text{ temperatures}\times3\text{ strains}$) were generated. 

In Step 3, static self-consistent calculations were performed on the training structures to obtain accurate energies \& forces for MTP training. These calculations were performed with a higher k-point density of at least $100/\text{\AA}^{-3}$, an energy cutoff of 520 eV and an electronic relaxation convergence condition of $5\times10^{-5}$ eV/atom, which were consistent with those used in MP \cite{jainCommentaryMaterialsProject2013a}. 

A main goal of this work is to evaluate the choice of the DFT functional on the training data and hence, the performance of the MTP generated. The initial structural relaxations and energy evaluations of symmetrically distinct LLTO and \ce{Li3YCl6} orderings were performed using the PBE \cite{perdewGeneralizedGradientApproximation1996a} functional. Subsequent structural relaxations, AIMD simulations and static energy valuations were performed using either the PBE functional or optB88 vdW functional \cite{klimesChemicalAccuracyVan2010, klimesVanWaalsDensity2011}, as shown in Figure \ref{fig:workflow}.

All DFT and AIMD simulations were carried out using fully-automated workflows \cite{dengDataDrivenFirstPrinciplesMethods2017} built on the Python Materials Genomics (pymatgen) \cite{ongPythonMaterialsGenomics2013a} library and FireWorks scientific workflow package \cite{jainFireWorksDynamicWorkflow2015a}.

\subsection{MTP Model Training and Verification}
\label{section: ML-IAP training and labeling}

The moment tensor potential (MTP) formalism has been extensively discussed in earlier works \cite{shapeevMomentTensorPotentials2016a, gubaevAcceleratingHighthroughputSearches2019, zuoPerformanceCostAssessment2020b} and successfully applied to many chemical systems, including metals \cite{shapeevMomentTensorPotentials2016a, novoselovMomentTensorPotentials2019, zuoPerformanceCostAssessment2020b}, boron \cite{podryabinkinAcceleratingCrystalStructure2019}, alloys \cite{gubaevAcceleratingHighthroughputSearches2019}, gas-phase reactions \cite{novikovAutomatedCalculationThermal2018} and cathode coating materials \cite{wangLithiumIonConduction2020}. Briefly, the MTP describes the local environment around each atom in terms of moment tensors $M_{\mu,\nu}$, defined as follows:
\begin{equation}
    M_{\mu,\nu}(\mathbf{n_i}) = \sum_{j}f_{\mu}(|\mathbf{r_{ij}}|,z_i,z_j)\underbrace{\mathbf{r_{ij}} \otimes... \otimes \mathbf{r_{ij}}}_\textrm{$\nu$ times} \label{eqn:mtp descriptor}
\end{equation}
Here, $\mathbf{n_i}$ denotes the atomic types as well as the relative positions of the $i^{th}$ atom and all of its neighboring atoms. $z_i$ and $z_j$ represent the atomic types (integers from 0 to n-1 for a system with n different types of atoms) of the $i^{th}$ atom and its $j^{th}$ neighbor, respectively, and $\mathbf{r_{ij}}$ is the position vector of the $j^{th}$ neighbor to the $i^{th}$ atom. The radial part of the atomic environment is given by the $f_{\mu}$ term, and the angular part is encoded by the outer product ($\otimes$) of the $\mathbf{r_{ij}}$ vectors, which is a tensor with rank $\nu$. The moment tensors $M_{\mu,\nu}$ are then contracted to basis functions $B_{\alpha}$, which are intrinsically invariant to atomic permutations, rotations and reflections. The energy of the system $E_{\rm{MTP}}$ is then expressed as a linear function of $B_{\alpha}$ as follows:
\begin{equation}
    E_{\rm{MTP}} = \sum_{i=1}^n\sum_{\alpha}\xi_{\alpha}B_{\alpha}(\mathbf{n_i}) \label{eqn:mtp energy}
\end{equation}
where $n$ and $\alpha$ are the total number of atoms inside the system and the total number of basis functions for each atom, respectively, and $\xi_{\alpha}$ are the coefficients fitted in the training process implemented in the MLIP package. Similarly, the forces and stresses can be expressed as the first and second derivatives of the $E_{\rm{MTP}}$ with respect to $\mathbf{r_{ij}}$ \cite{shapeevMomentTensorPotentials2016a, gubaevAcceleratingHighthroughputSearches2019}. An optimized MTP is then obtained by minimizing the errors in the predicted energies, forces and stresses with respect to the DFT training data. In this work, the energy, force and stress data points are assigned weights of 100:1:0, similar to previous works \cite{liQuantumaccurateSpectralNeighbor2018, dengElectrostaticSpectralNeighbor2019a, zuoPerformanceCostAssessment2020b, liComplexStrengtheningMechanisms2020a}.

Two key parameters control the performance trade-off of the MTP. The radius cutoff $R_{cut}$ determines the maximum interaction range between atoms. The larger the $R_{cut}$, the more the atomic interactions encoded in Equation \ref{eqn:mtp descriptor}. The completeness of the basis functions $B_{\alpha}$ is controlled by its maximum level ($lev_{max}$). The larger the $lev_{max}$, the larger the number of terms in the linear expansion in Equation \ref{eqn:mtp energy}, which in turn results in higher computational cost and a greater likelihood of over-fitting. In this work, the $R_{cut}$ was chosen to be 5.0 \AA, a typical value used in previously reported MTPs \cite{podryabinkinActiveLearningLinearly2017, gubaevAcceleratingHighthroughputSearches2019, wangLithiumIonConduction2020}, while the $lev_{max}$ were set as 18 for \ce{Li3YCl6} and \ce{Li7P3S11} and 16 for LLTO based on our convergence tests (see Figure S2-S5). For the fitting, a training:test split of 90:10 was used. In total, seven MTPs were fitted for LLTO, \ce{Li3YCl6} and \ce{Li7P3S11} according to the above discussed procedure. These MTPs are labeled with subscripts indicating the functionals used during the AIMD simulation (Step 2) and static energies and forces evaluations (Step 3). For instance, MTP$_{\mathrm{PBE,optB88}}$ refers to an MTP fitted using the snapshots extracted from AIMD simulations performed using the PBE functional, with energies and forces evaluated using the optB88 functional. It should be noted that MTP$_{\mathrm{optB88,optB88}}$ was fitted only for \ce{Li7P3S11} as a test case and because the results were highly similar to MTP$_{\mathrm{PBE,optB88}}$ (see later Results section), only the latter was fitted for the other two LSCs. Previously, the current authors have also used an alternative approach in which the long-ranged electrostatic interactions were accounted for separately via an Ewald summation of the formal oxidation states prior to fitting the residual interactions via the ML-IAP \cite{dengElectrostaticSpectralNeighbor2019a}. A similar ``electrostatic'' MTP (eMTP) for the LLTO LSC was also developed but the performance was similar to the MTP without separate accounting of electrostatics. It can be concluded that there is significant screening in these materials, and the radius cutoff used above was already sufficient to account for most of the electrostatic interactions (see Figure S6). 

All training, evaluations and simulations with MTP were performed using MLIP \cite{shapeevMomentTensorPotentials2016a, gubaevAcceleratingHighthroughputSearches2019}, LAMMPS \cite{plimptonFastParallelAlgorithms1995} and the open-source Materials Machine Learning (maml) Python package \cite{maml}.

\subsection{QHA Thermal Expansion}
Phonon calculations were performed using the supercells (outlined in Section \ref{Section: construction of supercell}) at nine fixed volumes (80\% to 120\% with 5\% intervals of the equilibrium volume at 0K from structural relaxations with PBE and optB88 functionals). Real-space force constants were calculated utilizing the density functional perturbation theory (DFPT) \cite{gonzeDynamicalMatricesBorn1997} method as implemented in VASP, while the real-space force constants from MTPs were generated with the finite displacement approach implemented in the Phonopy \cite{togoFirstPrinciplesPhonon2015a} package. Phonon frequencies were then calculated from the force constants, and the thermal expansion from 0K to 800K was calculated under the quasi-harmonic approximation (QHA).

\subsection{Diffusivity and Conductivity Calculations}

Classical MD simulations for each LSC were performed using the trained MTPs. Taking advantage of the lower computational cost and linear scaling of MTP calculations with respect to system size, larger supercells with all lattice parameters over 20 \AA~were used for these simulations. Based on benchmarks of the convergence of the ionic conductivity with cell sizes (Figure S7), simulation cells of $3\times3\times2$, $2\times2\times3$ and $3\times2\times2$ AIMD supercells are utilized for LLTO, \ce{Li3YCl6} and \ce{Li7P3S11}, respectively. The time step was set to 1 fs, and the total simulation time was at least 1 ns for all MD simulations. 

The tracer diffusivity ($D^{*}$) of Li ions was obtained by performing a linear fitting of the mean square displacement (MSD) of all diffusing Li ions with time, according to the Einstein relation \cite{vandervenRechargeableAlkaliIonBattery2020}:
\begin{equation}
        D^{*} = \frac{1}{2dNt} \sum_{i=1}^{N}[\Delta r_i(t)]^2
\end{equation}
where $d$ is the number of dimensions in which diffusion occurs ($d$=3 for all three electrolytes), $N$ is the total number of diffusing Li ions, and $\Delta r_i(t)$ is the displacement of the $i^{th}$ Li ion at time $t$.

The charge diffusivity ($D_{\sigma}$) of Li ions was calculated from the square net displacement of all diffusing Li ions, as described below \cite{vandervenRechargeableAlkaliIonBattery2020}:

\begin{equation}
% \color{red}
        D_{\sigma} = \frac{1}{2dNt} [\sum_{i=1}^{N}\Delta r_i(t)]^2
\end{equation}

The Haven ratio is then given by the following equation:
\begin{equation}
% \color{red}
    H_R = D^{*}/D_{\sigma}
\end{equation}

Finally, the ionic conductivity $\sigma (T)$ at temperature $T$ is given by the Nernst-Einstein equation \cite{vandervenRechargeableAlkaliIonBattery2020}:
    \begin{equation}
        \sigma (T) = \frac{\rho z^2 F^2}{RT}D_{\sigma}(T)
    \end{equation}
where $\rho$ is the molar density of diffusing ions in the unit cell, $z$, $F$ and $R$ are the charge of Li ions ($z=1$), the Faraday constant and the gas constant, respectively. Arrhenius plots were then generated to determine the temperature-dependent activation energies ($E_a$).

\section{Results}

\subsection{MTP Validation} 

\begin{table}[htbp]
  \centering
  \caption{Mean absolute errors (MAEs) on energies and forces predictions for fitted MTPs. The MAEs were calculated with respect to static energies and forces from the respective DFT functionals.}
   \makebox[\linewidth]{
    \begin{tabular}{clcccc}     \hline\hline
    \multicolumn{1}{c}{\multirow{2}[0]{*}{LSC}} & \multicolumn{1}{c}{\multirow{2}[0]{*}{MTP}} & \multicolumn{2}{c}{MAE$_{\mathrm{energies}}$ (meV/atom)} & \multicolumn{2}{c}{MAE$_{\mathrm{forces}}$ (eV/\AA)} \\\cline{3-6}
          &       & Training & Test & Training & Test \\\hline
    \multirow{2}[0]{*}{LLTO} & MTP$_{\mathrm{PBE, PBE}}$ & 1.40  & 1.44  & 0.12 & 0.12 \\
          & MTP$_{\mathrm{PBE, optB88}}$ & 1.39  & 1.24  & 0.10 & 0.10 \\\hline
    \multirow{2}[0]{*}{\ce{Li3YCl6}} & MTP$_{\mathrm{PBE, PBE}}$ & 0.96  & 1.11  & 0.04 & 0.04 \\
          & MTP$_{\mathrm{PBE, optB88}}$ & 1.00  & 1.06  & 0.04 & 0.04 \\\hline
    \multirow{3}[0]{*}{\ce{Li7P3S11}} & MTP$_{\mathrm{PBE, PBE}}$ & 1.77  & 1.92  & 0.09 & 0.08 \\
          & MTP$_{\mathrm{PBE, optB88}}$ & 1.70  & 1.78  & 0.08 & 0.08 \\
          & MTP$_{\mathrm{optB88, optB88}}$ & 1.79  & 2.07  & 0.09  & 0.09 \\\hline\hline
    \end{tabular}%
    }
  \label{Table: MLIAP evaluation on energies & forces}%
\end{table}%

Table \ref{Table: MLIAP evaluation on energies & forces} compares the mean absolute errors (MAEs) in energies and forces of the fitted MTPs. In all cases, the MAEs in energies are between 0.96 meV/atom and 2.07 meV/atom, while the MAEs in forces are below 0.20 eV/\AA. These MAEs are similar to or lower than those of other MTPs fitted in the literature \cite{zuoPerformanceCostAssessment2020b, wangLithiumIonConduction2020}, and a substantial improvement over traditional IAPs. The training and test MAEs are generally very similar, indicating that there is little likelihood of overfitting. The MAEs in energies and forces are also uniformly distributed with respect to the different temperatures that training structures were extracted from (see Figure S3-S5), indicating consistently high accuracy of our MTPs to reproduce DFT energies and forces at different temperatures. These results are consistent regardless of the DFT functionals (PBE or optB88) used to generate the training data. Further analysis also found that the local environments sampled by nanosecond NPT MD simulations using the fitted MTP are similar with those sampled by the AIMD training data (Figure S8) and the MAEs in forces are consistently low regardless of local environment (Figure S9). It should be noted that while it is possible that MD simulations under more extreme conditions, e.g., above 1200K, may sample local environments that are substantially different from the AIMD training data and possibly result in higher errors, such conditions are unlikely to be of interest for most applications of the fitted MTPs.

\subsection{Lattice Parameters}
\label{Section: structural relaxation}

\begin{table}[htbp]
  \centering
  \caption{Lattice parameters and densities of LSCs relaxed with the PBE and optB88 DFT functionals and the trained MTPs at 0K, in comparison with experimental lattice parameters and densities at room temperature for LLTO \cite{haradaLithiumIonConductivity1998}, \ce{Li3YCl6} \cite{asanoSolidHalideElectrolytes2018} and \ce{Li7P3S11} \cite{yamaneCrystalStructureSuperionic2007}. Values in brackets are the percentage differences between the computed values and the experimental measurements.}
  \makebox[\linewidth]{
  \begin{tabular}{ccccc}\hline\hline
          & a (\AA) & b (\AA) & c (\AA) & Density (g cm$^{-3}$) \\\hline
    \multicolumn{5}{l}{LLTO} \\
    DFT PBE & 3.96 (2.3\%) & 3.89 (0.5\%) & 3.91 (1.0\%) & 4.84 (-3.0\%) \\
    DFT optB88 & 3.95 (2.1\%) & 3.88 (0.3\%) & 3.90 (0.8\%) & 4.88 (-2.2\%) \\
    MTP$_{\mathrm{PBE, PBE}}$ & 3.96 (2.3\%) & 3.89 (0.5\%) & 3.90 (0.8\%) & 4.85 (-2.8\%)\\
    MTP$_{\mathrm{PBE, optB88}}$ & 3.95 (2.1\%) & 3.87 (0.0\%) & 3.89 (0.5\%) & 4.90 (-1.8\%) \\
    Experiment & 3.87  & 3.87  & 3.87  & 4.99 \\\hline
    \multicolumn{5}{l}{\ce{Li3YCl6}} \\
    DFT PBE & 11.17 (-0.3\%) & 11.17 (-0.3\%) & 6.22 (3.2\%) & 2.37 (-3.3\%) \\
    DFT optB88 & 11.01 (-1.7\%) & 11.01 (-1.7\%) & 6.02 (-0.2\%) & 2.52 (2.9\%) \\
    MTP$_{\mathrm{PBE, PBE}}$ & 11.18 (-0.2\%) & 11.18 (-0.2\%) & 6.27 (4.0\%) & 2.35 (-4.1\%)\\
    MTP$_{\mathrm{PBE, optB88}}$ & 11.04 (-1.4\%) & 11.04 (-1.4\%) & 6.08 (0.8\%) & 2.48 (1.2\%) \\
    Experiment & 11.20 & 11.20 & 6.03 & 2.45 \\\hline
    \multicolumn{5}{l}{\ce{Li7P3S11}} \\
    DFT PBE & 12.86 (2.9\%) & 6.19 (2.7\%) & 12.69 (1.3\%) & 1.87 (-5.6\%)\\
    DFT optB88 & 12.61 (0.9\%) & 6.08 (0.8\%) & 12.62 (0.7\%) & 1.95 (-1.5\%)\\
    MTP$_{\mathrm{PBE, PBE}}$ & 12.66 (1.3\%) & 6.33 (5.0\%) & 12.56 (0.2\%) & 1.88 (-5.1\%)\\
    MTP$_{\mathrm{PBE, optB88}}$ & 12.52 (0.2\%) & 6.14 (1.8\%) & 12.57 (0.3\%) & 1.95 (-1.5\%)\\
    MTP$_{\mathrm{optB88, optB88}}$ & 12.52 (0.2\%) & 6.12 (1.5\%) & 12.66 (1.0\%) & 1.96 (-1.0\%)\\
    Experiment & 12.50 & 6.03  & 12.53 & 1.98 \\\hline\hline
    \end{tabular}}
    \label{Table:structural relaxation}
\end{table}%

\begin{figure}[htbp]
    \centering
    \begin{subfigure}[b]{0.3\textwidth}
        \centering
	    \includegraphics[height=3.1cm]{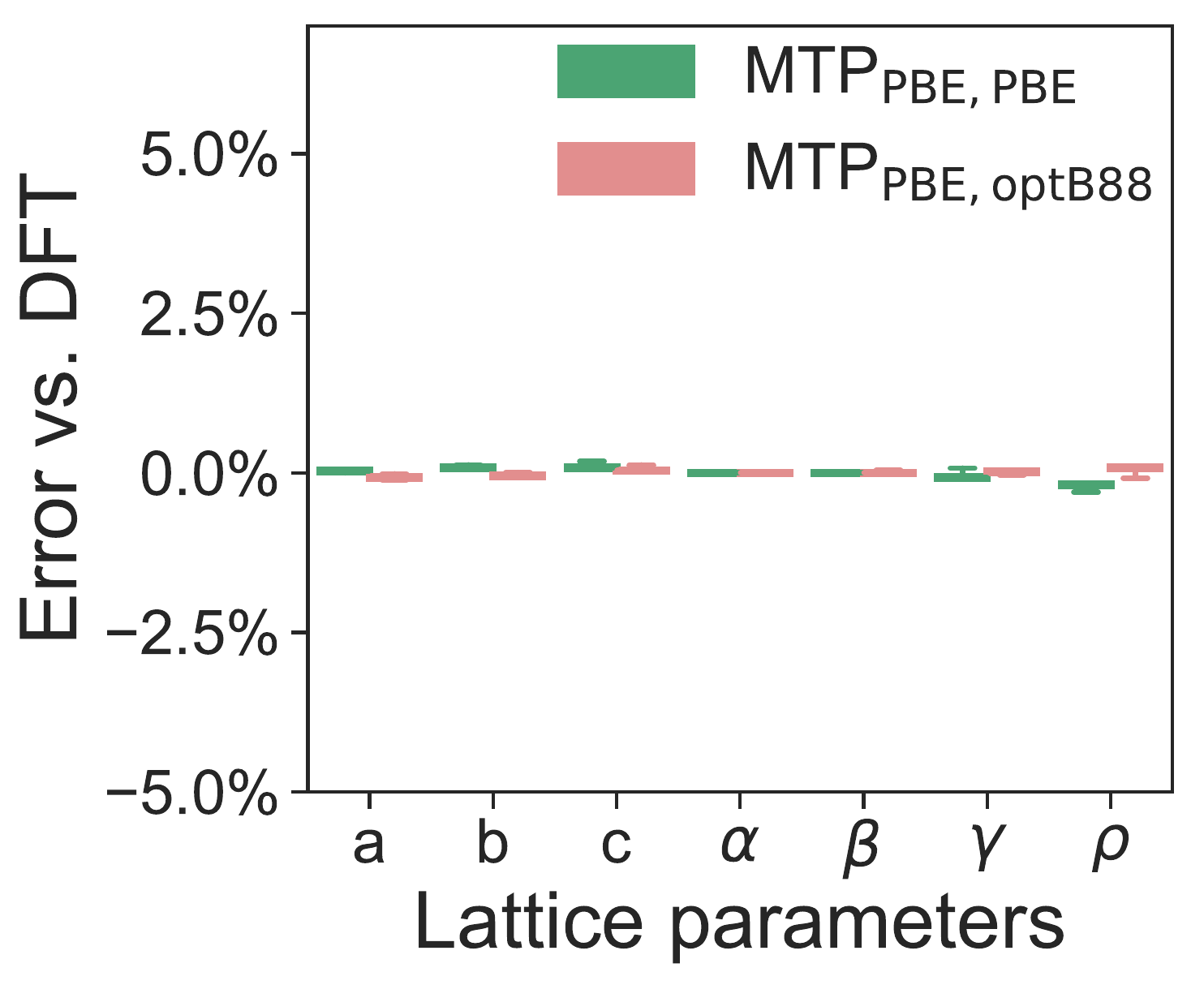}
    	\caption{\label{subfig:Lattice_LLTO}LLTO}
    \end{subfigure}
    \begin{subfigure}[b]{0.3\textwidth}
        \centering
	    \includegraphics[height=3.1cm]{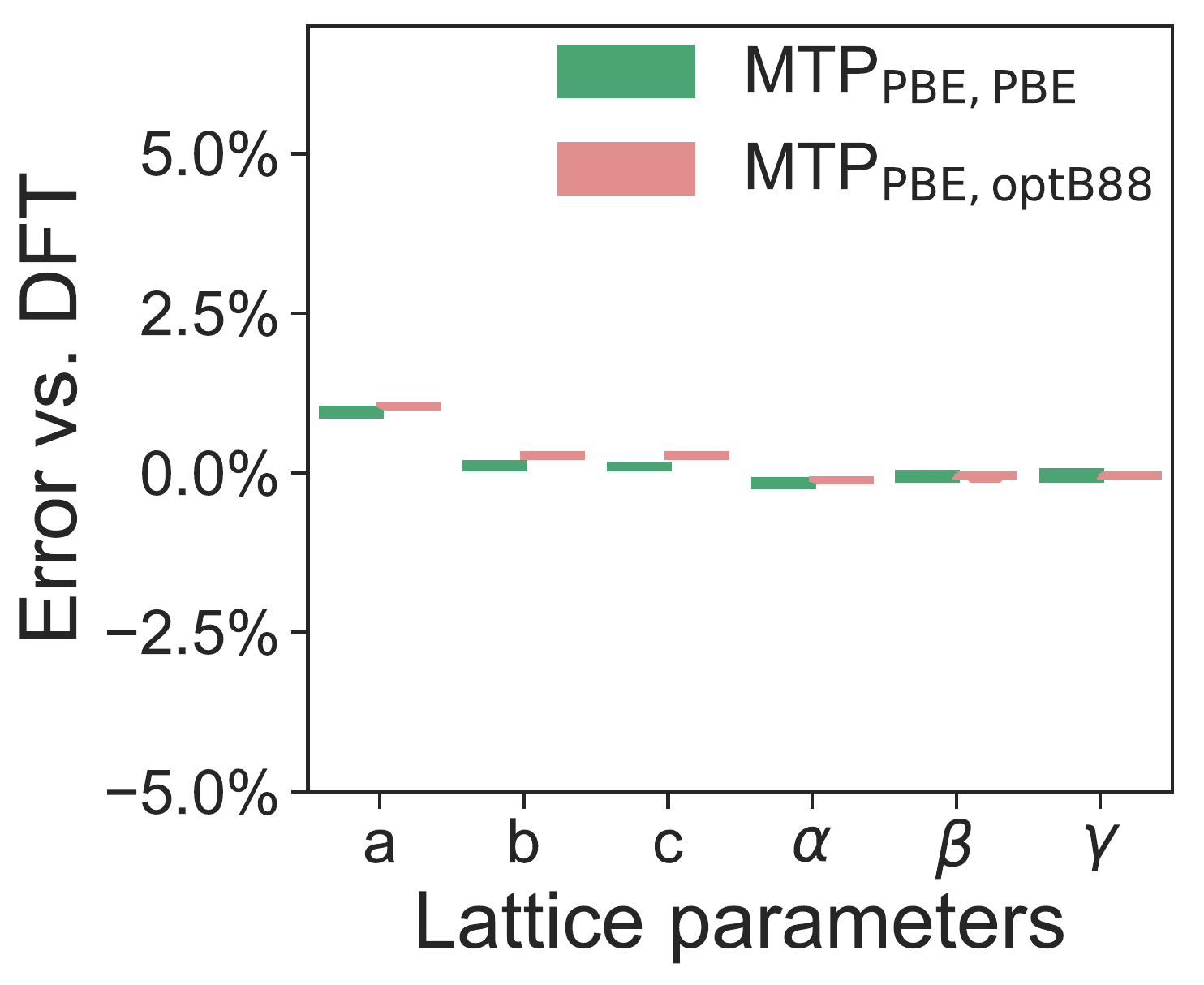}
    	\caption{\label{subfig:Lattice_LYC}\ce{Li3YCl6}}
    \end{subfigure}
    \begin{subfigure}[b]{0.3\textwidth}
        \centering
	    \includegraphics[height=3.1cm]{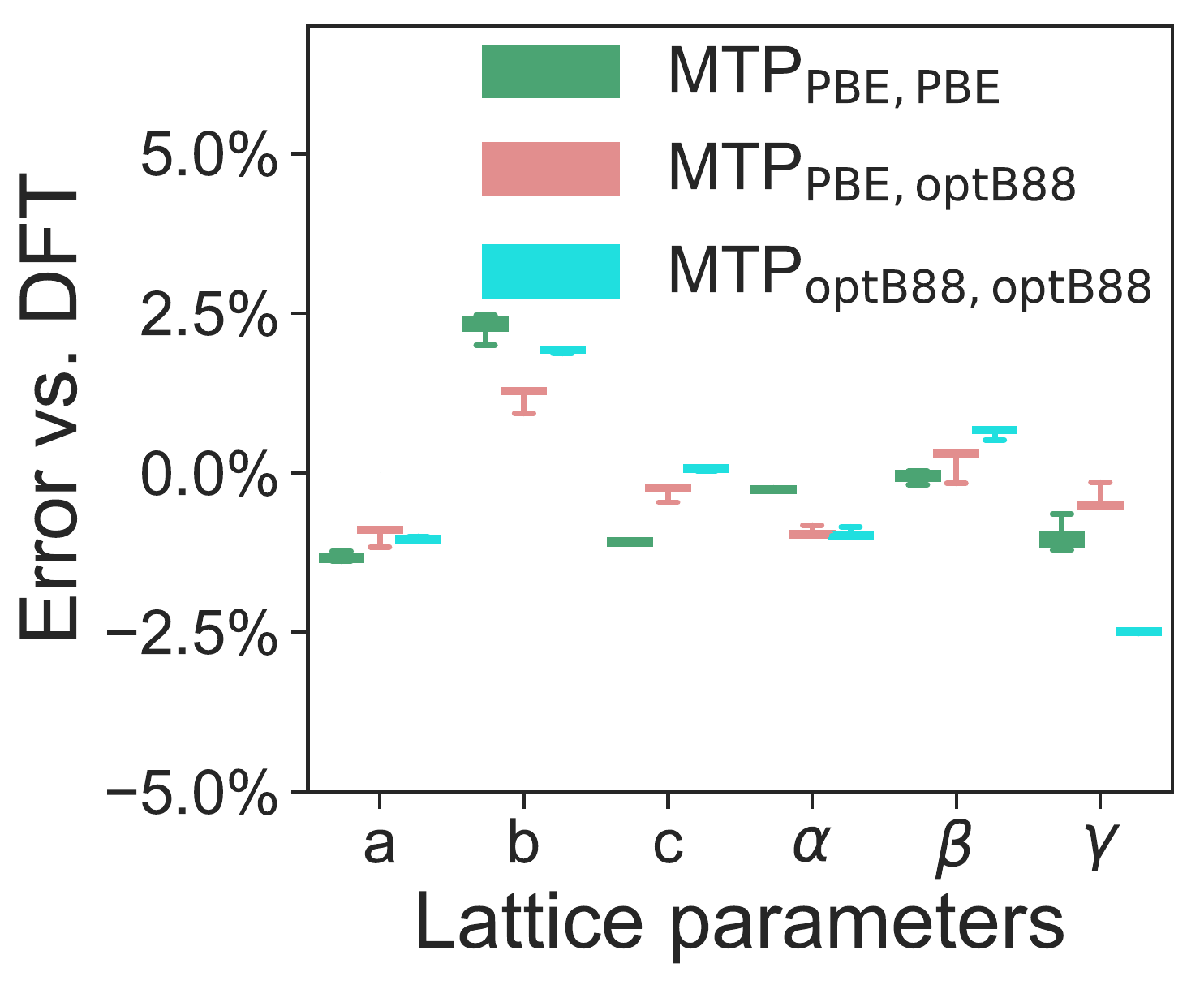}
    	\caption{\label{subfig:Lattice_Li7P3S11}\ce{Li7P3S11}}
    \end{subfigure}
    \caption{\label{fig:MTP lattice}Comparison of the lattice parameters predicted via relaxation with the trained MTPs and the DFT-relaxed values. For each MTP, the relaxation was performed using 36 strained structures constructed by applying strains of -0.15 to 0.15 with 0.05 intervals in six different modes to the DFT relaxed ground-state structure to assess the numerical stability of MTP relaxation.
  }
\end{figure}

Table~\ref{Table:structural relaxation} compares the lattice parameters and volumes for the three LSCs from DFT and MTP relaxations with experimental values. It can be seen that the use of the optB88 functional significantly improves the predicted lattice parameters and densities over the PBE functional for \ce{Li3YCl6} and \ce{Li7P3S11} LSCs, while yielding smaller improvements for LLTO LSC. The PBE computed densities underestimate the experimental densities by 3-6\%, consistent with the well-known propensity of PBE to underbind. The optB88 computed densities are within 1-3\% of the experimental densities due to the fact that optB88 functional is less repulsive at short interatomic separations \cite{klimesVanWaalsDensity2011}. As shown in Figure \ref{fig:MTP lattice}, the MTPs are generally able to reproduce the DFT lattice parameters to within 1-2.5\%. It should be noted that the errors with respect to DFT follows the order LLTO $<$ \ce{Li3YCl6} $\sim$ \ce{Li7P3S11}. We hypothesize that this can be attributed to the difference in the potential energy landscapes, \textit{i.e.,} \ce{Li3YCl6} and \ce{Li7P3S11} have shallower potential energy landscapes, which leads to smaller energy changes with lattice parameter variations. This can be seen to some degree in the equation of state plots (see later Figure \ref{fig:EOSQHA}), as the same percentage of change in volume leads to less change in the total energies of
\ce{Li3YCl6} and \ce{Li7P3S11}.

\subsection{Equations of State and Thermal Expansion}
\label{Section: EOS}
\begin{figure}[H]
    % \label{Figure:EOS}
    \centering
    \begin{subfigure}[b]{0.3\textwidth}
        \centering
	    \includegraphics[height=3.5cm]{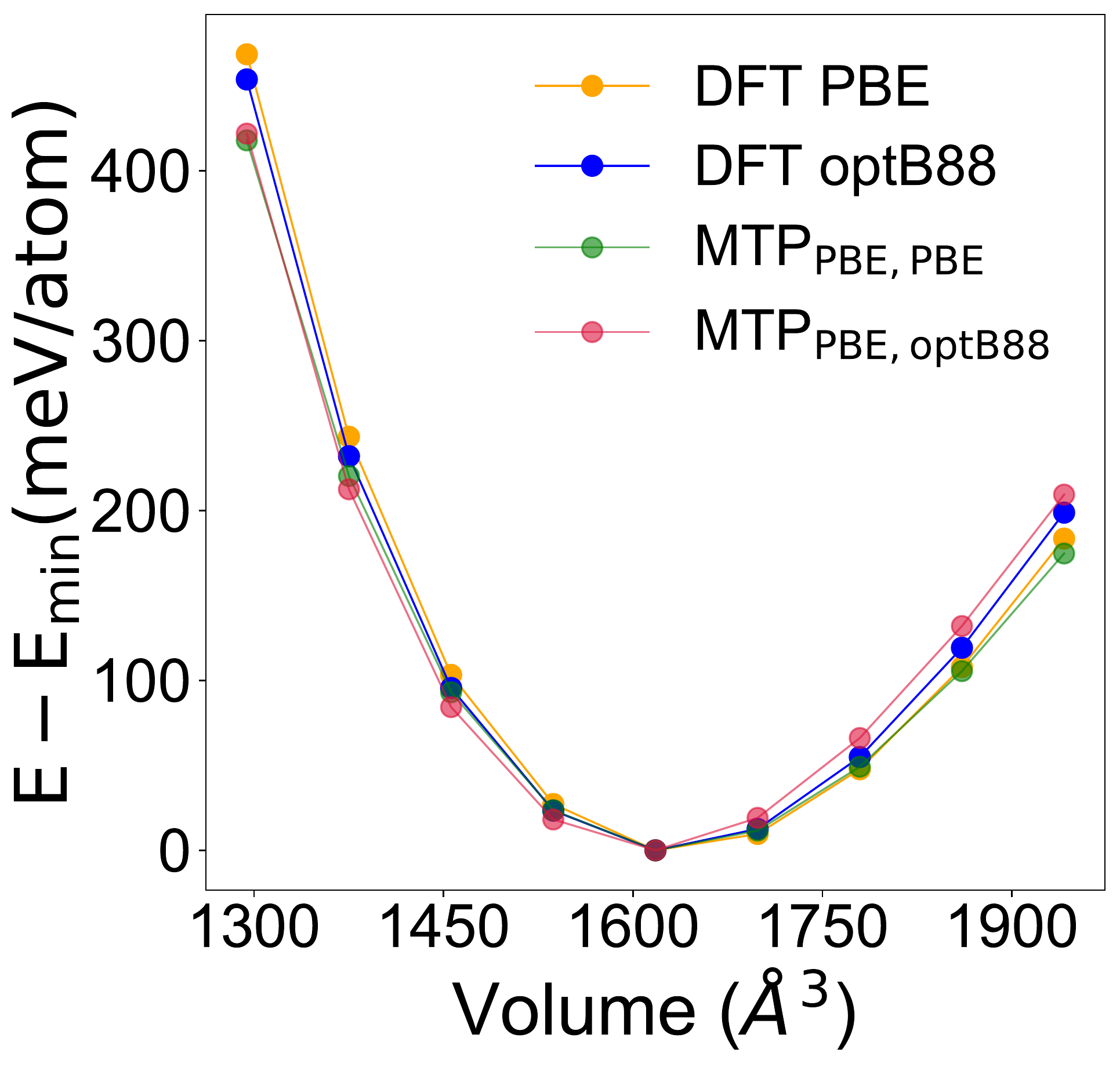}
    	\caption{\label{subfig:EOS_LLTO}EOS of LLTO}
    \end{subfigure}
    \begin{subfigure}[b]{0.3\textwidth}
        \centering
	    \includegraphics[height=3.5cm]{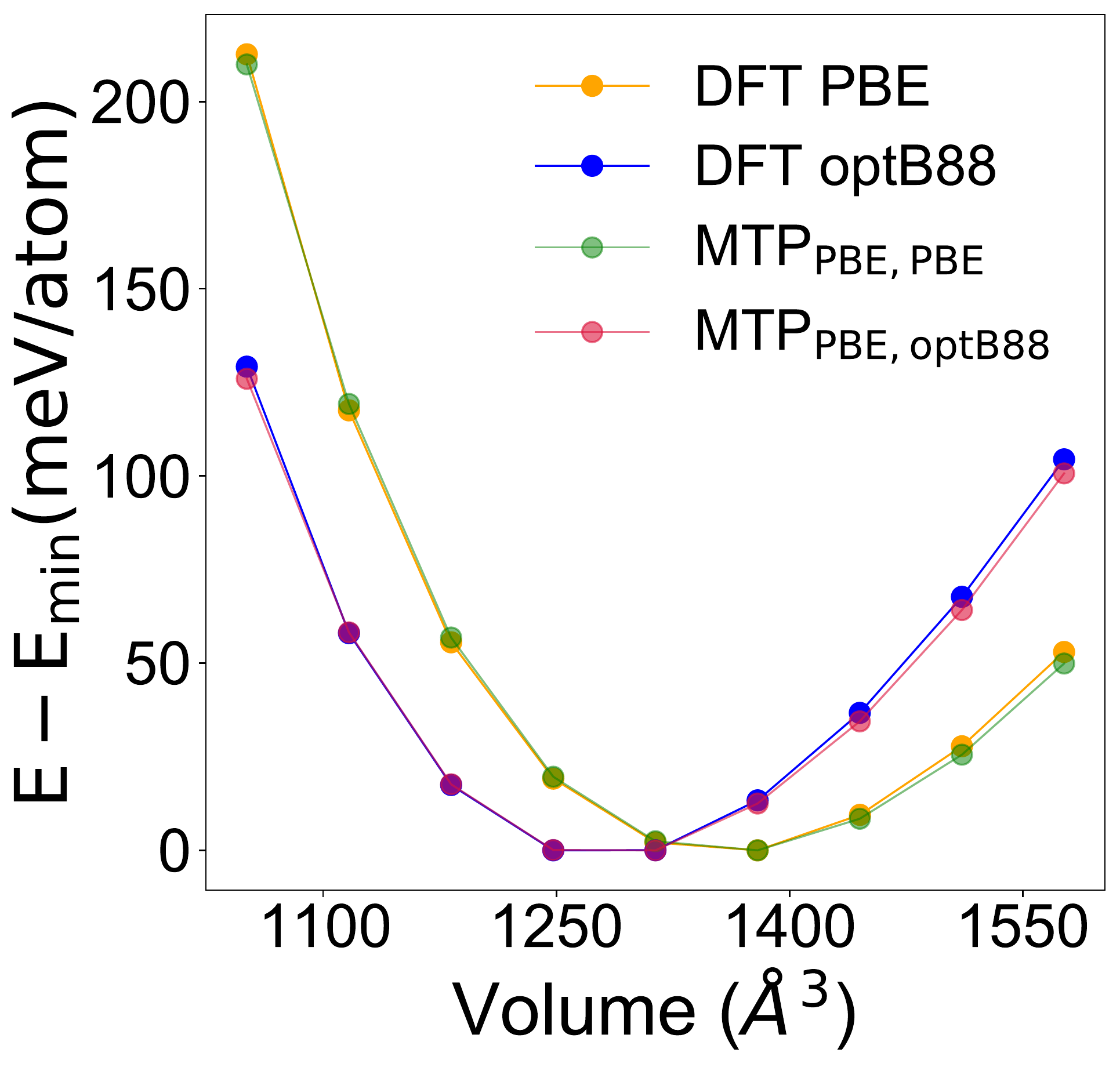}
    	\caption{\label{subfig:EOS_LYC}EOS of \ce{Li3YCl6}}
    \end{subfigure}
    \begin{subfigure}[b]{0.3\textwidth}
        \centering
	    \includegraphics[height=3.5cm]{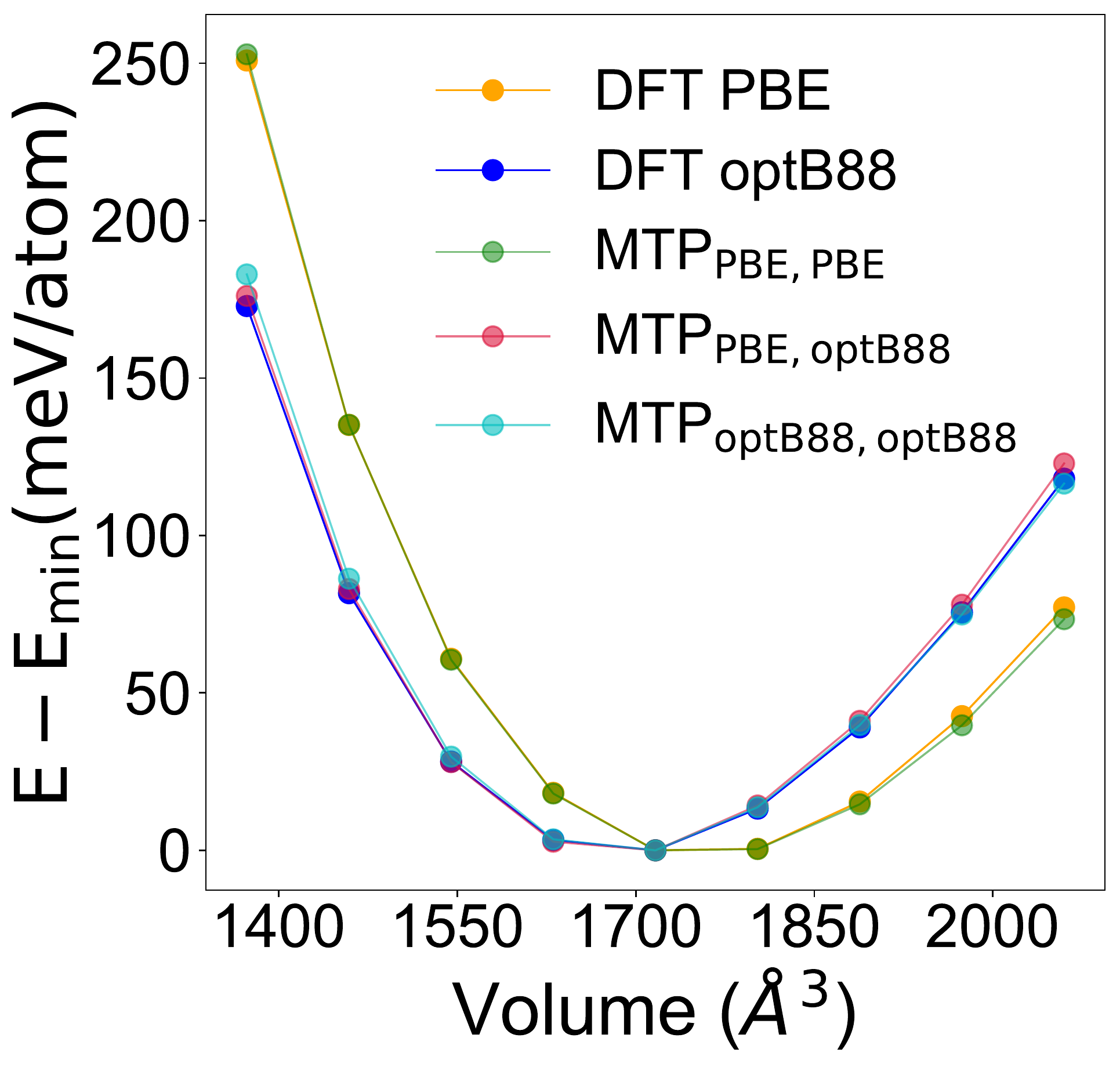}
    	\caption{\label{subfig:EOS_Li7P3S11}EOS of \ce{Li7P3S11}}
    \end{subfigure}
    \begin{subfigure}[b]{0.3\textwidth}
        \centering
	    \includegraphics[height=3.5cm]{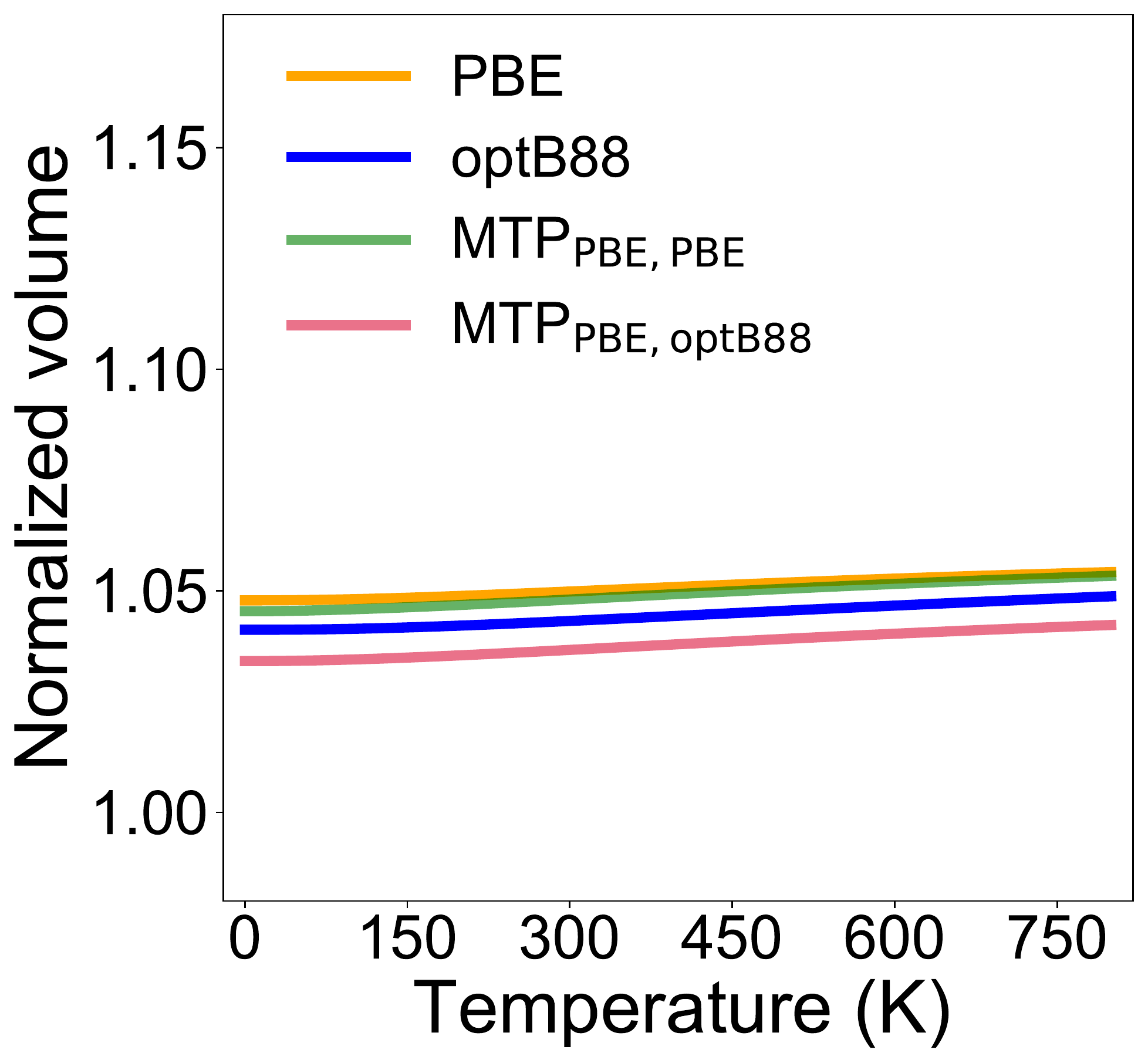}
    	\caption{\label{subfig:QHA_LLTO}$\beta$ of LLTO}
    \end{subfigure}
    \begin{subfigure}[b]{0.3\textwidth}
        \centering
	    \includegraphics[height=3.5cm]{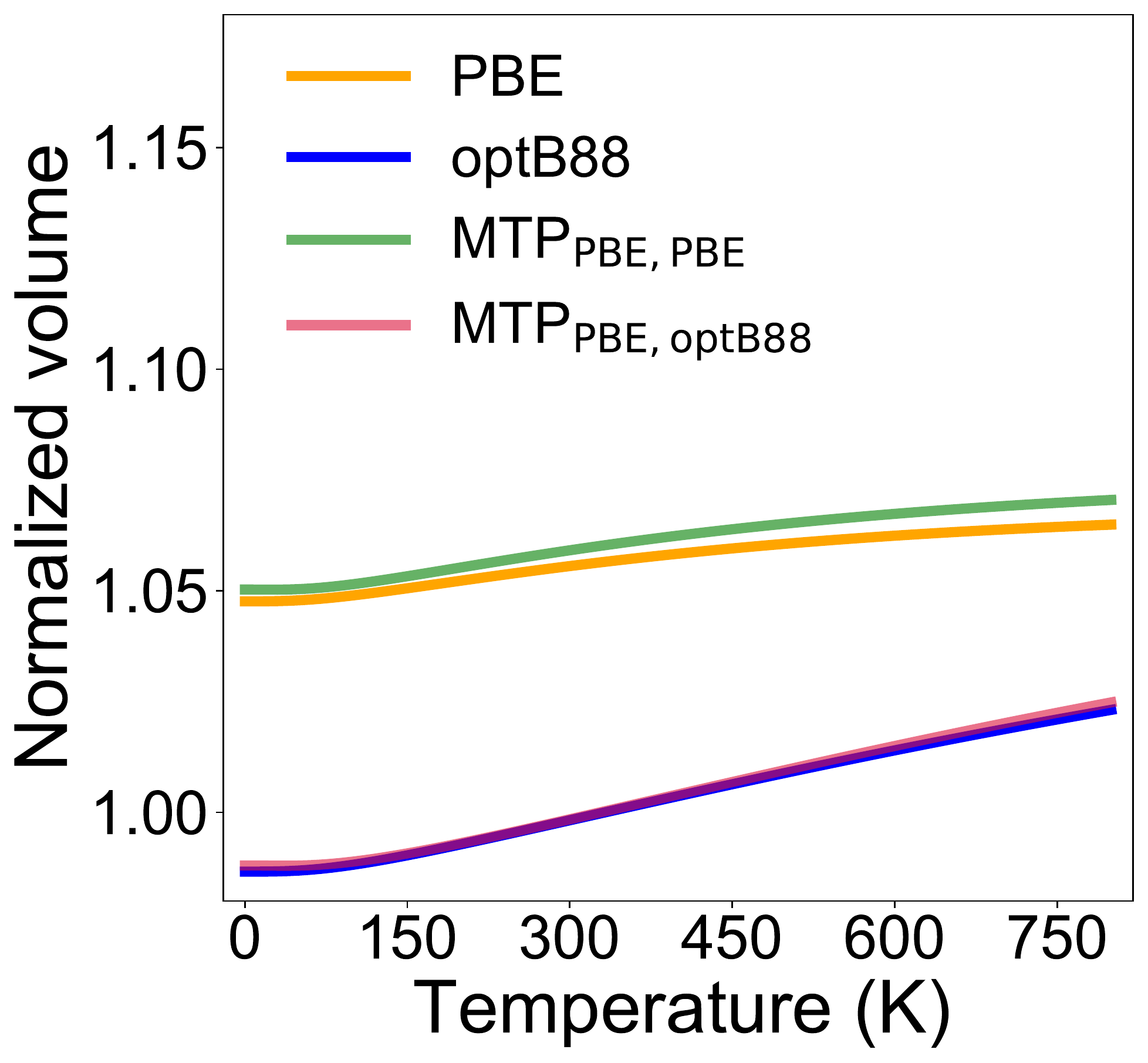}
    	\caption{\label{subfig:QHA_LYC}$\beta$ of \ce{Li3YCl6}}
    \end{subfigure}
    \begin{subfigure}[b]{0.3\textwidth}
        \centering
	    \includegraphics[height=3.5cm]{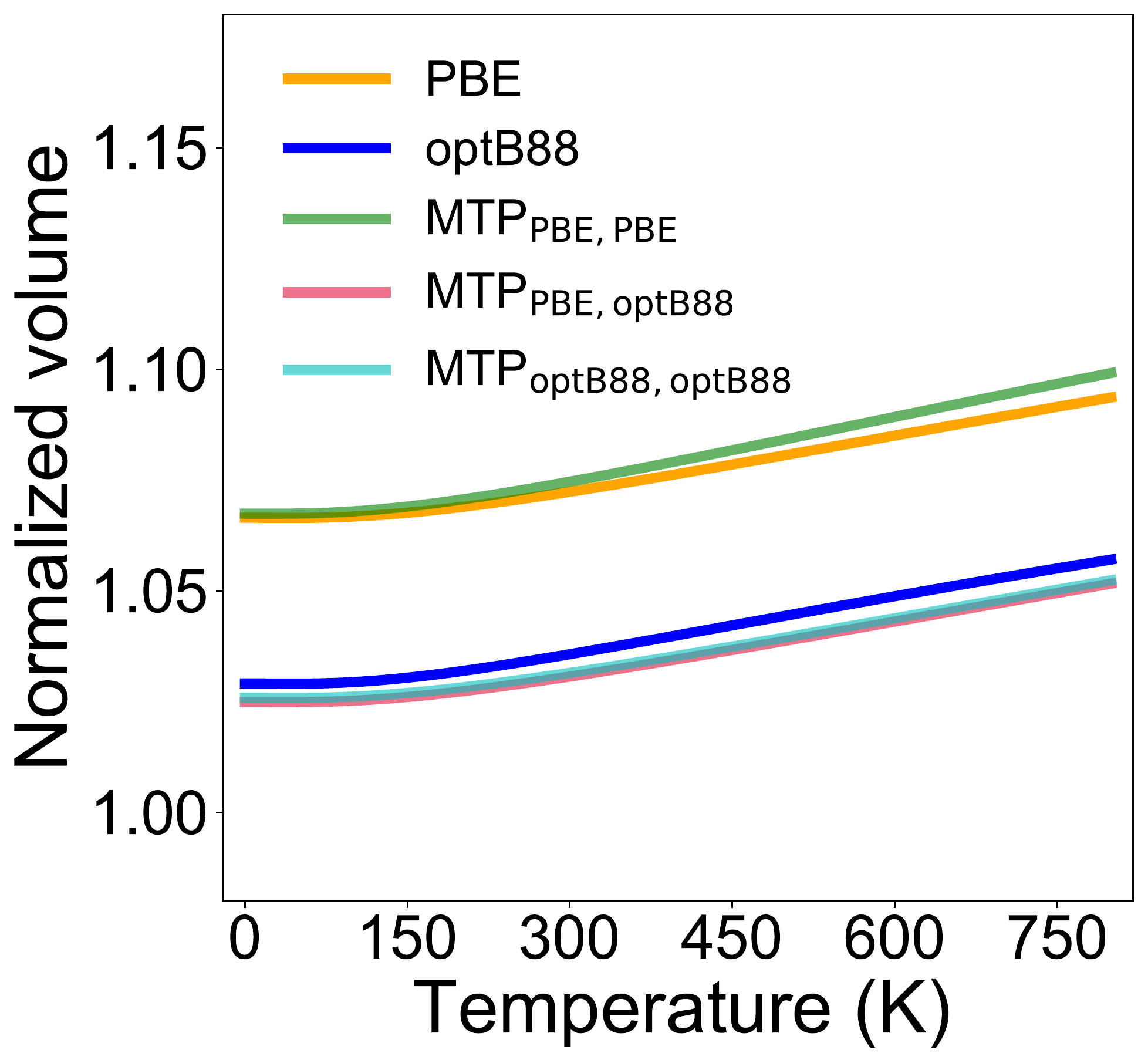}
    	\caption{\label{subfig:QHA_Li7P3S11}$\beta$ of \ce{Li7P3S11}}
    \end{subfigure}
    \caption{\label{fig:EOSQHA}Equations of state at 0K (a, b, c) and QHA thermal volume expansion $\beta$ (d, e, f) of LLTO, \ce{Li3YCl6} and \ce{Li7P3S11} calculated from DFT and MTPs. The volumes in the $\beta$ plots (d, e, f) are normalized with respect to the experimentally measured volumes at room temperatures \cite{haradaLithiumIonConductivity1998, asanoSolidHalideElectrolytes2018, yamaneCrystalStructureSuperionic2007}.}
\end{figure}

Figure \ref{fig:EOSQHA}a-c shows the equation of state (EOS) curves of the three LSCs calculated from DFT and MTPs. In general, the MTP computed EOSs agree well with the corresponding DFT EOSs, further attesting to the robustness of the MTP fitting procedure. In addition to the differences in equilibrium volumes discussed in the preceding section, we note that the optB88 DFT and MTP calculations predict a larger curvature for \ce{Li3YCl6} and \ce{Li7P3S11} LSCs, i.e., the optB88 functional predicts a higher bulk modulus than PBE. These results are consistent with the calculated QHA thermal volume expansion, plotted in Figure~\ref{fig:EOSQHA}d-f. The MTP QHA thermal expansion curves match closely with the corresponding DFT QHA thermal expansion curves. In general, the optB88 DFT and MTP volumes are much closer to the experimental volumes for respective temperatures between 0 and 800K. LLTO has the smallest MTP$_{\mathrm{PBE,optB88}}$ volume expansion coefficient ($\beta_{\mathrm{PBE,optB88}}$) of $1.08\times10^{-5}$ K$^{-1}$ from 300K to 800K, in excellent agreement with the reported low thermal expansion coefficient of $9.35\times10^{-6}$ K$^{-1}$ from X-ray diffraction analysis from 298K to 800K \cite{okumuraComputationalSimulationsLi2006}. \ce{Li3YCl6} and \ce{Li7P3S11} have much higher predicted $\beta_{\mathrm{PBE,optB88}}$ of $5.28\times10^{-5}$ K$^{-1}$ and $4.07\times10^{-5}$ K$^{-1}$, respectively. A slightly higher volume expansion is predicted for \ce{Li3YCl6} by optB88 compared to PBE. 

\subsection{Ionic Conductivity}
\label{Section: Arrhenius plot}
\begin{figure}[H]
    \centering
	    \includegraphics[height=6cm]{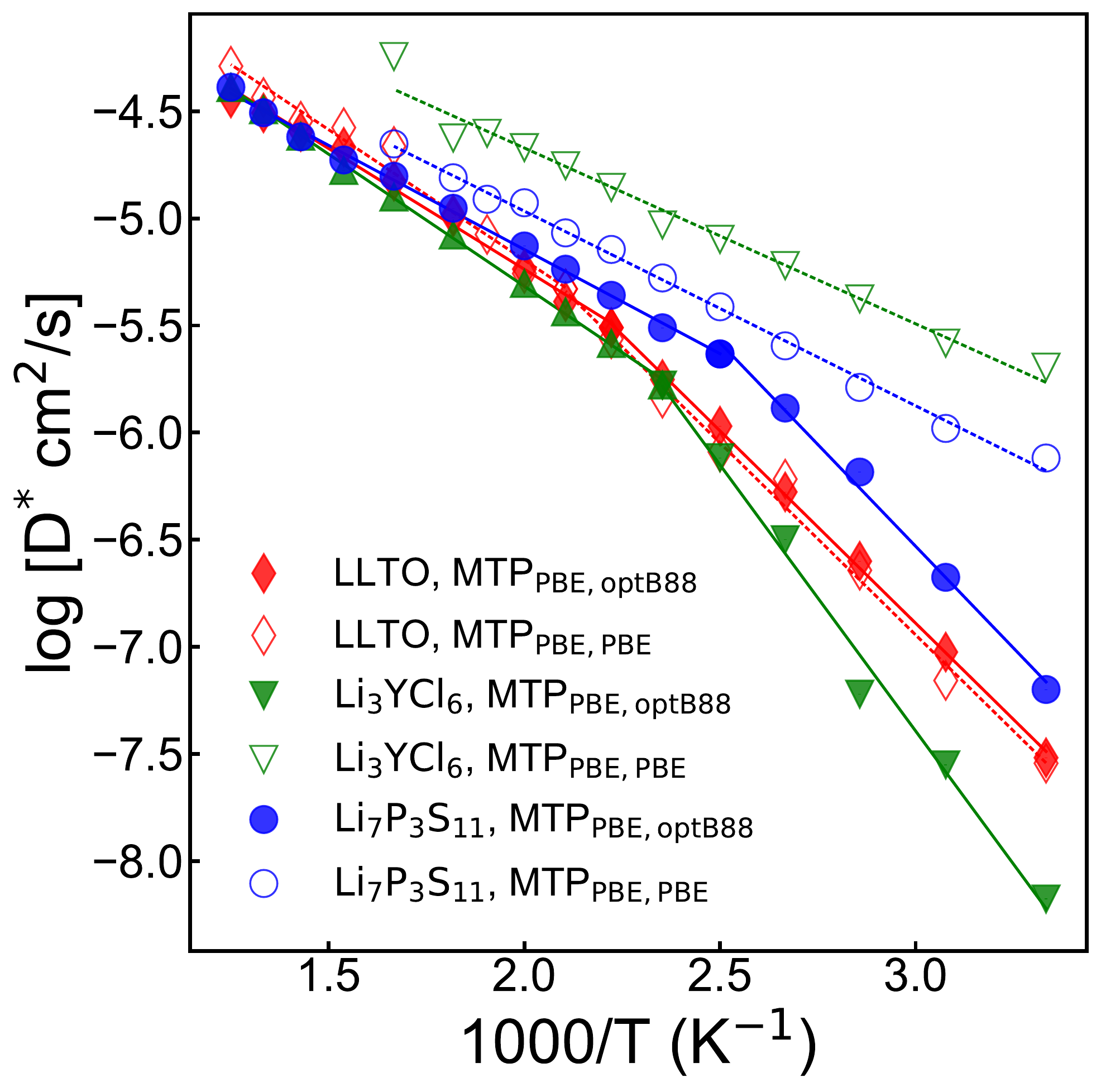}
    \footnotesize{
\begin{tabular}{ccccc}\hline\hline
          & AIMD  & MTP$_{\mathrm{PBE, PBE}}$ & MTP$_{\mathrm{PBE, optB88}}$ & Experiment \\  \hline
    \multicolumn{5}{l}{LLTO} \\
    \multirow{2}[0]{*}{$E_a (eV)$} & 0.26     & 0.36 (300-475K) & 0.36 (300-450K) & 0.40 (300K) \cite{inagumaHighIonicConductivity1993a} \\
          & (900-1800K) \cite{heOriginFastIon2017}     & 0.24 (475-800K) & 0.22 (450-800K) & 0.22 (600K) \cite{salkusDeterminationNonArrheniusBehaviour2011} \\
    $\sigma_{300K}$ (mS cm$^{-1}$) & -     & 0.97 ($\sigma^{*}$)  & 1.04 ($\sigma^{*}$), 1.53 ($\sigma$)  & 1-1.3 \cite{inagumaHighIonicConductivity1993a,haradaLithiumIonConductivity1998} \\\hline
    \multicolumn{5}{l}{\ce{Li3YCl6}}\\
    \multirow{2}[0]{*}{$E_a (eV)$} & 0.19 & \multirow{2}[0]{*}{0.16 (300-600K)} & 0.49 (300-425K) & \multirow{2}[0]{*}{0.40 (230-360K) \cite{asanoSolidHalideElectrolytes2018}} \\
          &     (500-900K) \cite{wangLithiumChloridesBromides2019}  &       & 0.24 (425-800K) &  \\
    $\sigma_{300 K}$ (mS cm$^{-1}$) & 14 \cite{wangLithiumChloridesBromides2019}    & 152.66 ($\sigma^{*}$) & 0.56 ($\sigma^{*}$), 1.64 ($\sigma$)  & 0.51 \cite{asanoSolidHalideElectrolytes2018} \\\hline
    \multicolumn{5}{l}{\ce{Li7P3S11}} \\
    \multirow{2}[0]{*}{$E_a (eV)$} & 0.19 & \multirow{2}[0]{*}{0.18 (300-600K)} & 0.38 (300-400K) & 0.29 (298-373K) \cite{buscheSituMonitoringFast2016} \\
          &  (400-1200K) \cite{chuInsightsPerformanceLimits2016a}     &       & 0.19 (400-800K) & 0.18 (300-600K) \cite{chuInsightsPerformanceLimits2016a} \\
    $\sigma_{300K}$ (mS cm$^{-1}$) & 57 \cite{chuInsightsPerformanceLimits2016a}    & 67.37 ($\sigma^{*}$) & 6.50 ($\sigma^{*}$), 7.51 ($\sigma$)  & 4-17 \cite{wenzelInterphaseFormationDegradation2016a,buscheSituMonitoringFast2016,chuInsightsPerformanceLimits2016a,seinoSulphideLithiumSuper2014a} \\
    \hline\hline
    \end{tabular}
    }
    \caption{\label{fig:Arrhenius}(Top) Arrhenius plot for LLTO, \ce{Li3YCl6} and \ce{Li7P3S11} from NPT/MD simulations using MTP$_{\mathrm{PBE, PBE}}$ (open markers) and MTP$_{\mathrm{PBE, optB88}}$ (solid markers). The diffusivities were obtained by averaging the mean square displacements from five independent MD simulations at each temperature for at least 1 ns. (Bottom) Room temperature ionic conductivities ($\sigma_{300K}$) and activation energies ($E_a$) for LLTO, \ce{Li3YCl6} and \ce{Li7P3S11}, obtained from MD simulations using MTP$_{\mathrm{PBE, PBE}}$ and MTP$_{\mathrm{PBE, optB88}}$. For MTP$_{\mathrm{PBE, optB88}}$, transitions between two quasi-linear Arrhenius regimes were observed for all three LSCs and the $E_a$ for each regime are reported separately. Available AIMD derived values as well as experimental references are also listed.}
\end{figure}

Figure \ref{fig:Arrhenius} shows the Arrhenius plots for the three LSCs from MD simulations performed using the MTPs and a summary of the derived activation energies ($E_a$) and conductivities at room temperature ($\sigma_{300K}$) in comparison with experiments and previous AIMD simulations. From the MTP$_{\mathrm{PBE, optB88}}$ Arrhenius plots (filled markers and solid lines), it is immediately apparent that all three LSCs do not exhibit a single linear Arrhenius regime, which is the common assumption made when extrapolating high-temperature (HT) AIMD simulations to room temperature. Transitions between a HT quasi-linear regime with lower $E_a$ and a low-temperature (LT) quasi-linear regime with higher $E_a$ occur at $\sim$ 450K, 425K and 400K for LLTO, \ce{Li3YCl6} and \ce{Li7P3S11}, respectively. In all cases, the MTP$_{\mathrm{PBE,optB88}}$ predicted $\sigma_{300K}$ and $E_a$ are in remarkably good agreement with previously reported experimental values for all three LSCs. In particular, while previous HT AIMD simulations predicted an extraordinarily high $\sigma^{*}_{300K}$ of 57 mS cm$^{-1}$ for \ce{Li7P3S11}, the MTP$_{\mathrm{PBE,optB88}}$ predicted $\sigma^{*}_{300K}$ and $\sigma_{300K}$ are only 6.50 and 7.51 mS cm$^{-1}$, much closer to the 4-17 mS cm$^{-1}$ that have been reported experimentally thus far \cite{wenzelInterphaseFormationDegradation2016a,buscheSituMonitoringFast2016,chuInsightsPerformanceLimits2016a,seinoSulphideLithiumSuper2014a}. We further note that the MTP$_{\mathrm{PBE,optB88}}$ HT $E_a$ are also consistent with those obtained from previous HT AIMD simulations for \ce{Li3YCl6} \cite{wangLithiumChloridesBromides2019} and \ce{Li7P3S11} \cite{chuInsightsPerformanceLimits2016a}. The \ce{Li3YCl6} LSC has only been experimentally studied at 230-360K. Our MTP$_{\mathrm{PBE,optB88}}$ MD simulations predict that \ce{Li3YCl6} would undergo a transition to a lower $E_a$ regime at around 425K; a prediction that would need to be verified by further experiments from the community.

The Haven ratios ($H_R$) at 300K are 0.68, 0.34 and 0.87 for LLTO, \ce{Li3YCl6} and \ce{Li7P3S11}, respectively. There are no existing reports of the Haven ratios for these LSCs to the authors' knowledge. However, LLTO has been experimentally reported to possess highly correlated motions at room temperature \cite{bohnkeFastLithiumionConducting2008}, consistent with the computed $H_R$ of 0.68. Further, our calculated $H_R$ for \ce{Li7P3S11} also lies in between the AIMD simulated $H_R$ (0.42, 0.53) \cite{marcolongoIonicCorrelationsFailure2017} and NMR measured $H_R$ (in the order of 1) \cite{kuhnTetragonalLi10GeP2S12Li7GePS82013} of \ce{Li10GeP2S12}, another sulfide LSC.  

We also note that the MTP$_{\mathrm{PBE, optB88}}$ and MTP$_{\mathrm{PBE, PBE}}$ yield fundamentally different results for \ce{Li3YCl6} and \ce{Li7P3S11}. In both cases, the MTP$_{\mathrm{PBE, PBE}}$ do not predict any transitions between quasi-linear Arrhenius regimes in the simulation temperature range of 300-600K. The activation energies $E_a$ and room temperature ionic conductivities $\sigma_{300K}$ are also severely underestimated and overestimated, respectively, compared to experiments, similar to prior AIMD simulations using the PBE functional. The poor performance of the PBE-based MTP and AIMD simulations can be traced to the substantial overestimation of the lattice parameters by the PBE functional, which can lead to lower activation barriers and higher ionic conductivities \cite{ongPhaseStabilityElectrochemical2013, kuhnNewUltrafastSuperionic2014}. It should be noted that we cannot rule out the possibility that the observed transitions between quasi-Arrhenius regimes are an artifact of the DFT functional used in generating the training data, i.e., optB88, but the generally improved agreement between the predicted room-temperature ionic conductivities and experimental values suggest that the transitions are a real phenomenon. It is our hope that future detailed experiments may shed further light on these predictions.

\subsection{Transitions in Diffusion Mechanisms}

\begin{figure}[htbp]
    \centering
    \begin{subfigure}[b]{0.3\textwidth}
        \centering
	    \includegraphics[height=4cm]{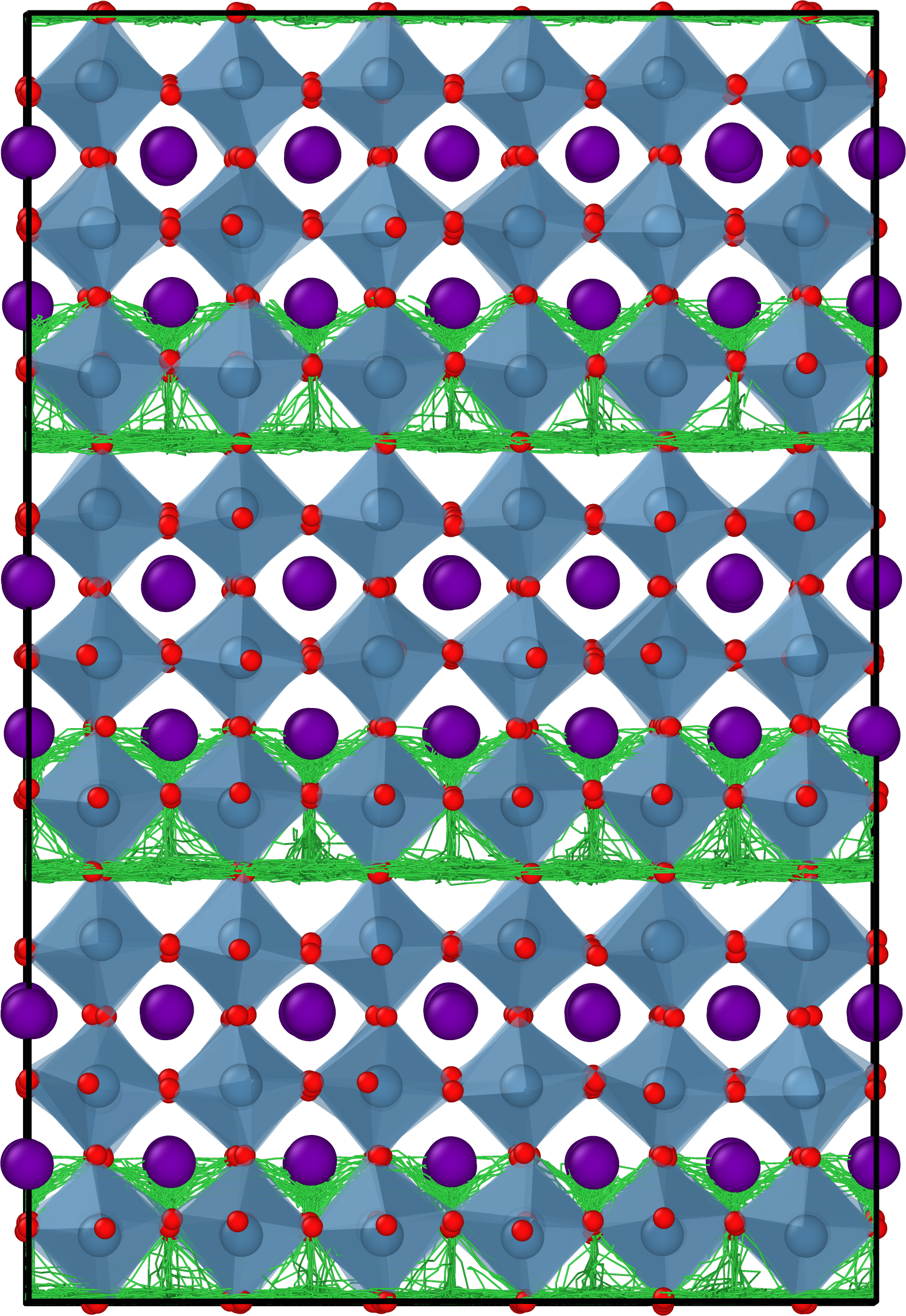}
    	\caption{\label{subfig:traj_LLTO_500K}LLTO at 500K}
    \end{subfigure}
    \begin{subfigure}[b]{0.3\textwidth}
        \centering
        \includegraphics[height=4cm]{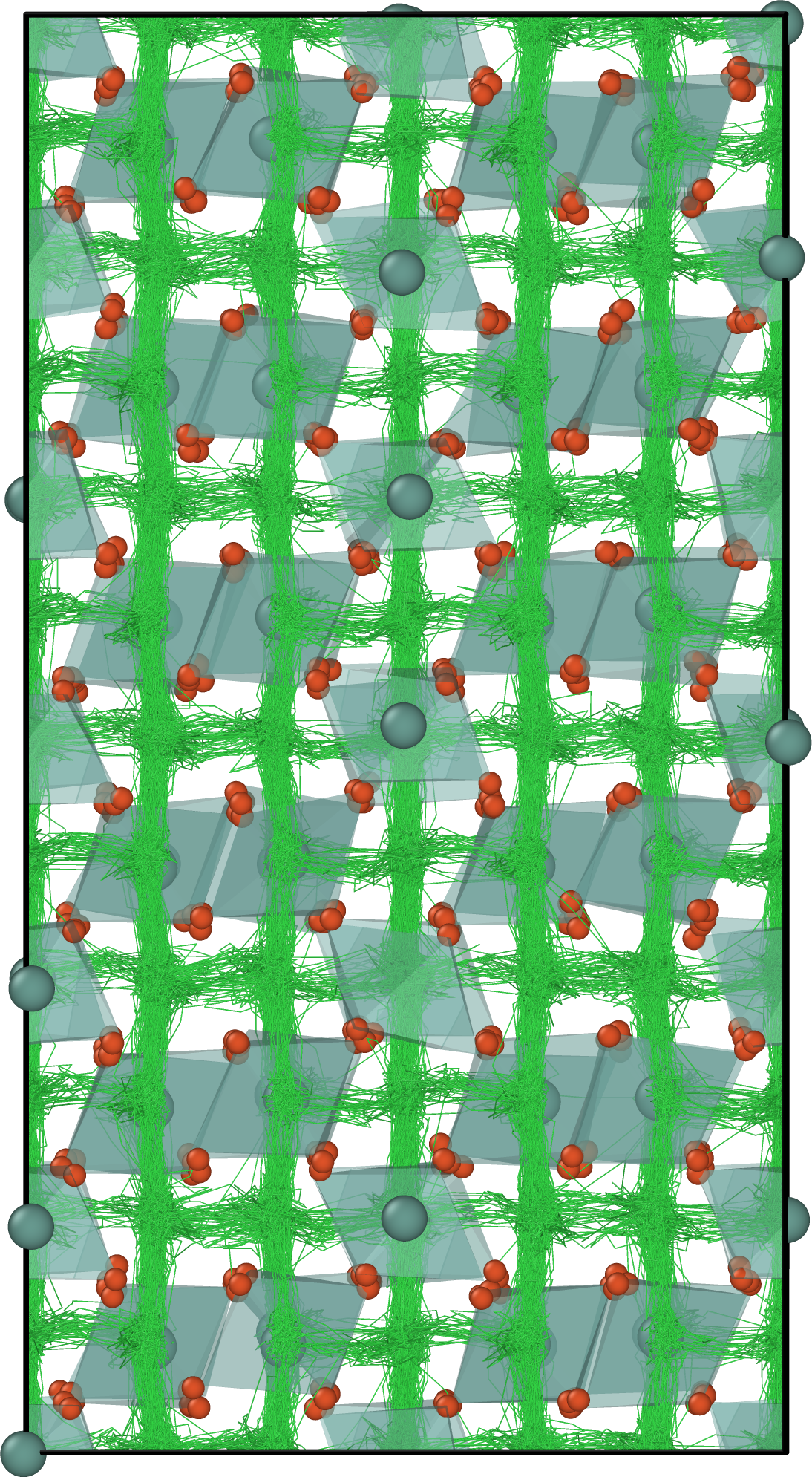}
    	\caption{\label{subfig:traj_LYC_500K}\ce{Li3YCl6} at 500K}
    \end{subfigure}
    \begin{subfigure}[b]{0.3\textwidth}
        \centering
        \includegraphics[height=4cm]{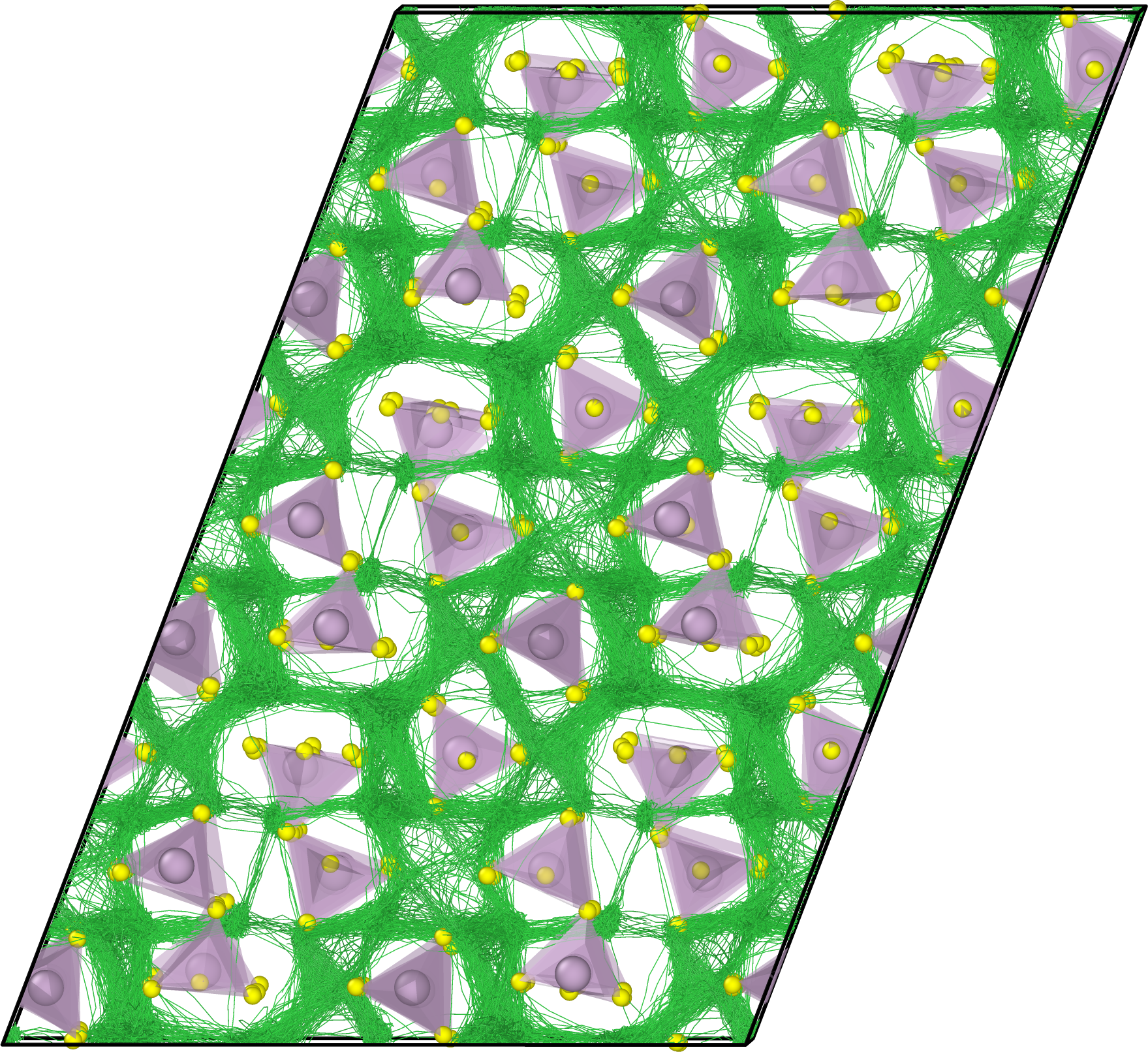}
    	\caption{\label{subfig:traj_Li7P3S11_500K}\ce{Li7P3S11} at 500K}
    \end{subfigure}
    \begin{subfigure}[b]{0.3\textwidth}
        \centering
	    \includegraphics[height=4cm]{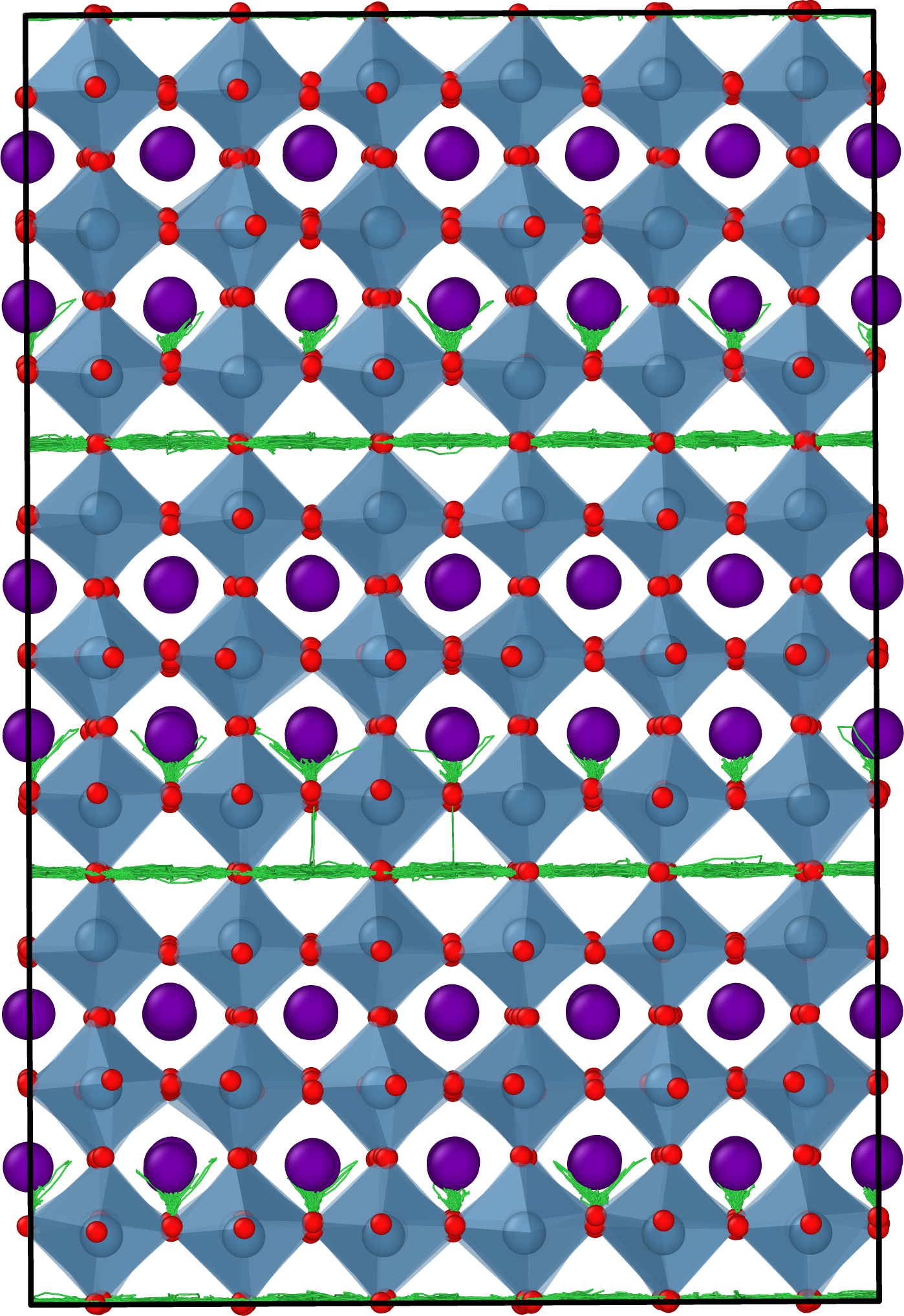}
    	\caption{\label{subfig:traj_LLTO_300K}LLTO at 300K}
    \end{subfigure}
    \begin{subfigure}[b]{0.3\textwidth}
        \centering
	    \includegraphics[height=4cm]{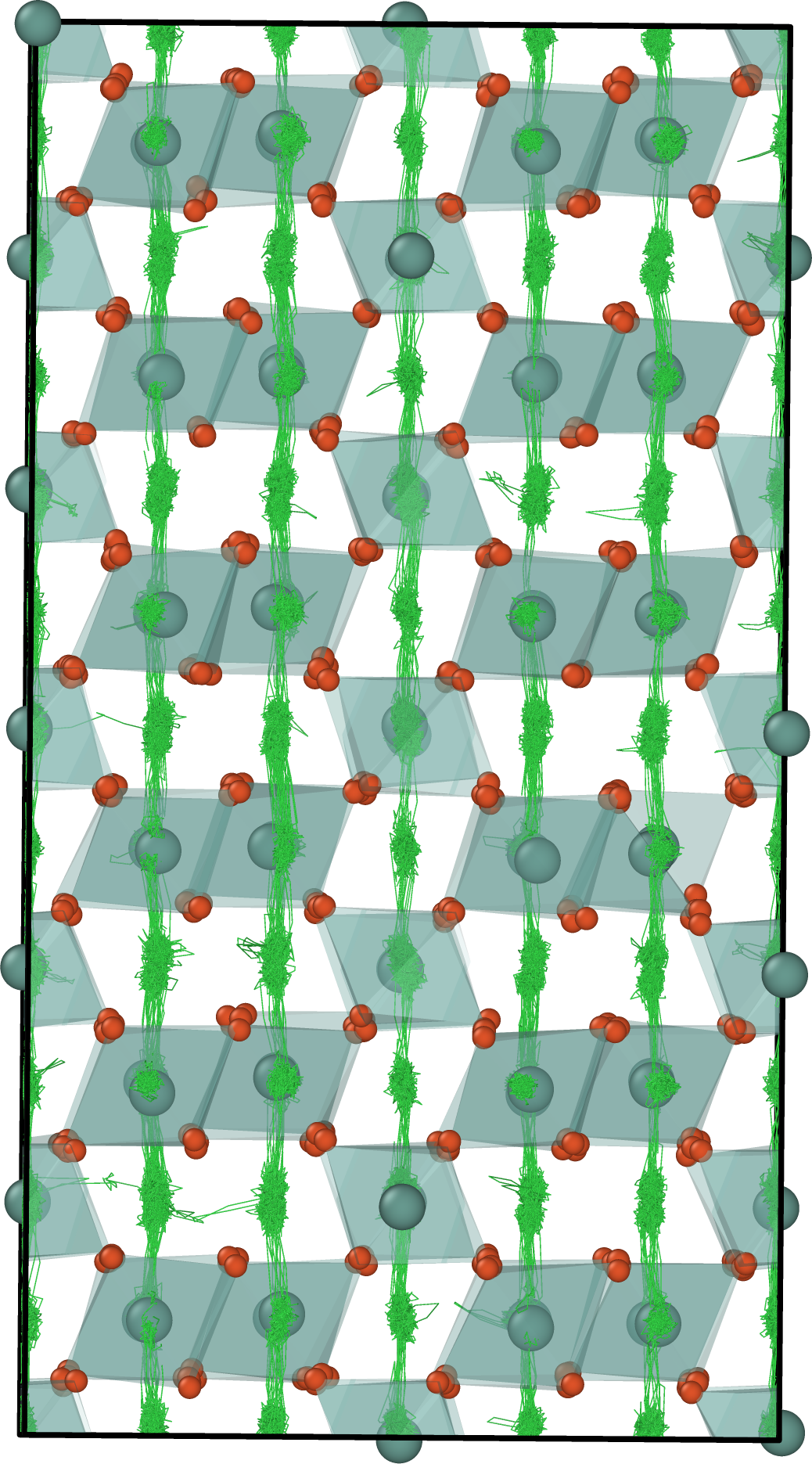}
    	\caption{\label{subfig:traj_LYC_300K}\ce{Li3YCl6} at 300K}
    \end{subfigure}
    \begin{subfigure}[b]{0.3\textwidth}
        \centering
        \includegraphics[height=4cm]{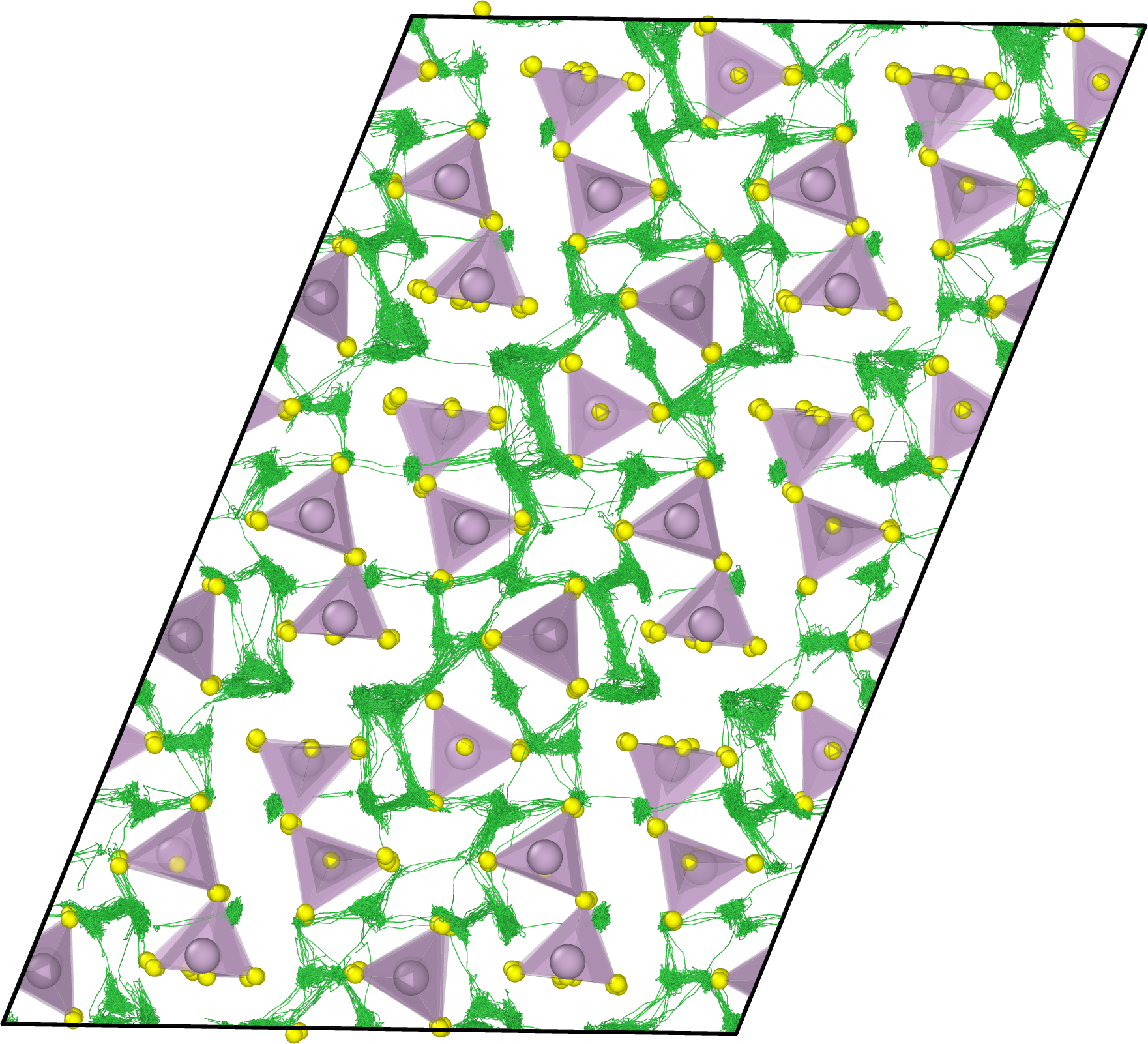}
    	\caption{\label{subfig:traj_Li7P3S11_300K}\ce{Li7P3S11} at 300K}
    \end{subfigure}
    \caption{Li trajectories (colored as green) from MTP$_{\mathrm{PBE,optB88}}$ MD simulations of the LLTO, \ce{Li3YCl6} and \ce{Li7P3S11} LSCs at room temperature (300K) and above the transition temperatures. For brevity, only the projection in the crystallographic a-c, b-c and a-c planes are shown for LLTO, \ce{Li3YCl6} and \ce{Li7P3S11}, respectively, at each temperature. The projections in the other crystallographic planes are provided in Figure S10-S12.}
    \label{fig:trajectories}
\end{figure}

To understand the reason behind observed transitions between quasi-linear Arrhenius regimes, we have extracted the trajectories from the MTP$_{\mathrm{PBE,optB88}}$ MD simulations of the three LSCs at room temperature and above the transition temperature and plotted them in Figure \ref{fig:trajectories} and Figure S10-S12. In all three LSCs, a substantial change in the number and variety of diffusion pathways is observed. For LLTO, Li diffusion at 300K occurs primarily along b and c directions inside the La-poor layer, which agrees well with experimental observations at room temperature \cite{maMesoscopicFrameworkEnables2016}, but additional diffusion pathways between planes are activated above the transition temperature of 450K (Figure \ref{subfig:traj_LLTO_500K}). Similarly, \ce{Li3YCl6} exhibits quasi-1D diffusion at 300K and 3D diffusion above the transition temperature of 425K. For \ce{Li7P3S11}, Li diffusion is already 3D at 300K but occurs along well-defined pathways. Above the transition temperature of 400K, additional pathways are activated. The decrease in activation energies $E_a$ can be directly traced to the increase in the variety and dimensionality of diffusion in the three LSCs. The most significant reduction in $E_a$ ($\sim$ 0.25 eV) is observed for \ce{Li3YCl6}, which transitions from a quasi-1D to 3D conductor at around 425K.

\section{Discussion}
%This should explore the significance of the results of the work, not repeat them. A combined Results and Discussion section is often appropriate. Avoid extensive citations and discussion of published literature.

Briefly, the above results have shown that the discrepancy between computed and experimentally measured ionic conductivities in the literature can be traced to two effects. First, the choice of the DFT functional can lead to substantial errors in the lattice parameters, which can have a large effect on the predicted activation barriers and ionic conductivities. Second, most AIMD simulations of LSCs in the literature were performed at high temperatures in the NVT ensemble to obtain sufficient hop statistics. Not only does this lead to further errors in the lattice parameters, there is also a strong likelihood that these simulations do not capture transitions in quasi-linear Arrhenius regimes occurring at lower temperatures. 

These issues can be addressed through carefully-trained ML-IAPs. With a generalizable stepwise workflow for the construction of MTPs for LSCs (see Figure \ref{fig:workflow}), we have consistently generated training structures sampling a range of local environments (see Figure S8) and fitted MTPs to study LLTO, \ce{Li3YCl6} and \ce{Li7P3S11} LSCs. Our results suggest that the choice of DFT functional used in generating the initial and snapshot training structures is relatively unimportant, but the choice of the DFT functional used in the energy and force evaluations to generate the training data is critically important. Here, we show that the use of the optB88 vdW functional for static energy and force calculations and the ns-scale ML-IAP MD at lower temperatures significantly improve the agreement in lattice parameters, activation energies and room-temperature ionic conductivities of the LSCs with experiments.

These results have broad implications for ML-IAP development strategy. The typical approach in the literature thus far has been to use the same DFT functional in both the generation of training structures as well as energy/force evaluations. Decoupling these two choices allow one to use a relatively cheap computational method such as PBE functional or even other empirical potentials, for the most expensive step of generating training structures, while using a more expensive but accurate computational method, e.g., SCAN or HSE, for the static energy and force evaluations. While the generation of the DFT training data dominates the computational effort in developing the ML-IAP, it should be noted that the ML-IAPs themselves are many orders of magnitude computationally less expensive than DFT calculations, and more importantly, scales linearly with respect to the number of atoms (see Table S3). It is this linear scaling property, while retaining close to DFT accuracy in energies and forces, that enables the simulations of large systems at long time scales in this work.

Our results also have significant implications for LSC development. For all three LSCs investigated, a transition between diffusive regimes is observed at relatively low temperatures (400-450K). In most cases, the activation barriers and ionic conductivities that have been experimentally measured correspond to the low-temperature regime. One potential avenue for further enhancing the ionic conductivities of these and other LSCs is to attempt to stabilize the high-temperature, lower activation energy diffusion regimes at room temperature. This may be achieved by quenching from higher temperatures, compositional modifications (e.g., dopants or substitutions), or mechanical modifications (e.g., introducing strain). By developing ML-IAPs using the approach outlined in this work, MD simulations can provide critical guidance on potential pathways for further LSC optimization.

\section{Conclusions}

To conclude, we have shown that MTPs trained using optB88 energies and forces can successfully reproduce the experimental lattice parameters, activation energies and room temperature ionic conductivities of the LLTO, \ce{Li3YCl6} and \ce{Li7P3S11} LSCs. In all three LSCs, MD simulations using the trained MTP identify a transition between quasi-linear Arrhenius regimes occurring at relatively low temperatures. These results not only highlight the fundamental limitations in using high-temperature, short time scale AIMD simulations to predict room-temperature properties of materials, but also suggest a potential pathway and strategy to predictive LSC design through the use of machine learning interatomic potentials.

\section{Acknowledgements}
%Acknowledgements and Reference heading should be left justified, bold, with the first letter capitalized but have no numbers. Text below continues as normal.
This work was primarily supported by Nissan Motor Co., Ltd., and Nissan North America Inc. under Award Number 20202446. The authors also acknowledge software infrastructure (pymatgen and atomate) supported by the Materials Project, funded by the US Department of Energy, Office of Science, Office of Basic Energy Sciences, Materials Sciences and Engineering Division under contract no. DE-AC02-05-CH11231: Materials Project program KC23MP, and computing resources provided by the National Energy Research Scientific Computing Center (NERSC), the Triton Shared Computing Cluster (TSCC) at the University of California, San Diego, and the Extreme Science and Engineering Discovery Environment (XSEDE) under grant ACI-1548562.

\section{Data Availability}
The datasets generated and/or analyzed during the current study are available from the corresponding author on reasonable request.
% \section*{References}
\bibliography{Ref_MTP_Electrolytes_LLTO_LYC_Li7P3S11}

\end{document}

\begin{frontmatter}

\title{SUPPLEMENTARY INFORMATION\\Bridging the Gap Between Simulated and Experimental Ionic Conductivities in Lithium Superionic Conductors}
\author[mymainaddress]{Ji Qi}
% \ead{j1qi@eng.ucsd.edu}
\author[mysecondaryaddress]{Swastika Banerjee}
\author[mysecondaryaddress]{Yunxing Zuo}
\author[mysecondaryaddress]{Chi Chen}
\author[mysecondaryaddress]{Zhuoying Zhu}
\author[mysecondaryaddress]{H.C. Manas Likhit}
\author[mysecondaryaddress]{Xiangguo Li}
\author[mysecondaryaddress]{Shyue Ping Ong\corref{mycorrespondingauthor}}
\cortext[mycorrespondingauthor]{This is to indicate the corresponding author.}
\ead{ongsp@eng.ucsd.edu}

\address[mymainaddress]{Materials Science and Engineering Program, University of California San Diego, 9500 Gilman Dr, Mail Code 0448, La Jolla, CA 92093-0448, United States}
\address[mysecondaryaddress]{Department of NanoEngineering, University of California San Diego, 9500 Gilman Dr, Mail Code 0448, La Jolla, CA 92093-0448, United States}

\end{frontmatter}

\clearpage

\begin{figure}[htb!]
\renewcommand{\thefigure}{S\arabic{figure}}
\center\includegraphics[height=7cm]{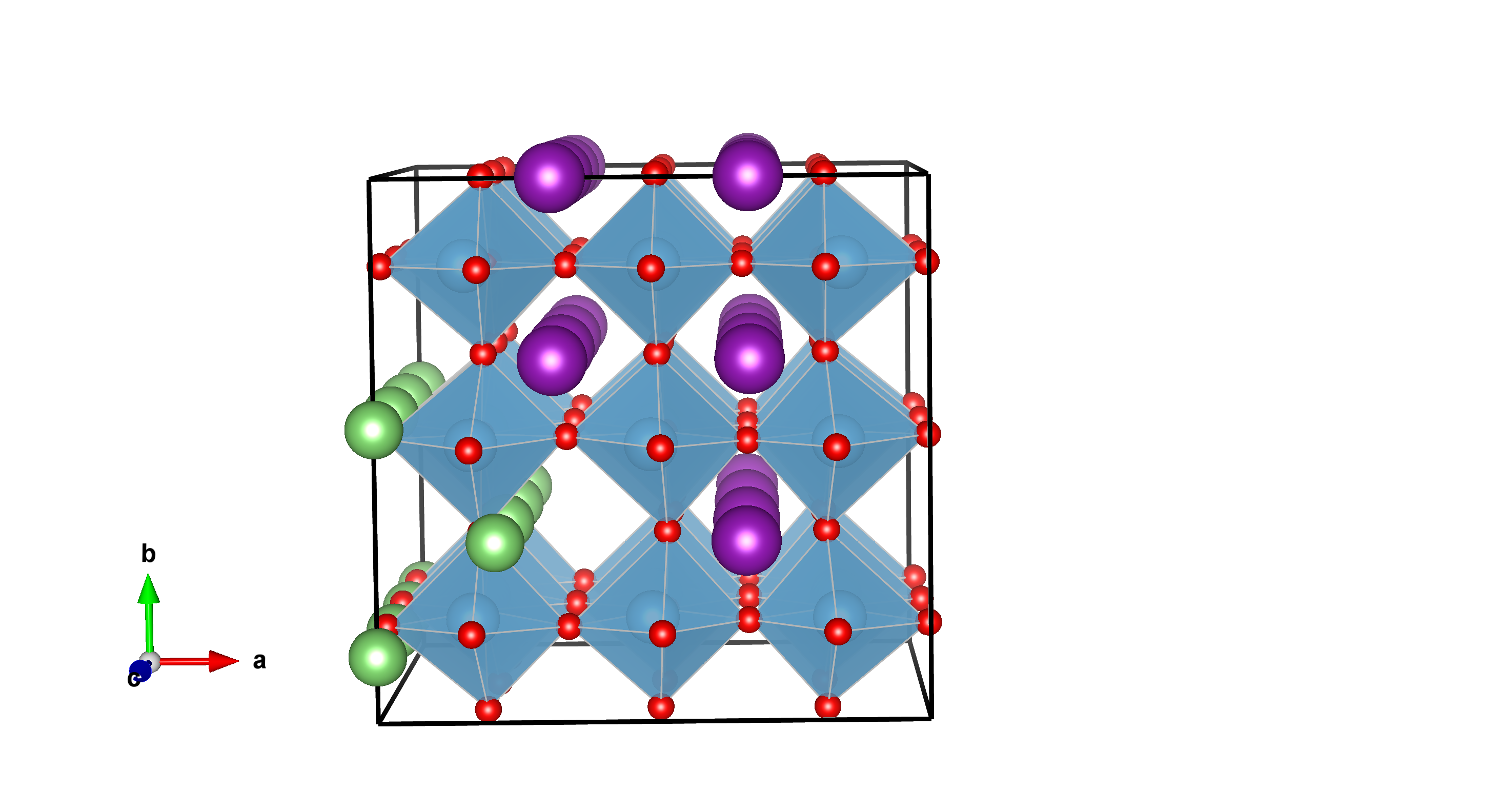}
\caption{Lowest energy ordering of \ce{Li_{0.33}La_{0.56}TiO3}. The $3\times3\times3$ perovskite has formula \ce{Li9La15Ti27O81}.}
\end{figure}

\begin{table}[htb!]
\renewcommand{\thetable}{S\arabic{table}}
  \centering
  \caption{Reported crystal structure of \ce{Li3YCl6} from XRD in ref \citenum{asanoSolidHalideElectrolytes2018}. Adjusted occupancies used for enumeration of symmetrically distinct orderings were listed in the last column.}
    \begin{tabular}{ccccccc}
    \hline\hline
    \multicolumn{1}{c}{\multirow{2}[0]{*}{Atom}} & \multicolumn{1}{c}{\multirow{2}[0]{*}{Wyckoff position}} & \multicolumn{3}{c}{Atomic coordinates} & \multicolumn{1}{c}{\multirow{2}[0]{*}{Occ.}} & \multirow{2}[0]{*}{Adjusted Occ.} \\\cline{3-5}
          &       & \multicolumn{1}{c}{x} & \multicolumn{1}{c}{y} & \multicolumn{1}{c}{z} &       &  \\\hline
    Li    & 6g    & 0.3335 & 0     & 0     & 0.73  & 0.83 \\
    Li    & 6h    & 0.322 & 0     &  1/2  & 0.59  & 0.67 \\
    Y     & 1a    & 0     & 0     & 0     & 0.925 & 1 \\
    Y     & 2d    &  1/3  &  2/3  & 0.5119 & 0.829 & 1 \\
    Y     & 2d    &  1/3  &  2/3  & -0.007 & 0.121 & 0 \\
    Cl    & 6i    & 0.1144 & -0.1144 & 0.7706 & 1     & 1 \\
    Cl    & 6i    & 0.2213 & -0.2213 & 0.2643 & 1     & 1 \\
    Cl    & 6i    & 0.4469 & -0.4469 & 0.7536 & 0.91  & 1 \\\hline\hline
    \end{tabular}%
  \label{table:enumeration_LYC}%
\end{table}%

\begin{table}[htbp]
\renewcommand{\thetable}{S\arabic{table}}
  \centering
  \caption{The formation energies ($E_{f}$) and energy above the convex hull ($E_{hull}$) for all three LSCs calculated with the PBE and optB88 functionals.}
    \begin{tabular}{ccccc}
    \hline\hline
    \multirow{2}[3]{*}{LSC} & \multicolumn{2}{c}{$E_{hull}$ (eV/atom)} & \multicolumn{2}{c}{$E_{f}$ (eV/atom)} \\
\cline{2-5}          & optB88 & PBE   & optB88 & PBE \\
    \hline
    LLTO  & 0.042 & 0.039 & -3.326 & -3.108 \\
    \ce{Li3YCl6} & 0.003 & 0.026 & -2.105 & -1.982 \\
    \ce{Li7P3S11} & 0.012 & 0.019 & -0.902 & -0.823 \\
    \hline\hline
    \end{tabular}%
  \label{tab:addlabel}%
\end{table}%

\begin{figure}[htb!]
\renewcommand{\thefigure}{S\arabic{figure}}
    \centering
    \begin{subfigure}[b]{0.49\textwidth}
        \centering
        \includegraphics[height=3.4cm]{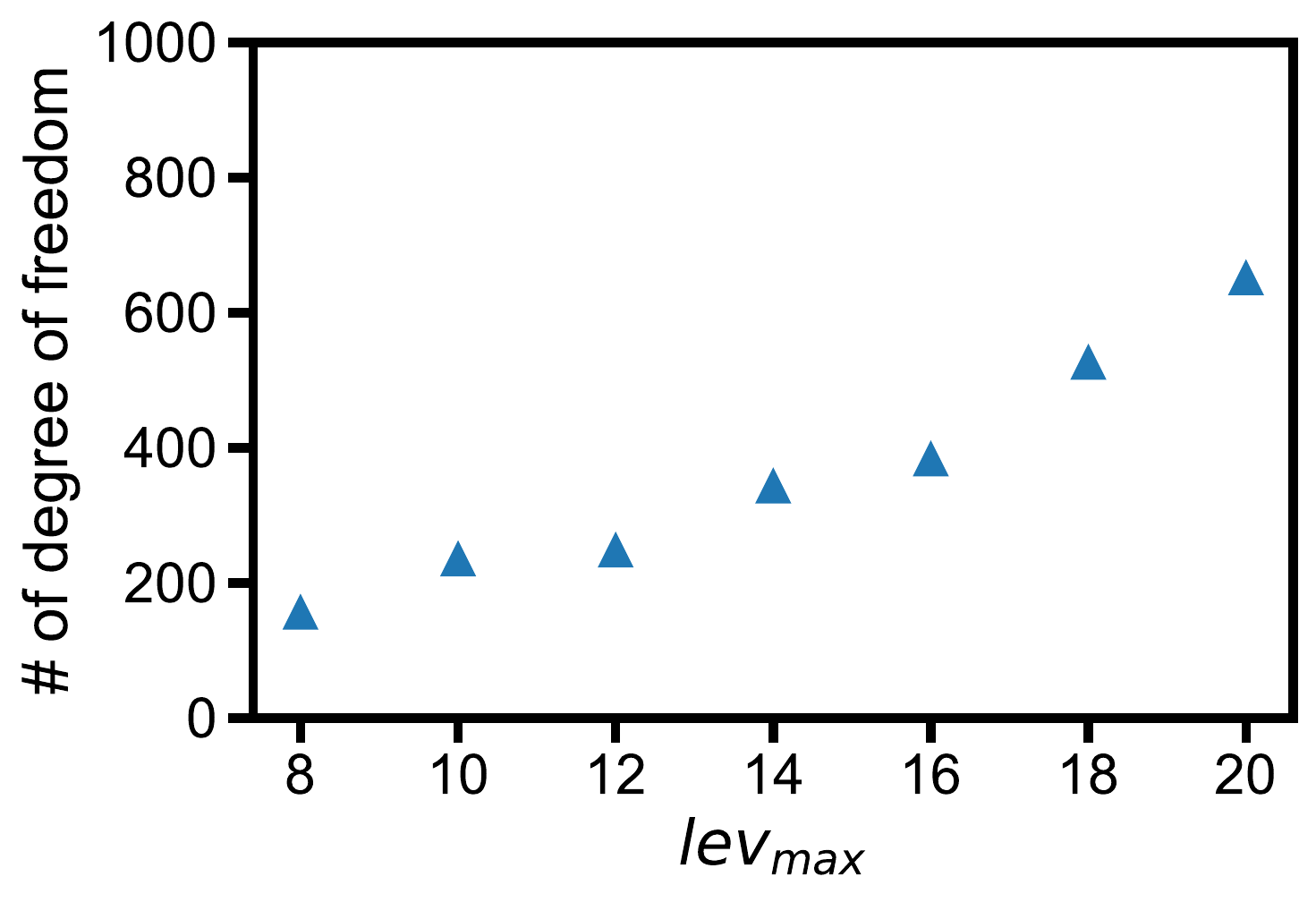}
    	\caption{}
    \end{subfigure}
    \begin{subfigure}[b]{0.49\textwidth}
        \centering
        \includegraphics[height=3.4cm]{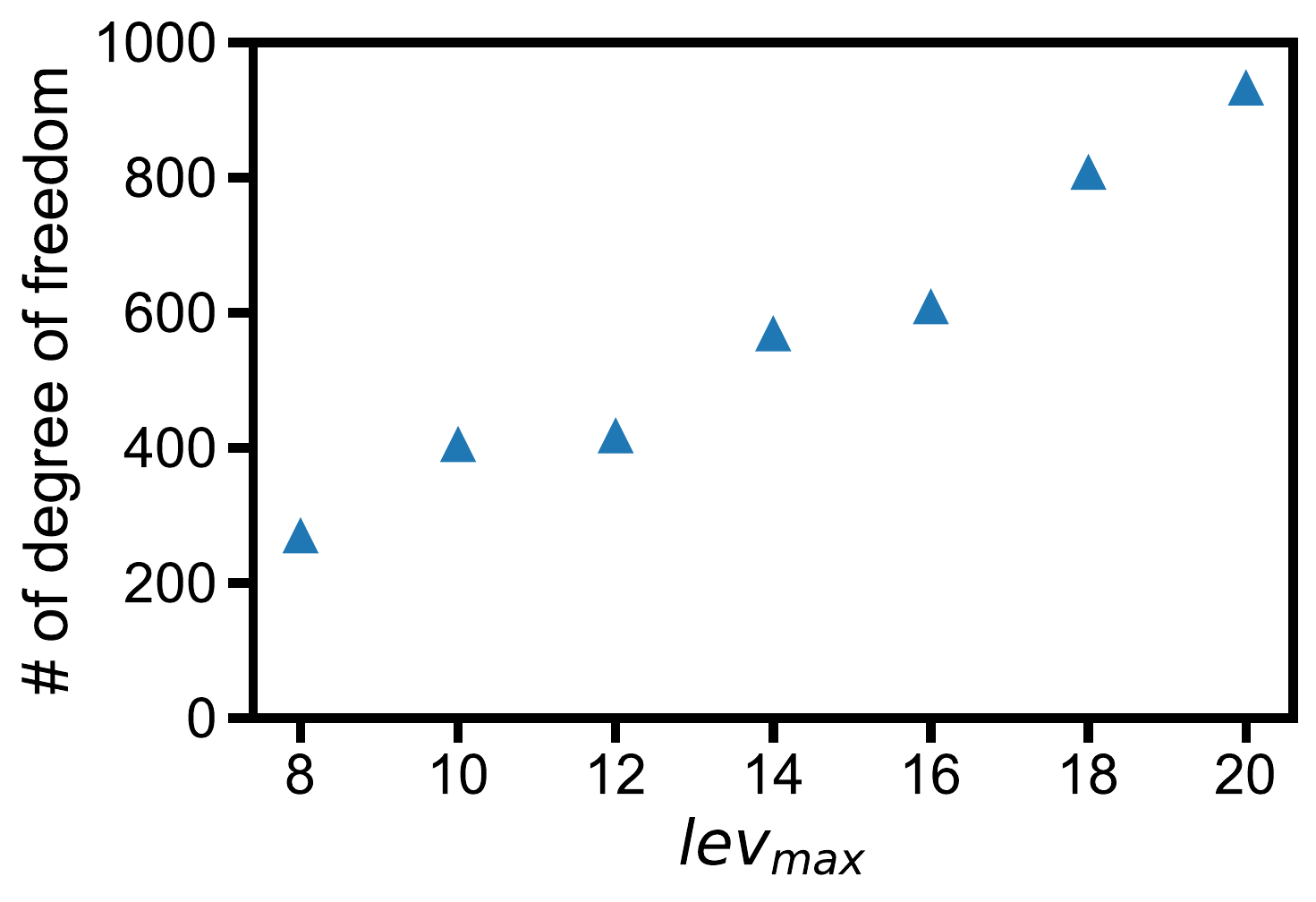}
    	\caption{}
    \end{subfigure}
    \begin{subfigure}[b]{0.49\textwidth}
        \centering
	    \includegraphics[height=3.4cm]{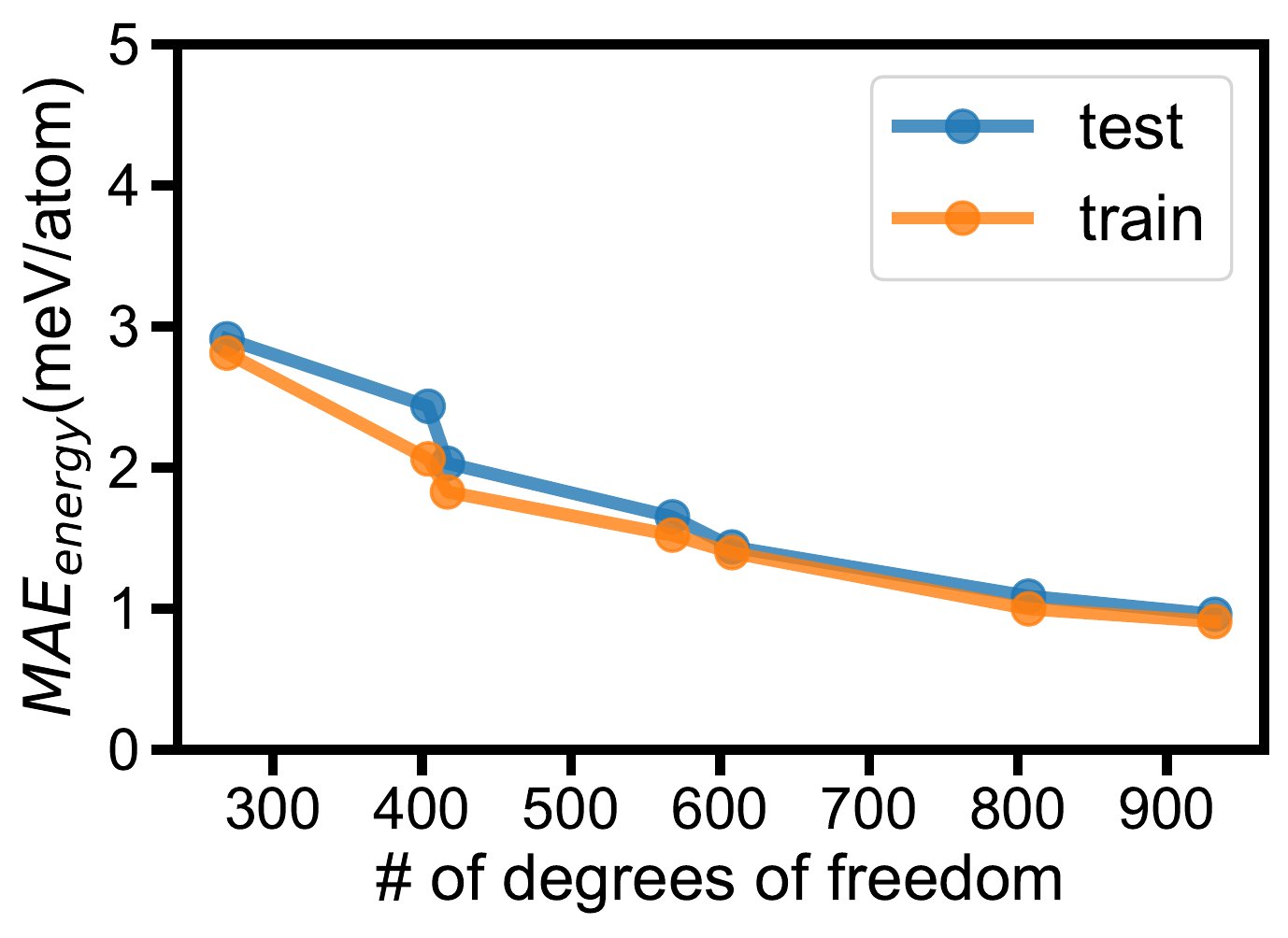}
    	\caption{}
    \end{subfigure}
    \begin{subfigure}[b]{0.49\textwidth}
        \centering
	    \includegraphics[height=3.4cm]{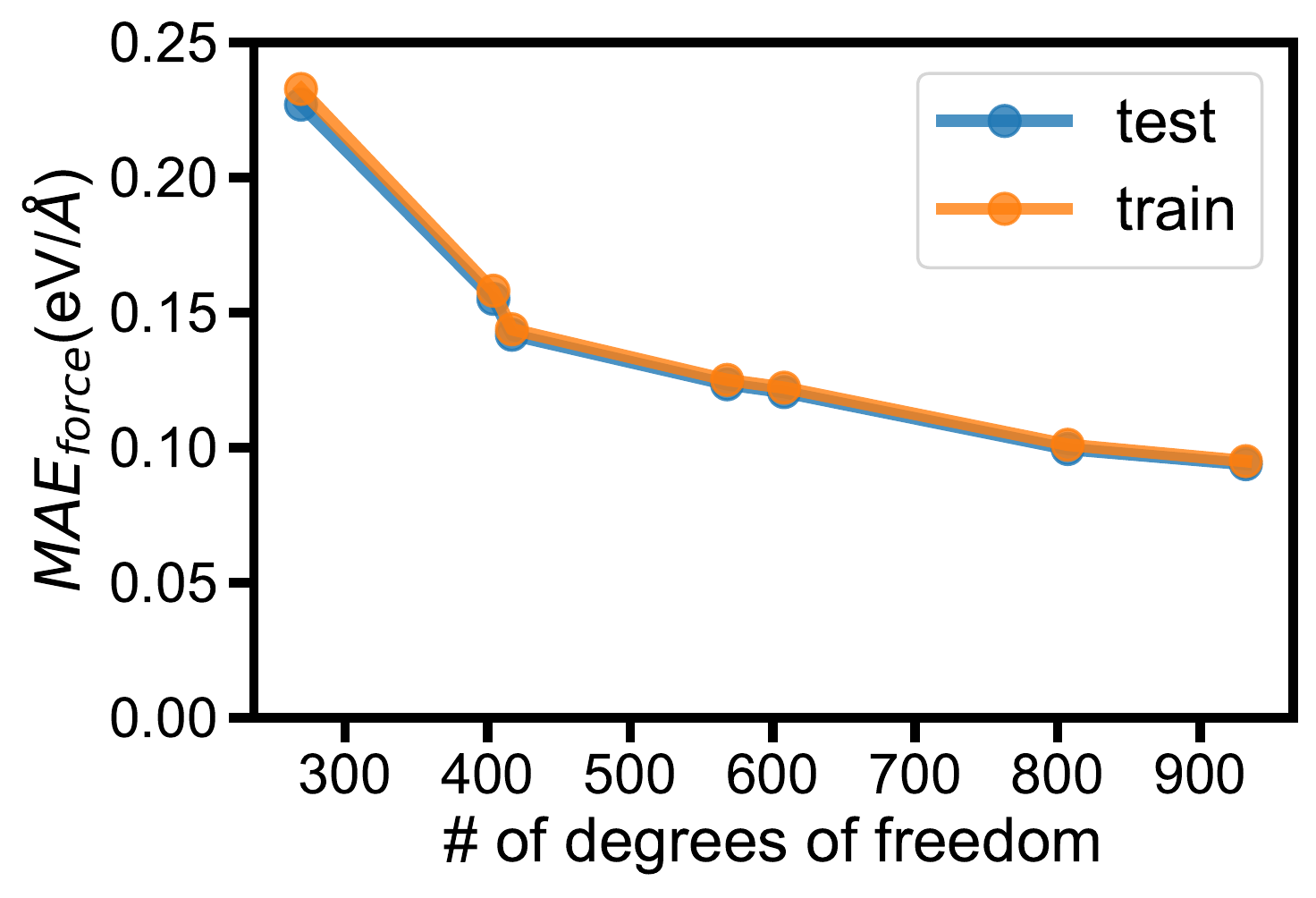}
    	\caption{}
    \end{subfigure}
    \begin{subfigure}[b]{0.49\textwidth}
        \centering
	    \includegraphics[height=3.4cm]{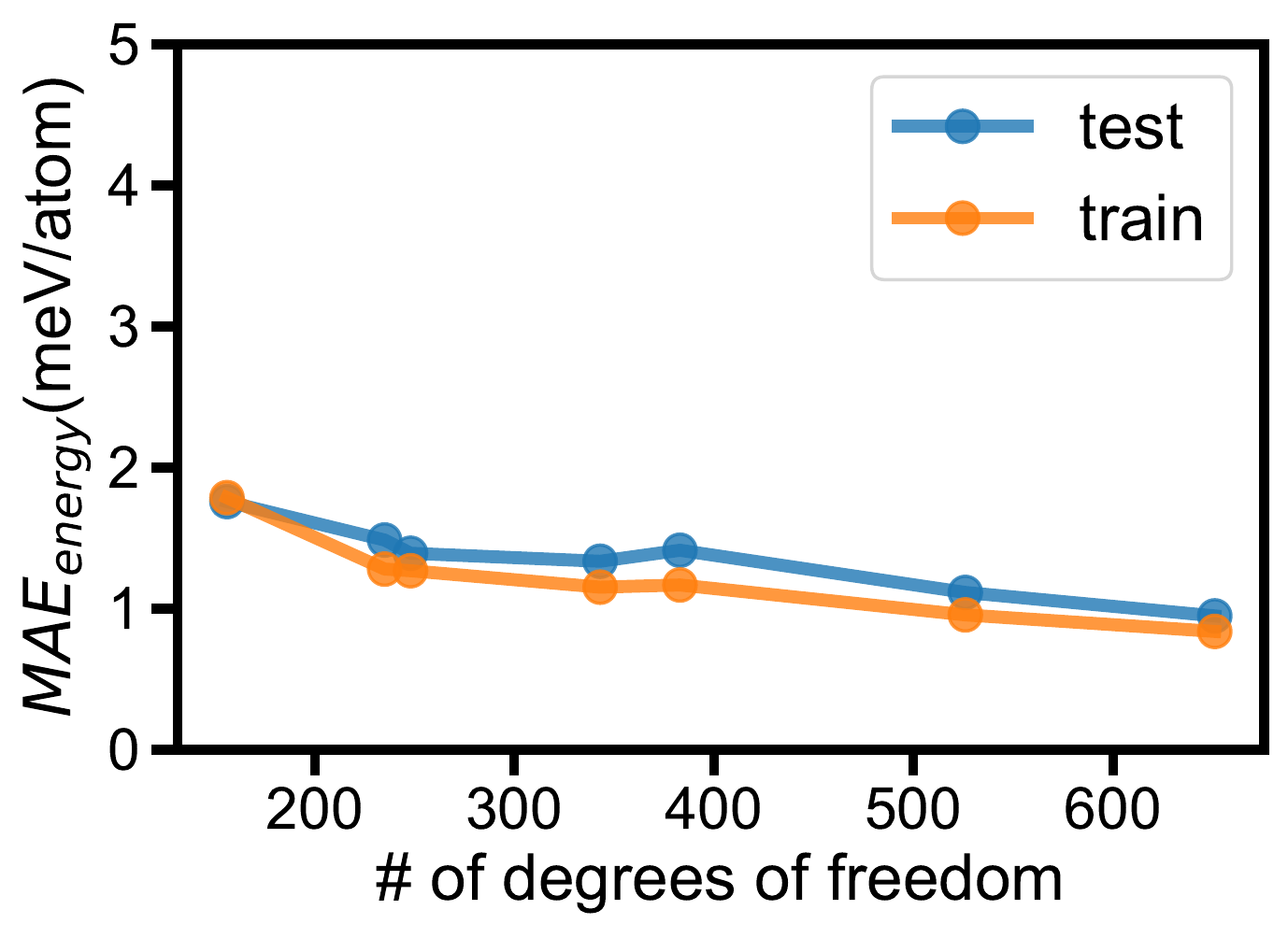}
    	\caption{}
    \end{subfigure}
    \begin{subfigure}[b]{0.49\textwidth}
        \centering
	    \includegraphics[height=3.4cm]{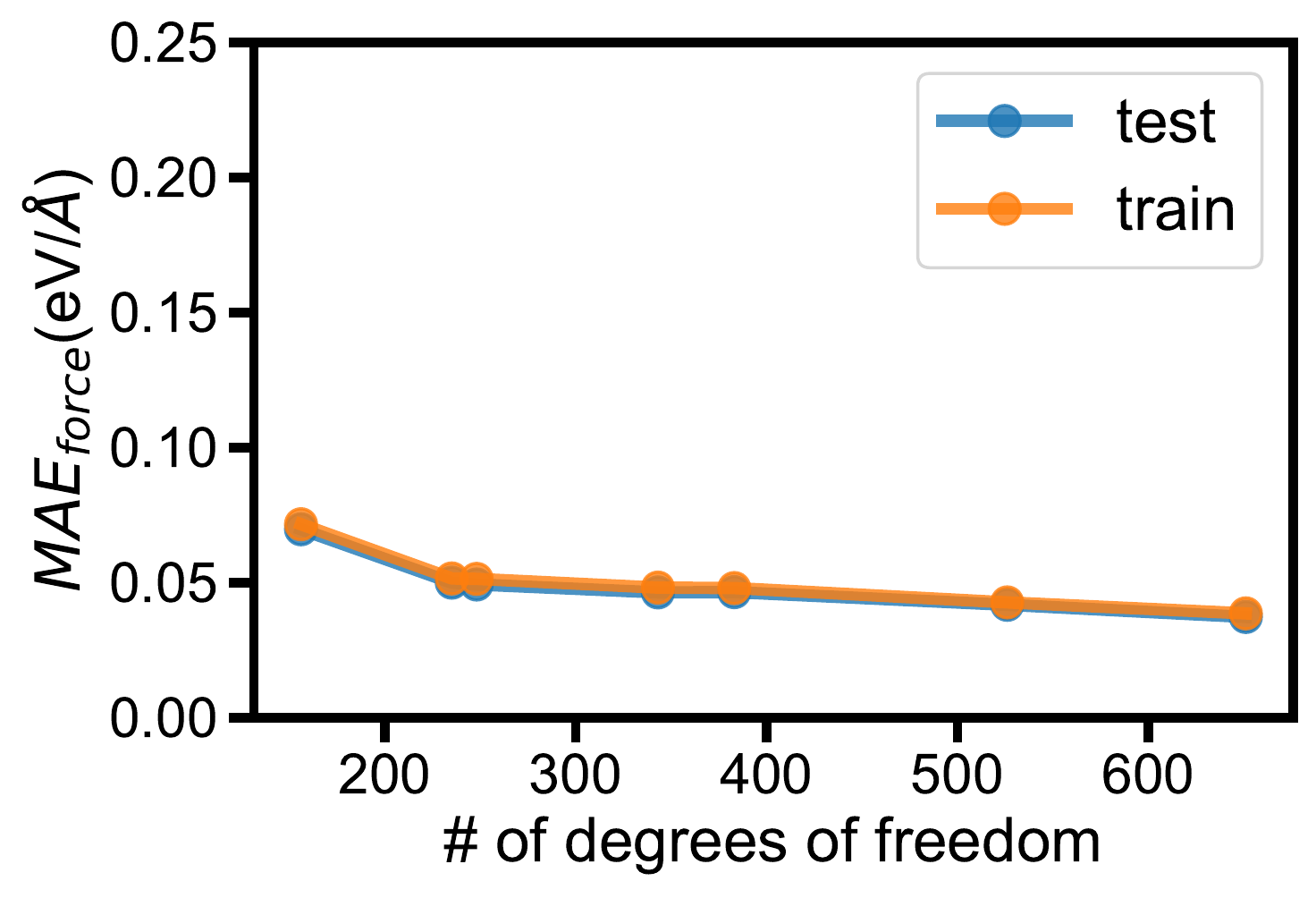}
    	\caption{}
    \end{subfigure}
    \begin{subfigure}[b]{0.49\textwidth}
        \centering
	    \includegraphics[height=3.4cm]{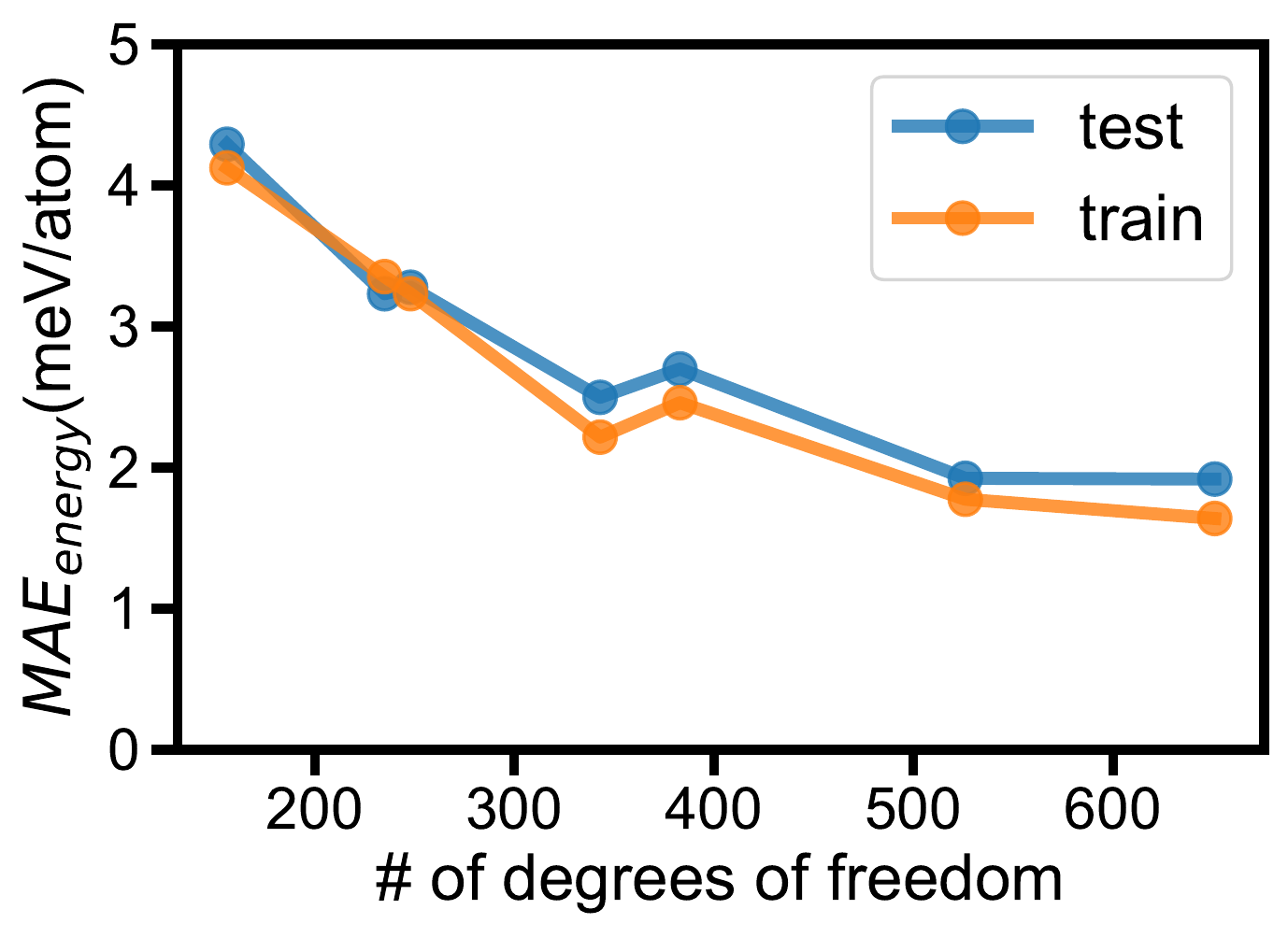}
    	\caption{}
    \end{subfigure}
    \begin{subfigure}[b]{0.49\textwidth}
        \centering
	    \includegraphics[height=3.4cm]{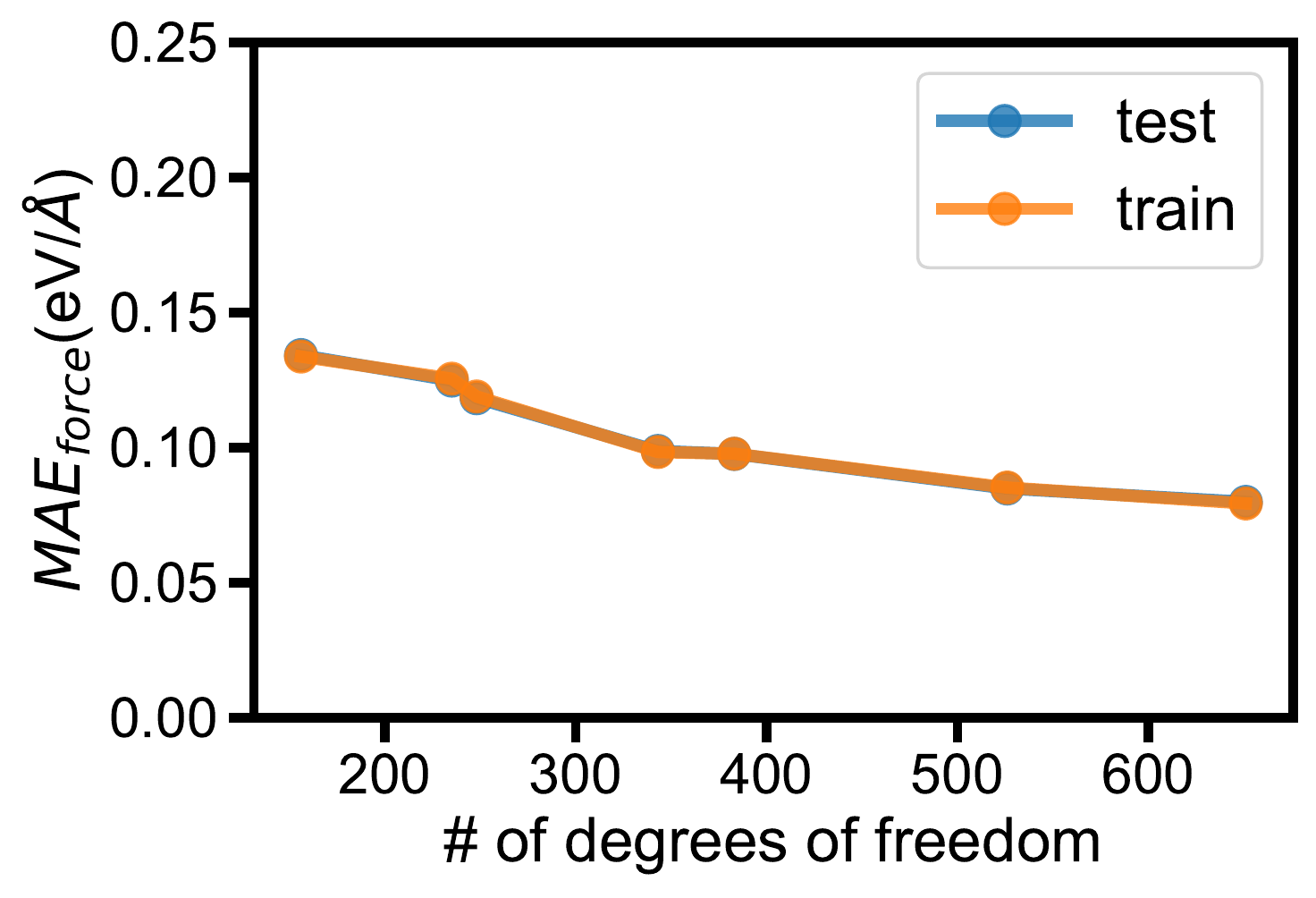}
    	\caption{}
    \end{subfigure}
    \caption{Convergence tests of MTP with respect to $lev_{max}$ for all three LSCs. $lev_{max}$ versus number of degrees of freedom for (a) 3-element MTPs (\ce{Li3YCl6} and \ce{Li7P3S11}) and (b) 4-element MTPs (LLTO). Plots of the MAEs in MTP predicted (c) LLTO energies, (d) LLTO forces, (e) \ce{Li3YCl6} energies, (f) \ce{Li3YCl6} forces, (g) \ce{Li7P3S11} energies and (h) \ce{Li7P3S11} forces with respect to number of degrees of freedom.}
    \label{fig:convergencetests}
\end{figure}

\begin{figure}[htb!]
\renewcommand{\thefigure}{S\arabic{figure}}
    \begin{subfigure}[b]{0.49\textwidth}
        \centering
	    \includegraphics[height=4.5cm]{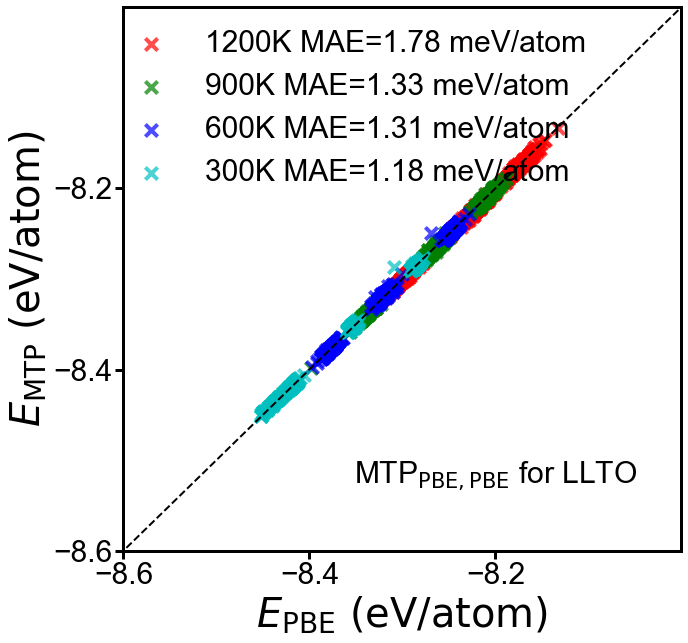}
    	\caption{}
    \end{subfigure}
    \begin{subfigure}[b]{0.49\textwidth}
        \centering
	    \includegraphics[height=4.5cm]{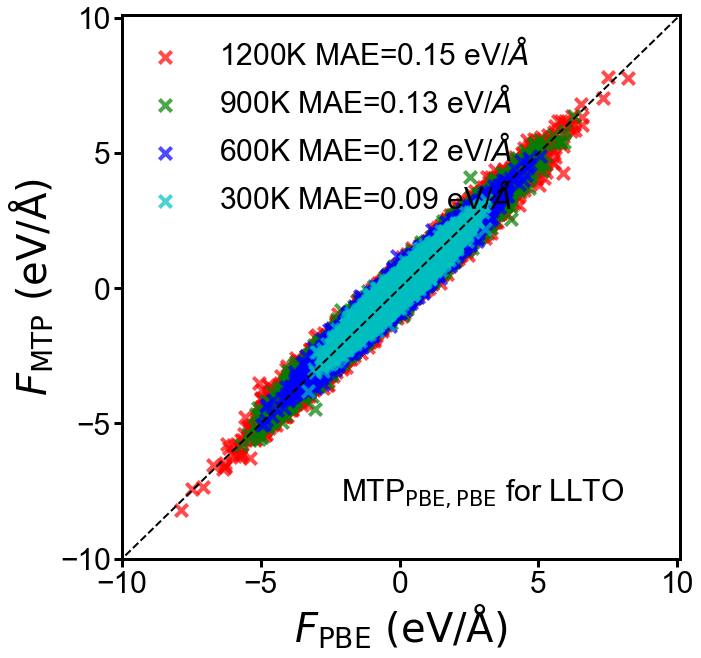}
    	\caption{}
    \end{subfigure}
    \begin{subfigure}[b]{0.49\textwidth}
        \centering
	    \includegraphics[height=4.5cm]{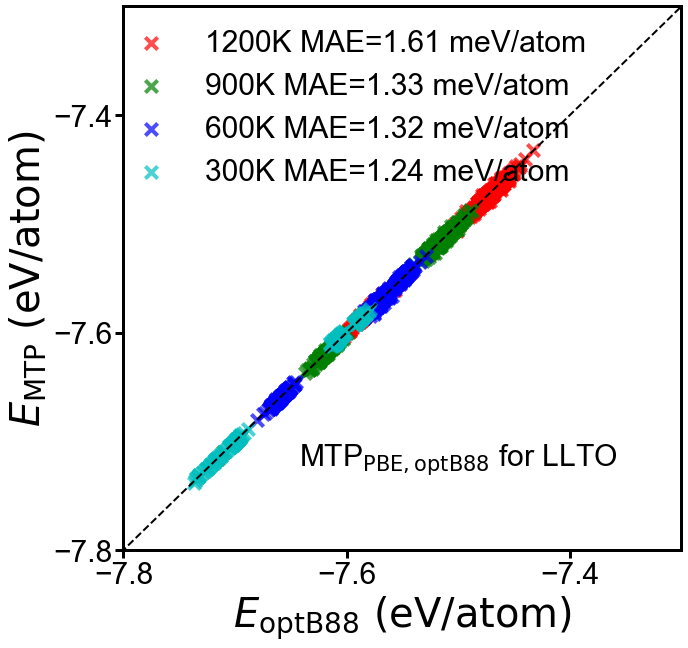}
    	\caption{}
    \end{subfigure}
    \begin{subfigure}[b]{0.49\textwidth}
        \centering
	    \includegraphics[height=4.5cm]{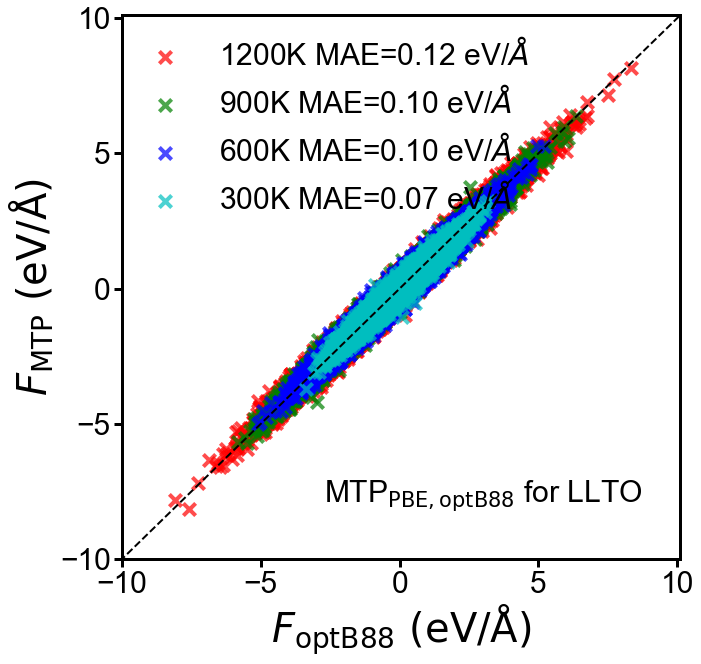}
    	\caption{}
    \end{subfigure}
    \caption{Plots of the MTP predicted versus DFT energies and forces for LLTO with $lev_{max}=16$. (a) MTP$_{\mathrm{PBE,PBE}}$ energies vs PBE energies. (b) MTP$_{\mathrm{PBE,PBE}}$ forces versus PBE forces. (c) MTP$_{\mathrm{PBE,optB88}}$ energies vs optB88 energies. (d) MTP$_{\mathrm{PBE,optB88}}$ forces vs optB88 forces.}
\end{figure}

\begin{figure}[htb!]
\renewcommand{\thefigure}{S\arabic{figure}}
    \begin{subfigure}[b]{0.49\textwidth}
        \centering
	    \includegraphics[height=5cm]{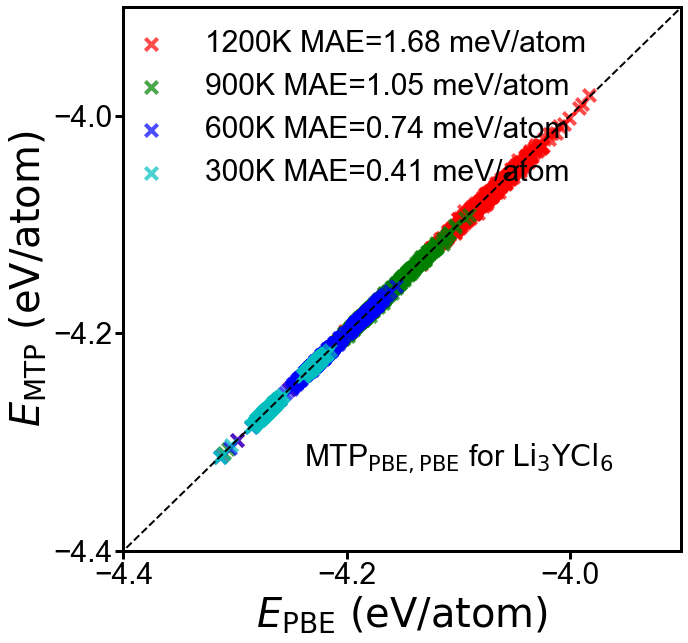}
    	\caption{}
    \end{subfigure}
    \begin{subfigure}[b]{0.49\textwidth}
        \centering
	    \includegraphics[height=5cm]{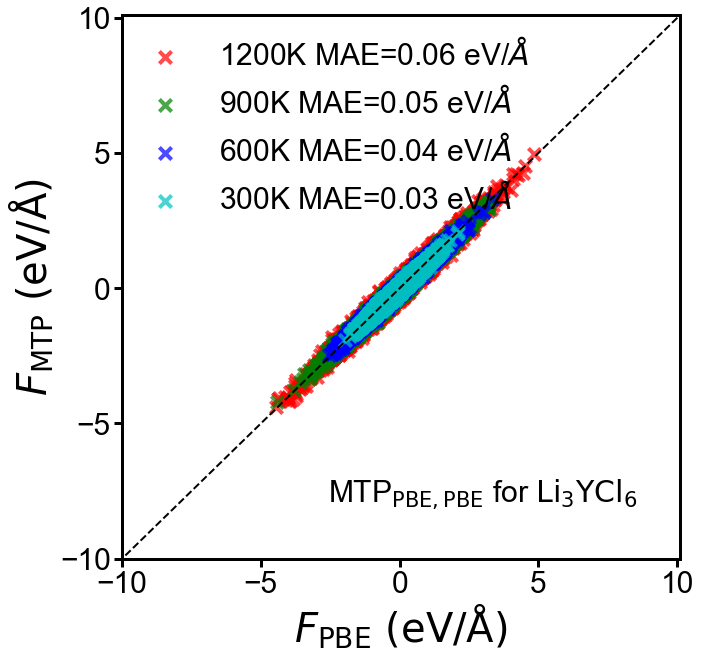}
    	\caption{}
    \end{subfigure}
    \begin{subfigure}[b]{0.49\textwidth}
        \centering
	    \includegraphics[height=5cm]{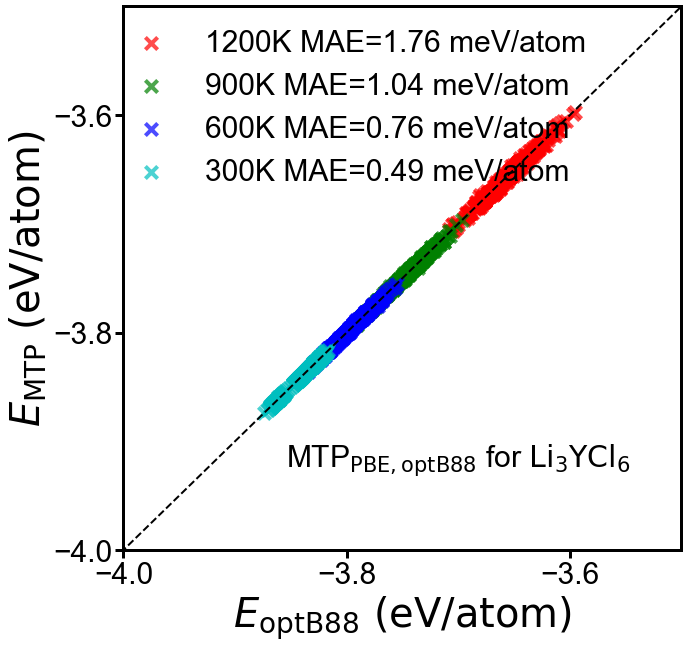}
    	\caption{}
    \end{subfigure}
    \begin{subfigure}[b]{0.49\textwidth}
        \centering
	    \includegraphics[height=5cm]{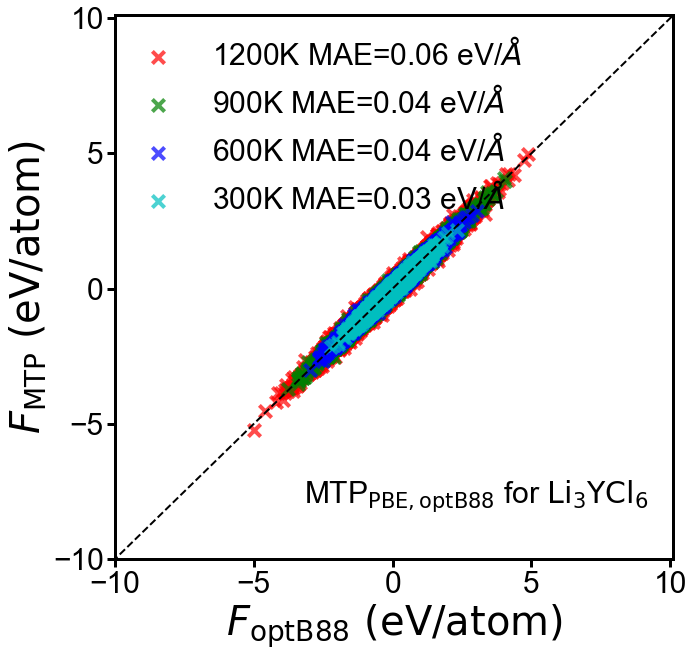}
    	\caption{}
    \end{subfigure}
    \caption{Plots of the MTP predicted versus DFT energies and forces for \ce{Li3YCl6} with $lev_{max}=18$. (a) MTP$_{\mathrm{PBE,PBE}}$ energies vs PBE energies. (b) MTP$_{\mathrm{PBE,PBE}}$ forces versus PBE forces. (c) MTP$_{\mathrm{PBE,optB88}}$ energies vs optB88 energies. (d) MTP$_{\mathrm{PBE,optB88}}$ forces vs optB88 forces.}
\end{figure}

\begin{figure}[htb!]
\renewcommand{\thefigure}{S\arabic{figure}}
    \centering
    \begin{subfigure}[b]{0.49\textwidth}
        \centering
	    \includegraphics[height=4.5cm]{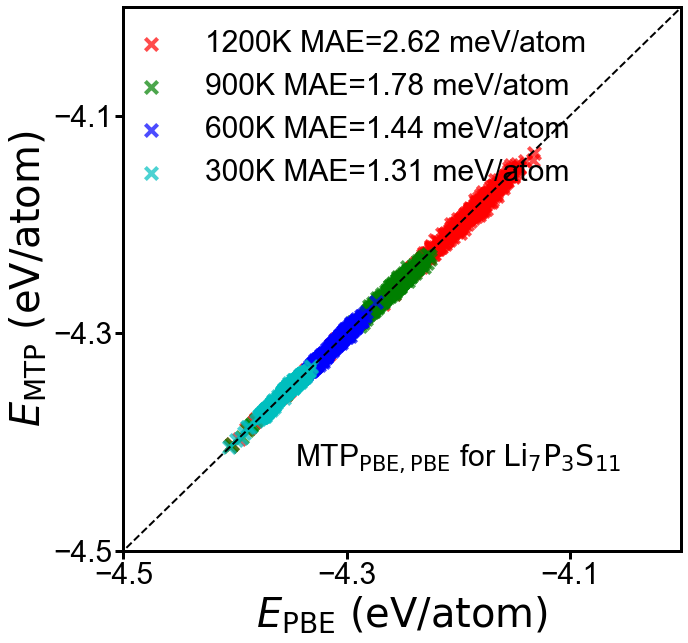}
    	\caption{}
    \end{subfigure}
    \begin{subfigure}[b]{0.49\textwidth}
        \centering
	    \includegraphics[height=4.5cm]{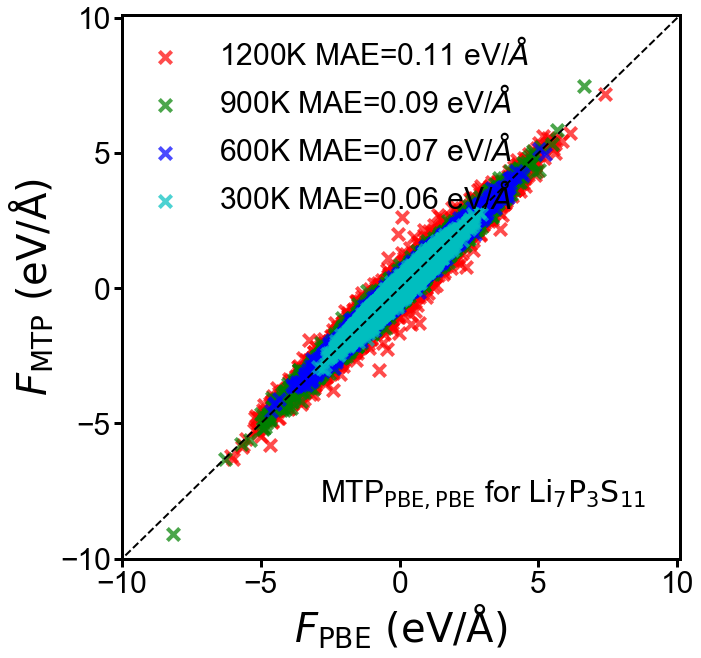}
    	\caption{}
    \end{subfigure}
    \begin{subfigure}[b]{0.49\textwidth}
        \centering
	    \includegraphics[height=4.5cm]{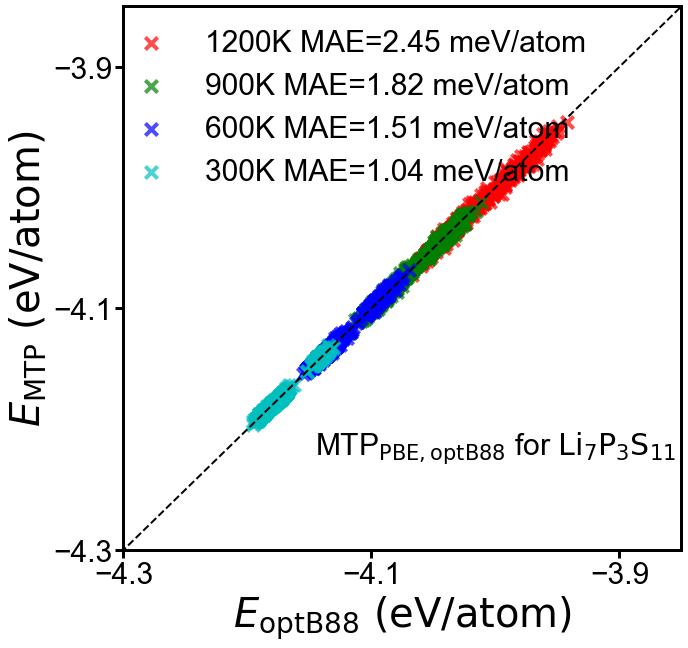}
    	\caption{}
    \end{subfigure}
    \begin{subfigure}[b]{0.49\textwidth}
        \centering
	    \includegraphics[height=4.5cm]{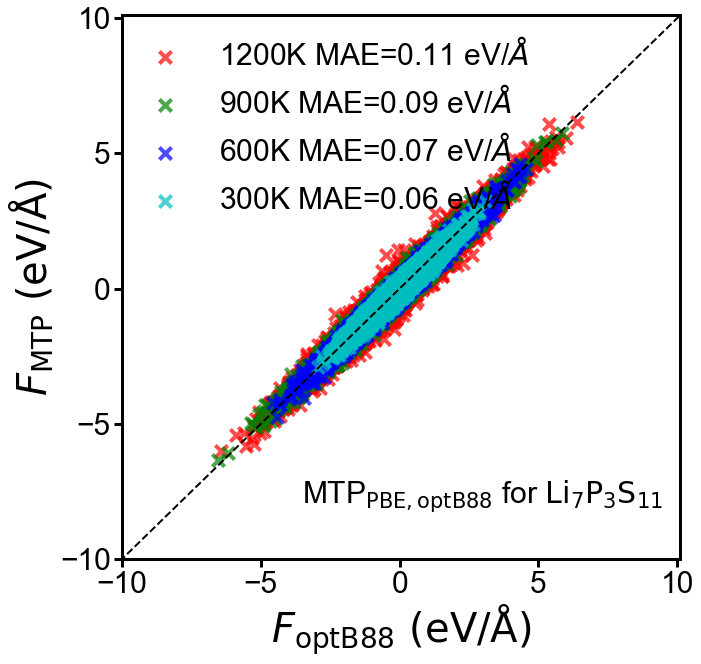}
    	\caption{}
    \end{subfigure}
    \begin{subfigure}[b]{0.49\textwidth}
        \centering
	    \includegraphics[height=4.5cm]{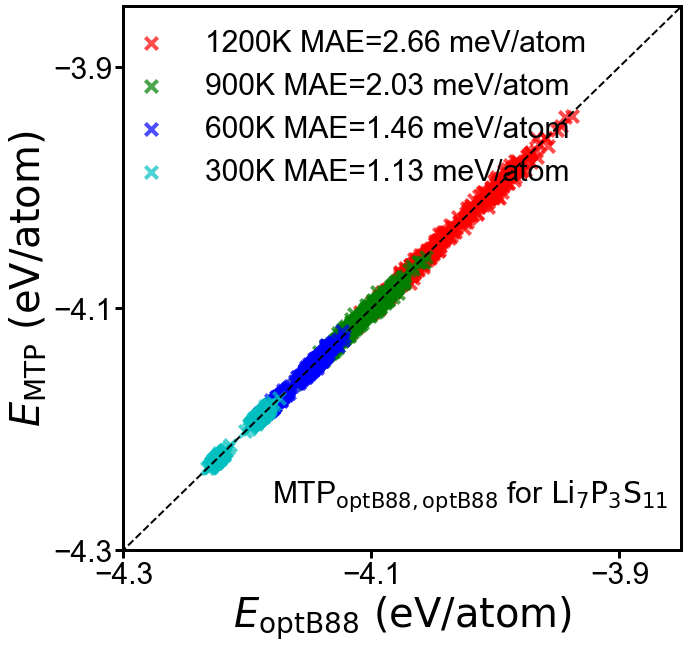}
    	\caption{}
    \end{subfigure}
    \begin{subfigure}[b]{0.49\textwidth}
        \centering
	    \includegraphics[height=4.5cm]{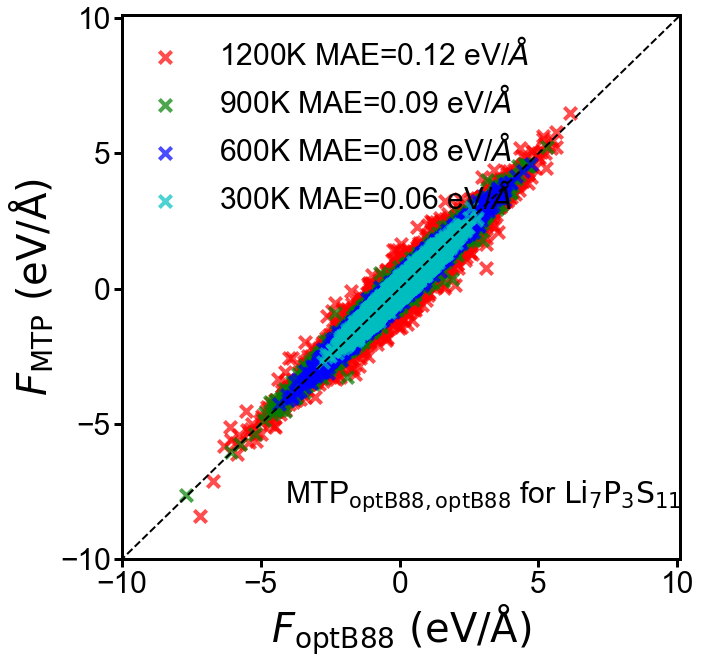}
    	\caption{}
    \end{subfigure}
    \caption{Plots of the MTP predicted versus DFT energies and forces for \ce{Li7P3S11} with $lev_{max}=18$. (a) MTP$_{\mathrm{PBE,PBE}}$ energies vs PBE energies. (b) MTP$_{\mathrm{PBE,PBE}}$ forces versus PBE forces. (c) MTP$_{\mathrm{PBE,optB88}}$ energies vs optB88 energies. (d) MTP$_{\mathrm{PBE,optB88}}$ forces vs optB88 forces. (e) MTP$_{\mathrm{optB88,optB88}}$ energies vs optB88 energies. (f) MTP$_{\mathrm{optB88,optB88}}$ forces vs optB88 forces.}
\end{figure}

\begin{figure}[htb!]
\renewcommand{\thefigure}{S\arabic{figure}}
    \begin{subfigure}[b]{0.31\textwidth}
        \centering
	    \includegraphics[height=3.5cm]{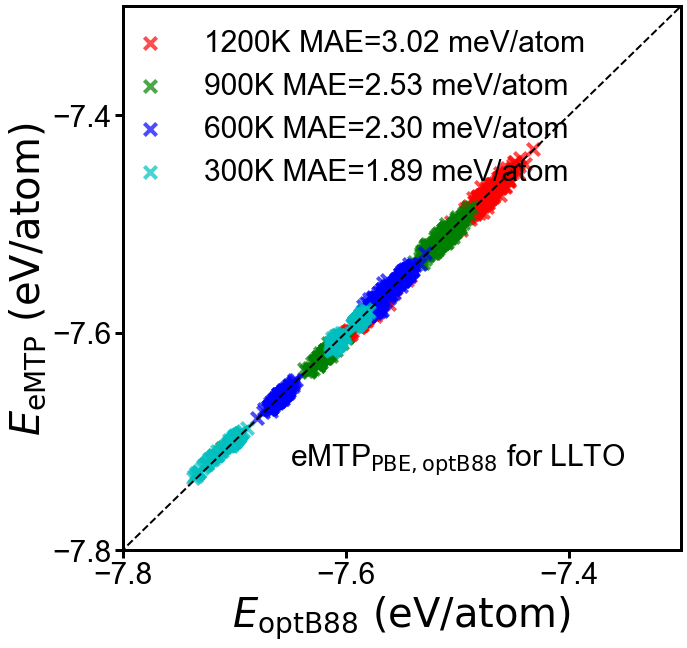}
    	\caption{Energies}
    \end{subfigure}
    \begin{subfigure}[b]{0.31\textwidth}
        \centering
	    \includegraphics[height=3.5cm]{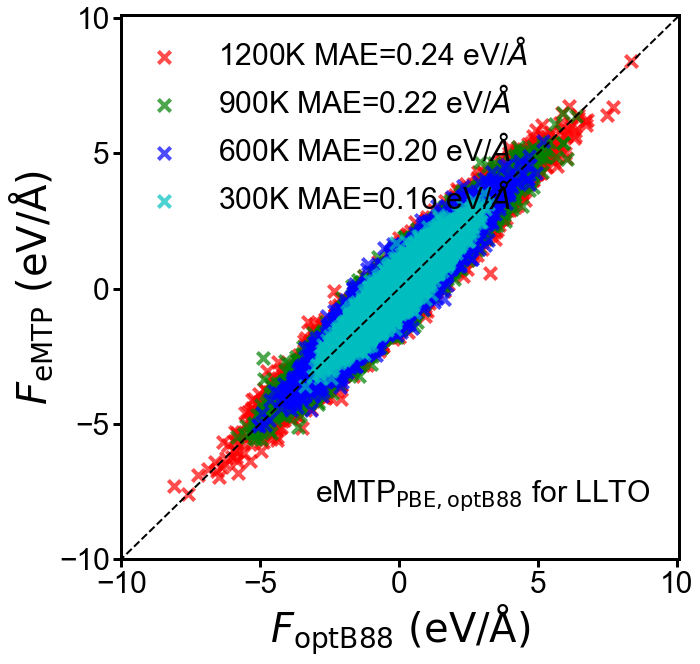}
    	\caption{Forces}
    \end{subfigure}
    \begin{subfigure}[b]{0.31\textwidth}
        \centering
	    \includegraphics[height=3.5cm]{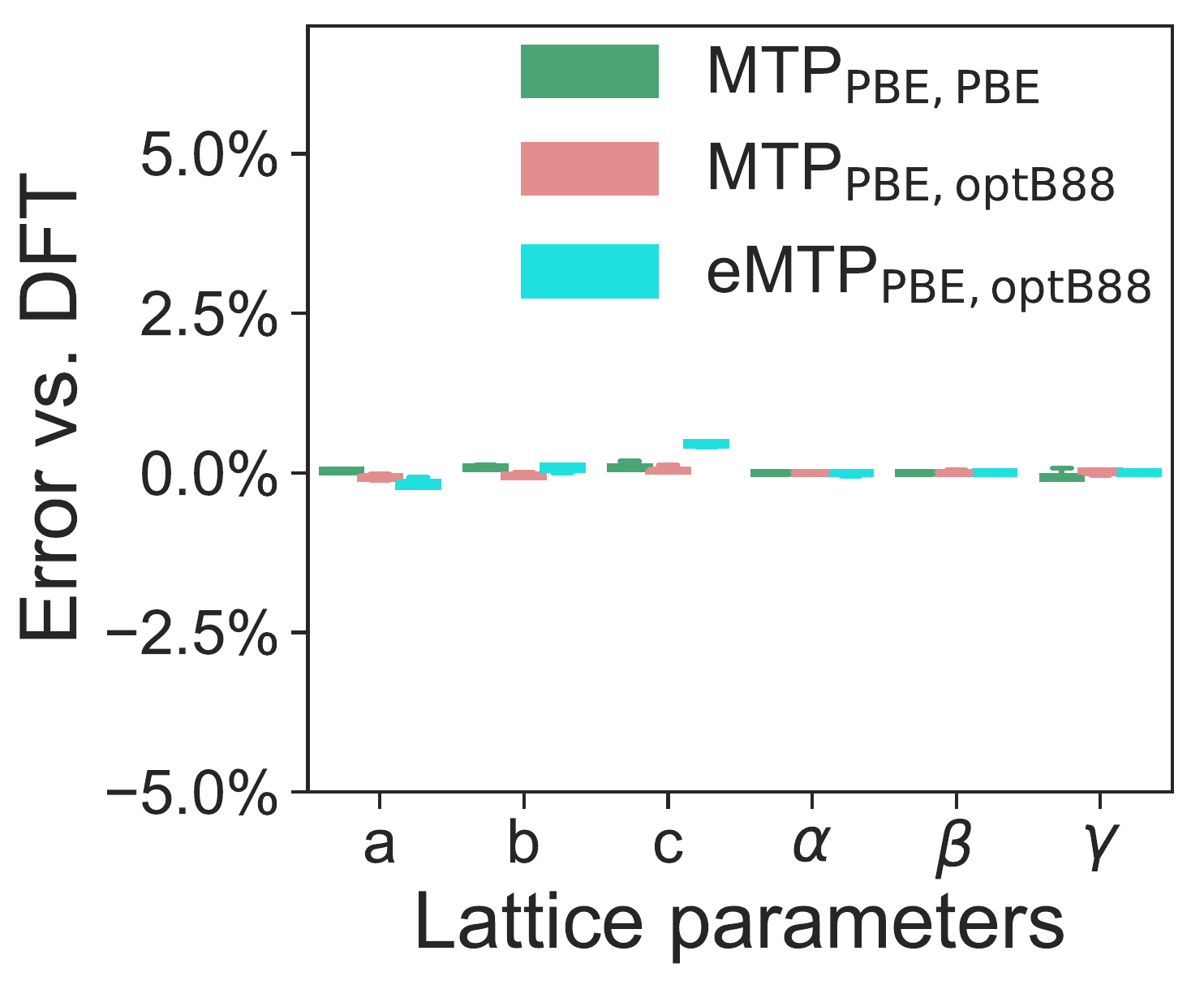}
    	\caption{Lattice parameters}
    \end{subfigure}
    \caption{Plots of the eMTP$_{\mathrm{PBE,optB88}}$ predicted versus DFT (a) energies, (b) forces and (c) lattice parameters for LLTO with $lev_{max}=16$. The eMTP$_{\mathrm{PBE,optB88}}$ performs very similarly to the MTP$_{\mathrm{PBE,optB88}}$ (without separate accounting of the electrostatics) in terms of energy and force errors as well as lattice parameter predictions.}
\end{figure}

\begin{figure}[htb!]
\renewcommand{\thefigure}{S\arabic{figure}}
    \begin{subfigure}[b]{0.3\textwidth}
        \centering
	    \includegraphics[height=3.5cm]{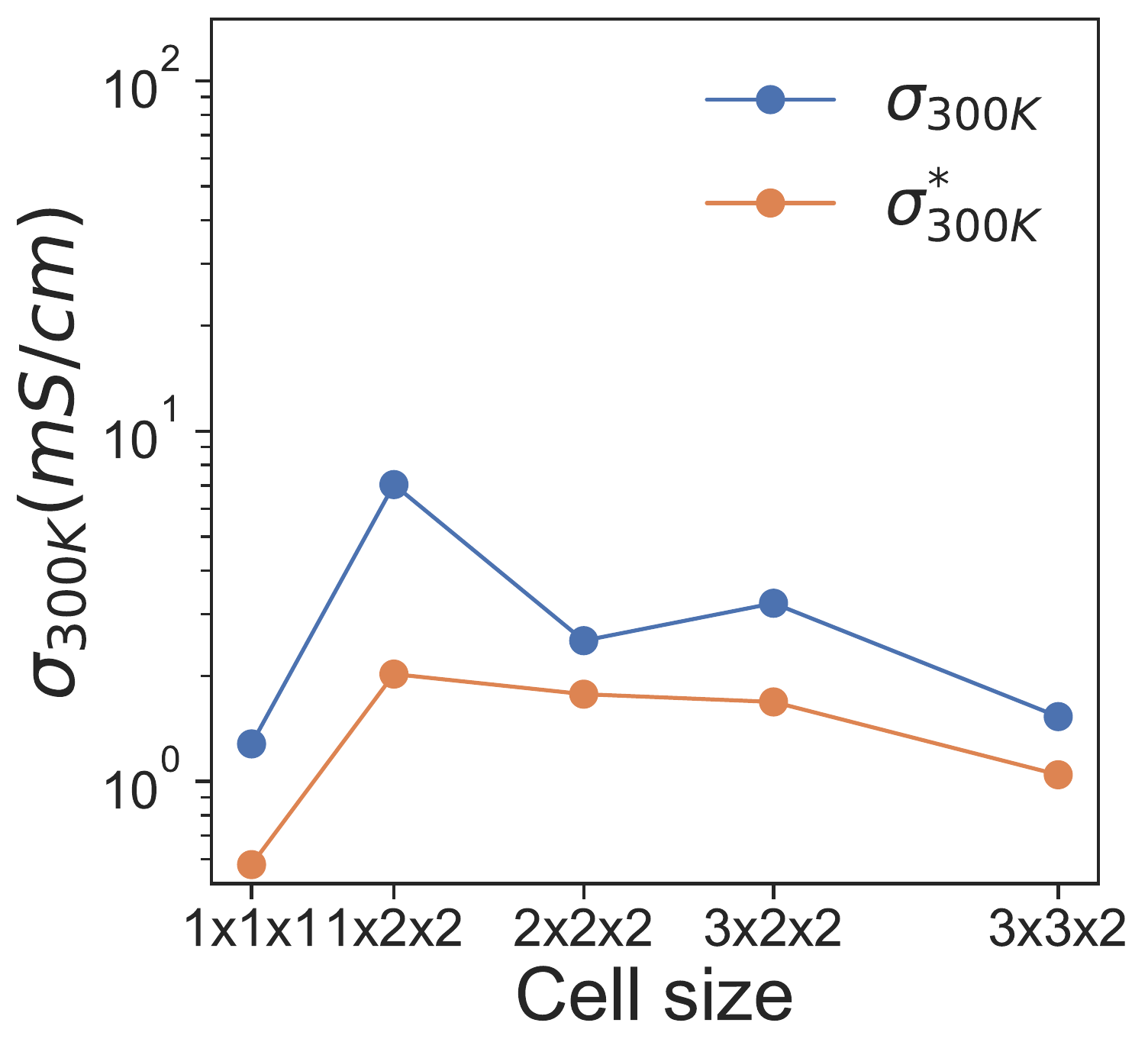}
    	\caption{\ce{LLTO}}
    \end{subfigure}
    \begin{subfigure}[b]{0.3\textwidth}
        \centering
	    \includegraphics[height=3.5cm]{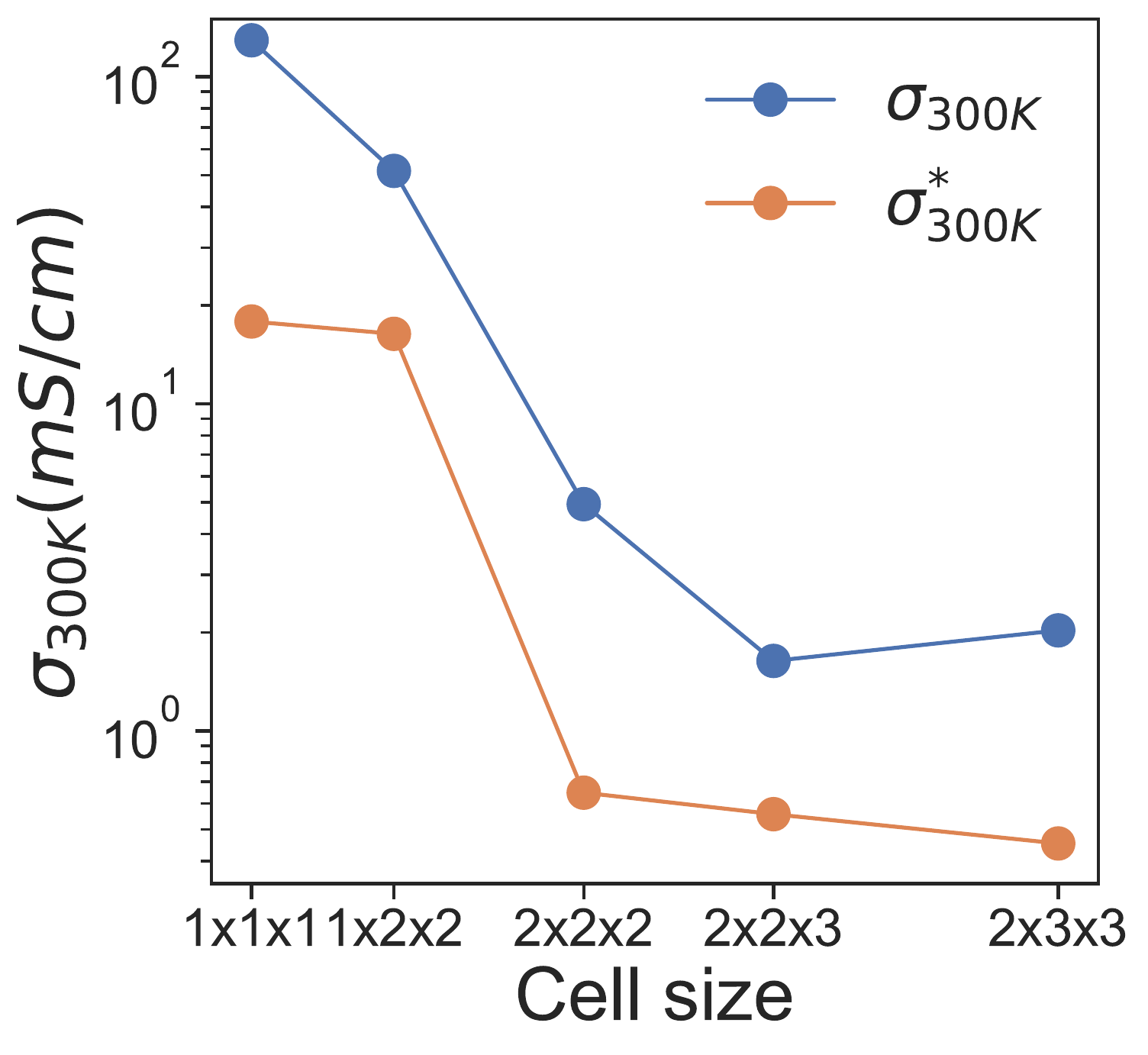}
    	\caption{\ce{Li3YCl6}}
    \end{subfigure}
    \begin{subfigure}[b]{0.3\textwidth}
        \centering
	    \includegraphics[height=3.5cm]{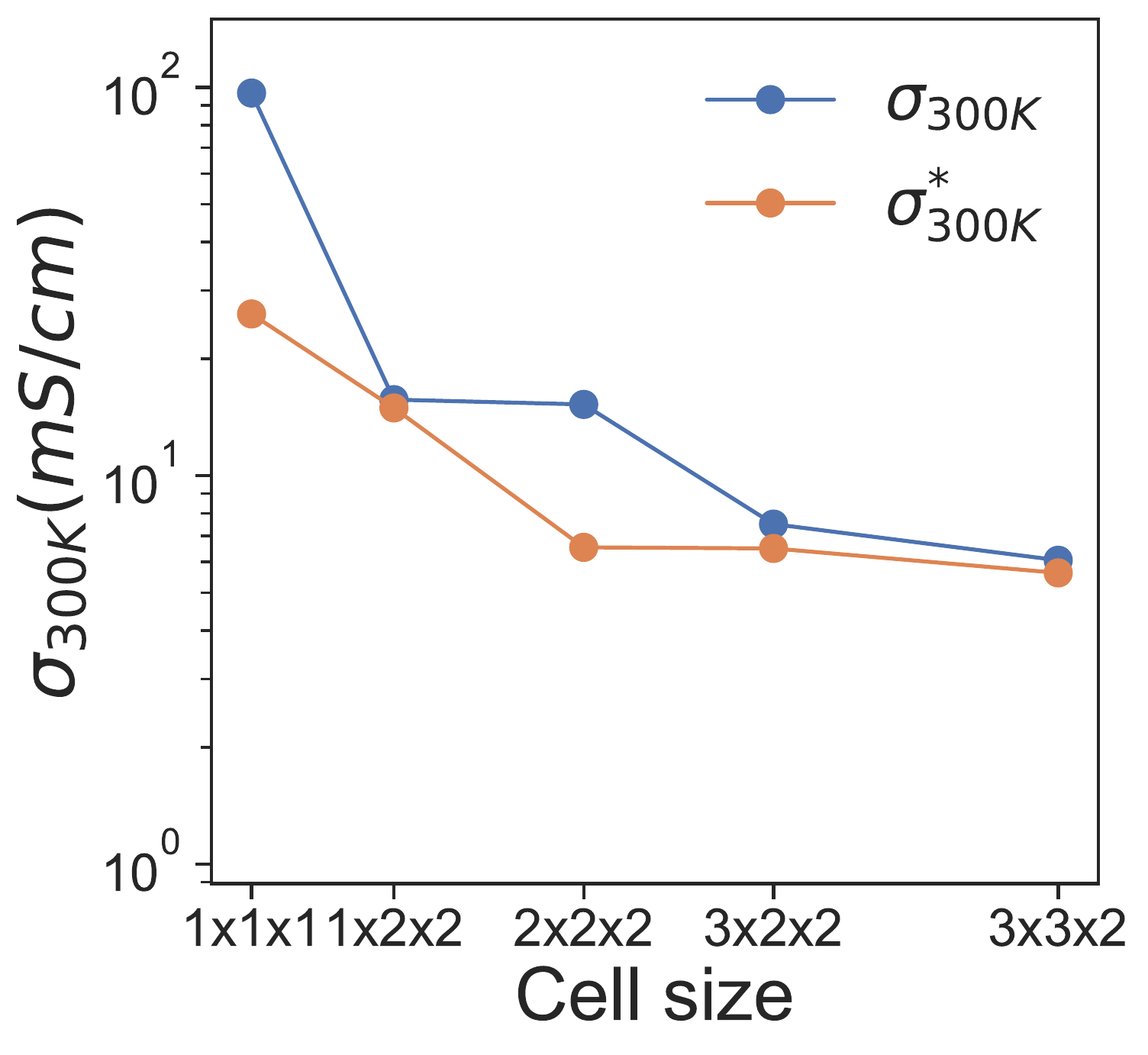}
    	\caption{\ce{Li7P3S11}}
    \end{subfigure}
    \caption{The effect of simulation cell size on the NPT/MD simulated charge ionic conductivity ($\sigma_{300K}$) and tracer ionic conductivity ($\sigma^{*}_{300K}$) at 300K with MTP$_{\mathrm{PBE,optB88}}$ of LLTO, \ce{Li3YCl6} and \ce{Li7P3S11}.}
\end{figure}

\begin{figure}[htb!]
\renewcommand{\thefigure}{S\arabic{figure}}
    \begin{subfigure}[b]{0.32\textwidth}
        \centering
	    \includegraphics[height=3.2cm]{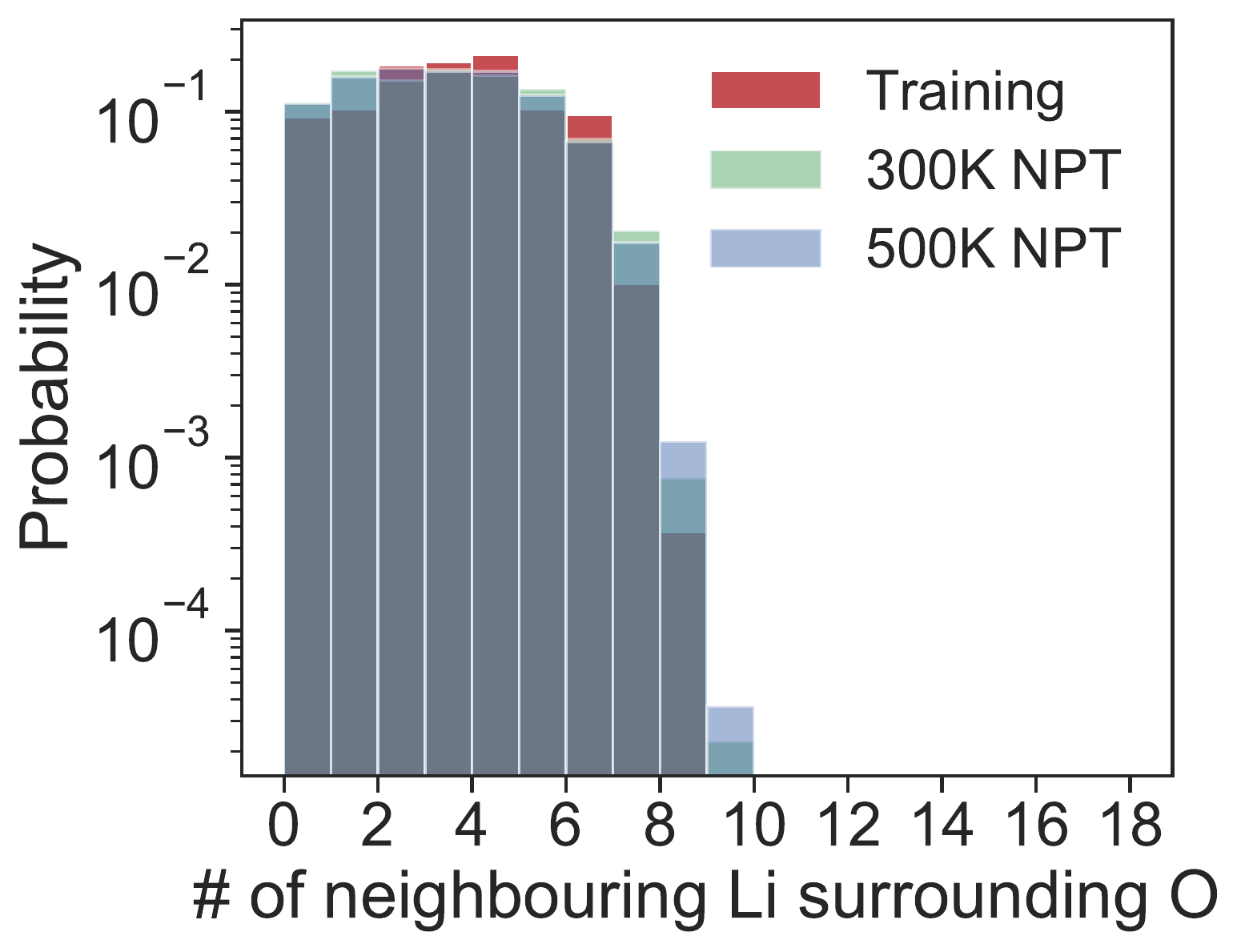}
    	\caption{\ce{LLTO}}
    \end{subfigure}
    \begin{subfigure}[b]{0.33\textwidth}
        \centering
	    \includegraphics[height=3.2cm]{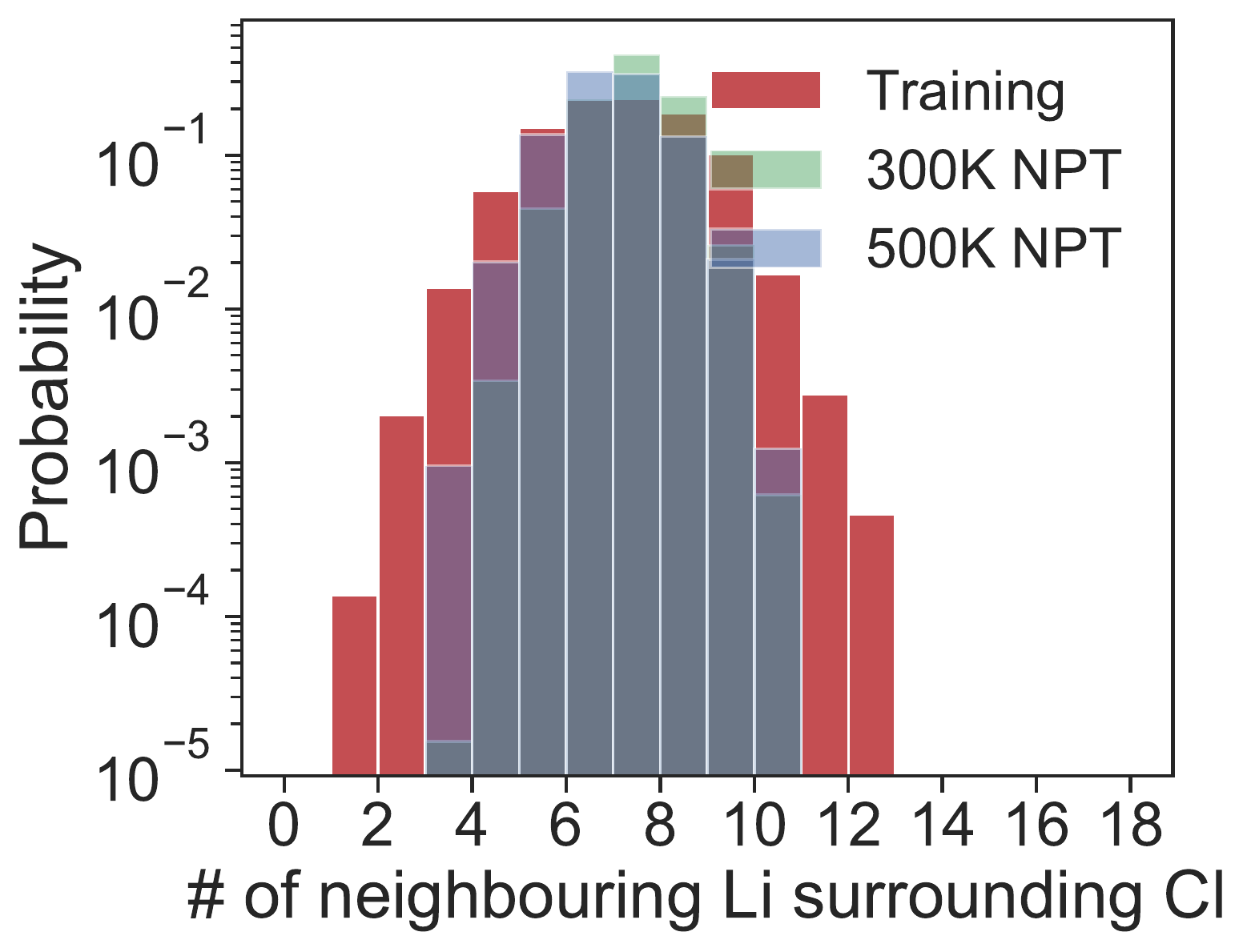}
    	\caption{\ce{Li3YCl6}}
    \end{subfigure}
    \begin{subfigure}[b]{0.32\textwidth}
        \centering
	    \includegraphics[height=3.2cm]{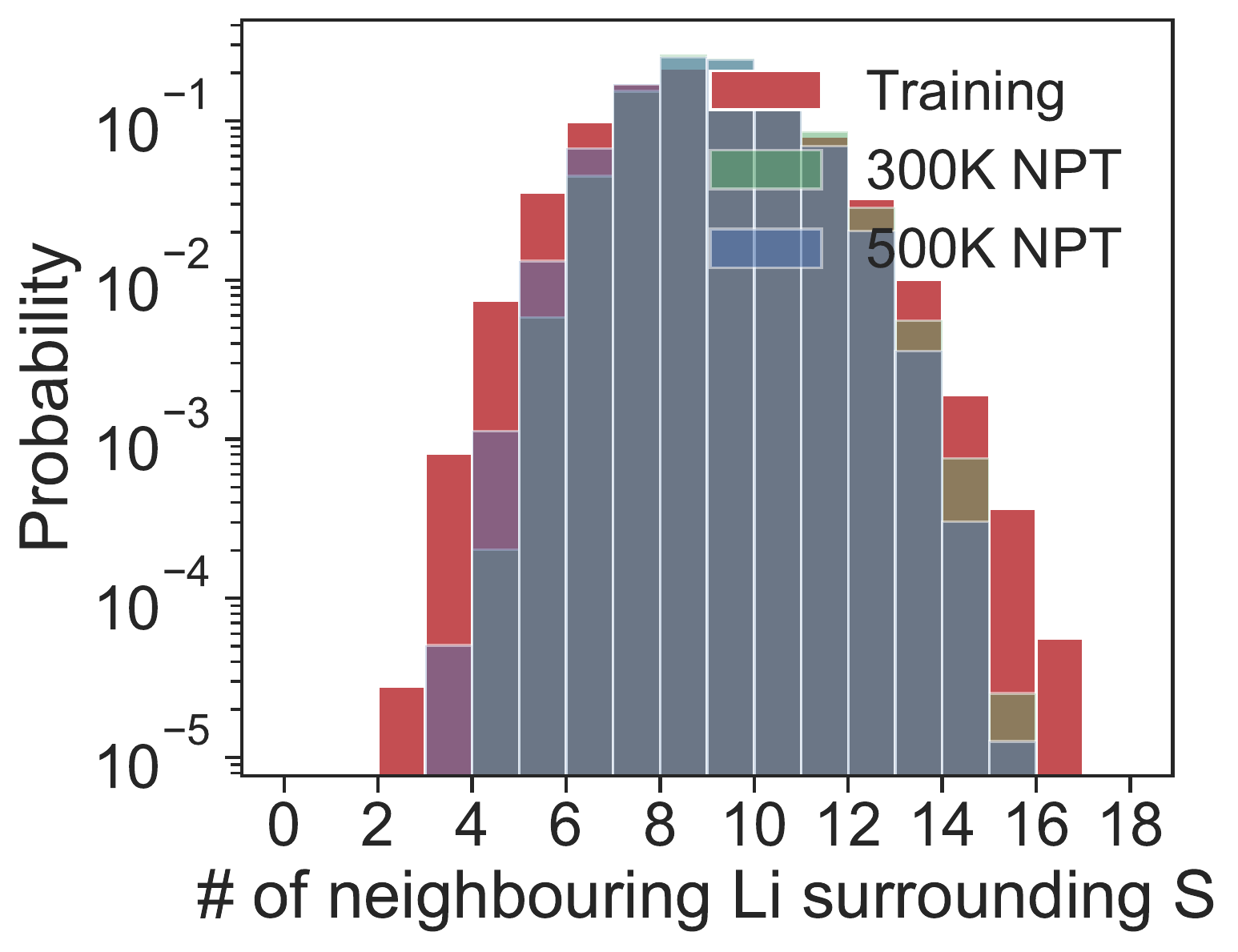}
    	\caption{\ce{Li7P3S11}}
    \end{subfigure}
           \begin{subfigure}[b]{0.33\textwidth}
        \centering
	    \includegraphics[height=3.2cm]{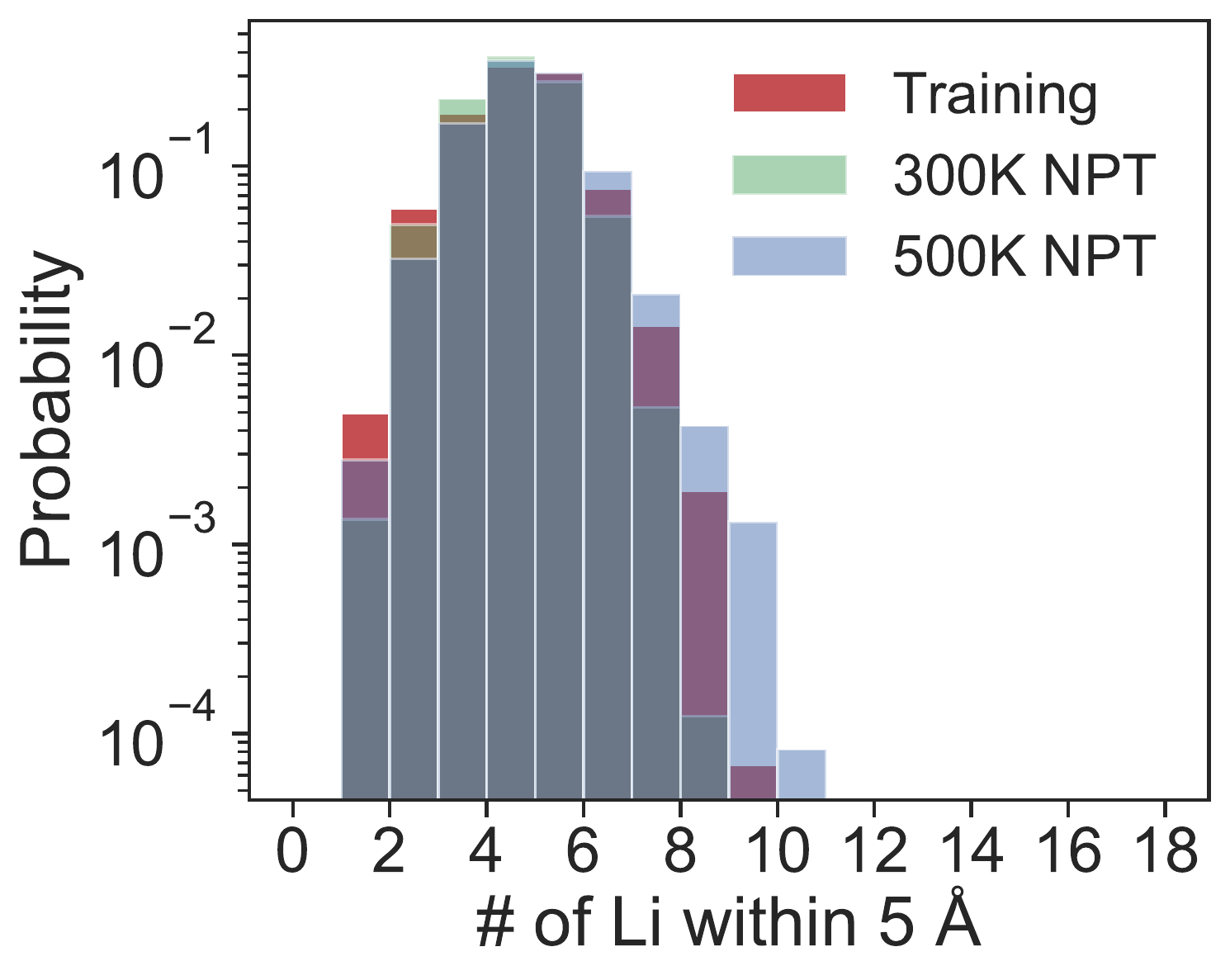}
    	\caption{\ce{LLTO}}
    \end{subfigure}
    \begin{subfigure}[b]{0.33\textwidth}
        \centering
	    \includegraphics[height=3.2cm]{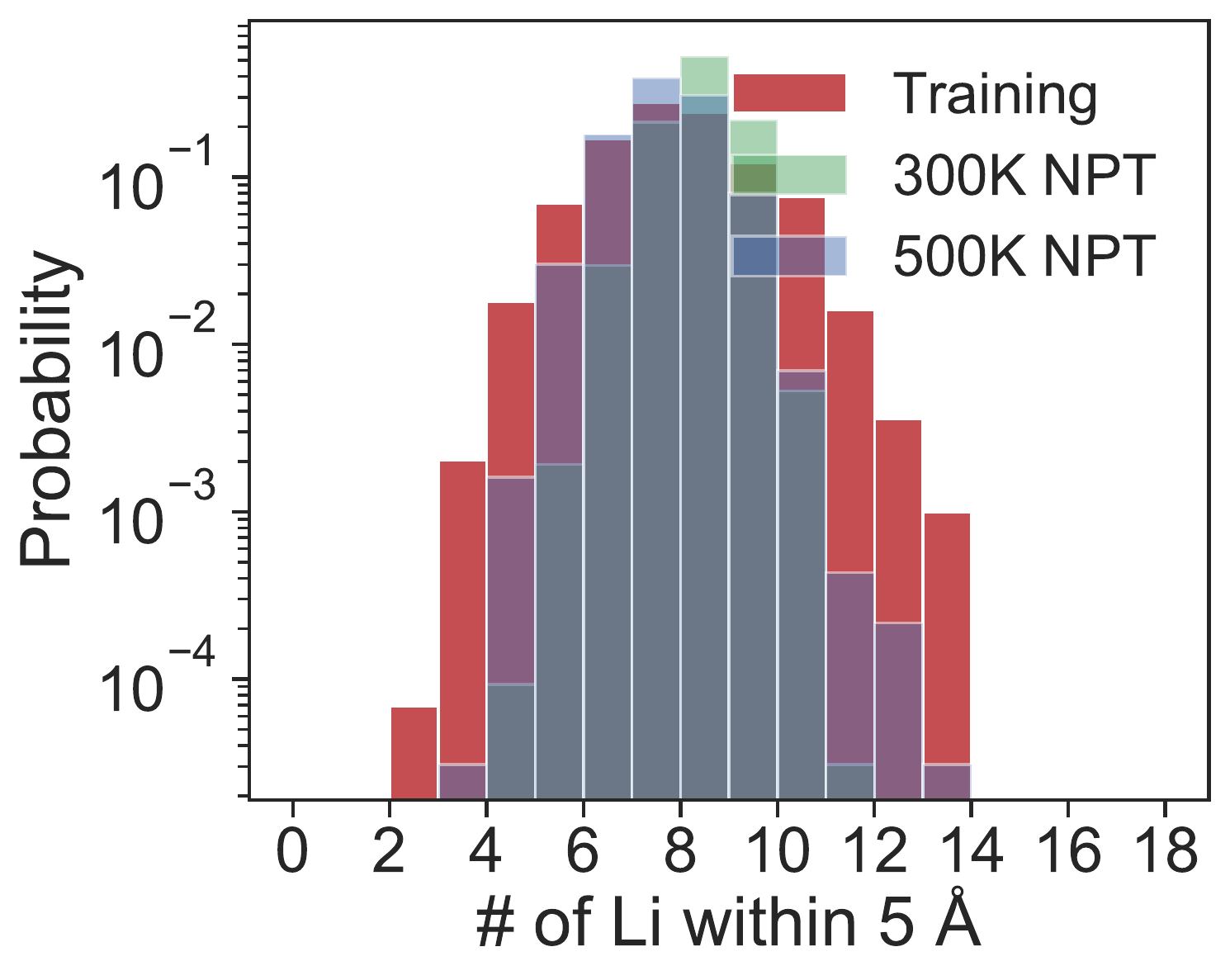}
    	\caption{\ce{Li3YCl6}}
    \end{subfigure}
    \begin{subfigure}[b]{0.3\textwidth}
        \centering
	    \includegraphics[height=3.2cm]{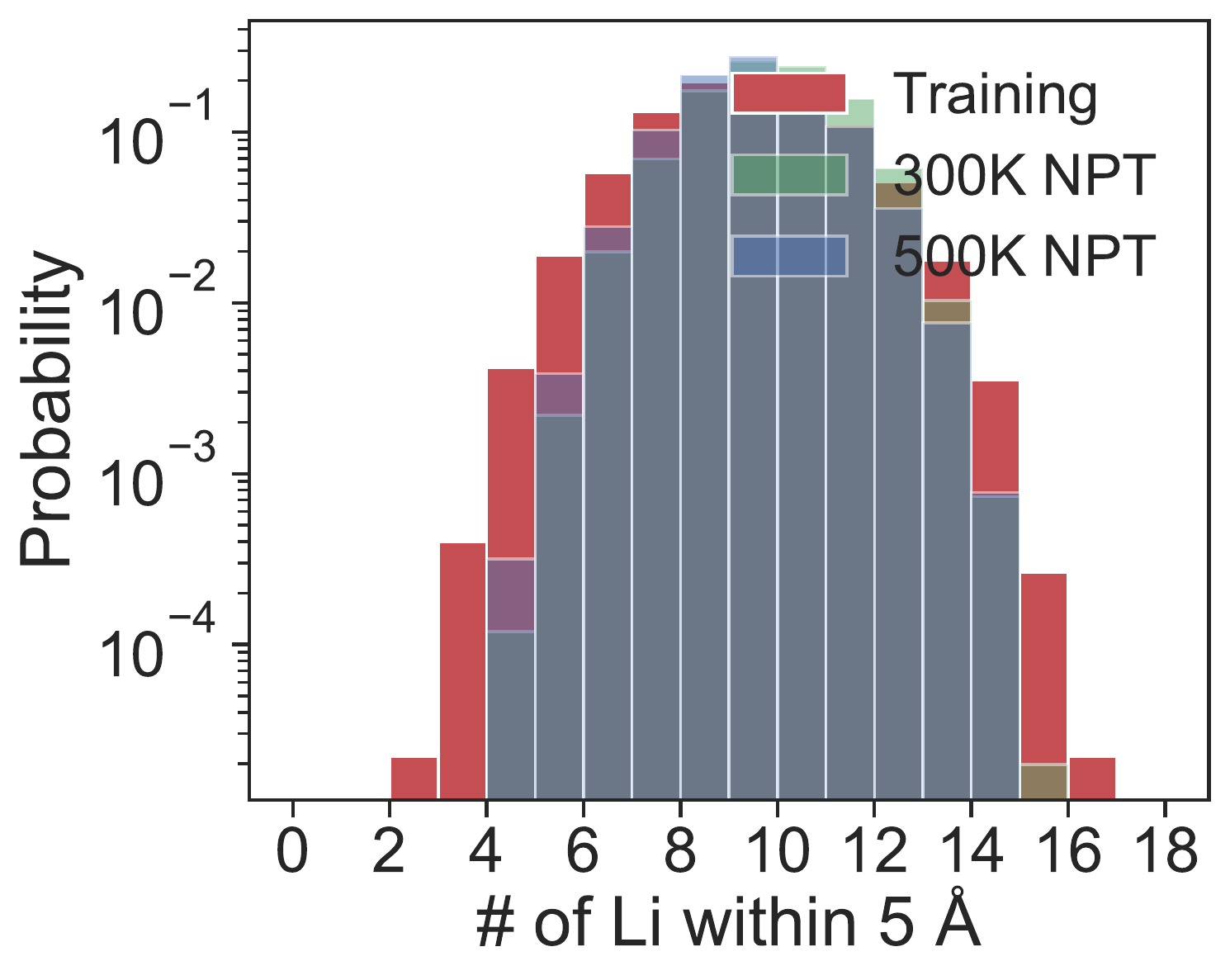}
    	\caption{\ce{Li7P3S11}}
    \end{subfigure}
    \caption{(a)-(c) The distribution of the numbers of neighbouring Li surrounding anions and (d)-(f) the distribution of the numbers of Li within the cutoff radius of 5 \AA~sampled by training structures, 300K and 500K NPT trajectories. For each LSC, the NPT trajectories at each temperature contain 150 structures extracted from 1500ps with 10ps intervals of the MD simulations performed to study diffusivities.}
\end{figure}

\begin{figure}[htb!]
\renewcommand{\thefigure}{S\arabic{figure}}
    \begin{subfigure}[b]{0.32\textwidth}
        \centering
	    \includegraphics[height=3.2cm]{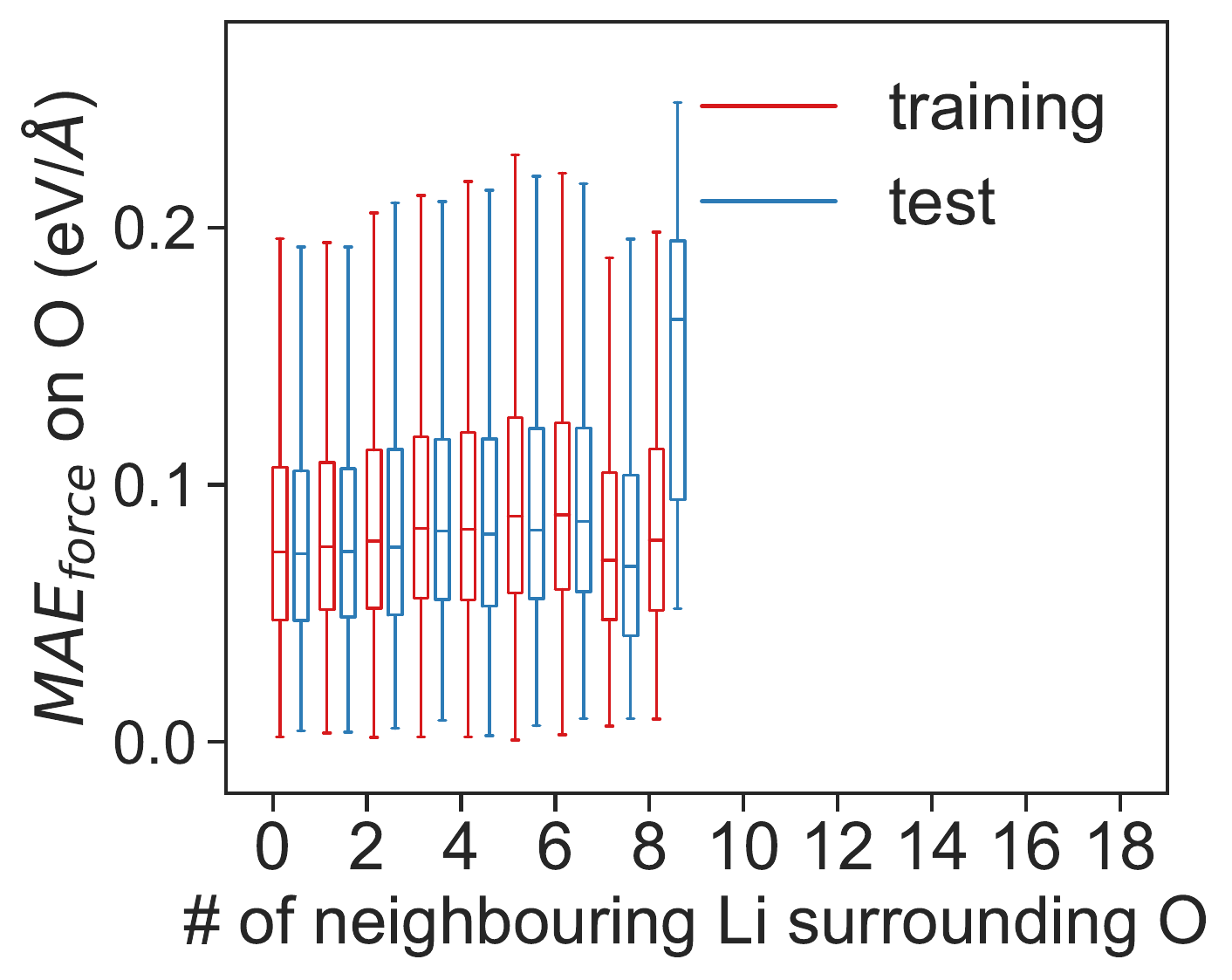}
    	\caption{\ce{LLTO}}
    \end{subfigure}
    \begin{subfigure}[b]{0.33\textwidth}
        \centering
	    \includegraphics[height=3.2cm]{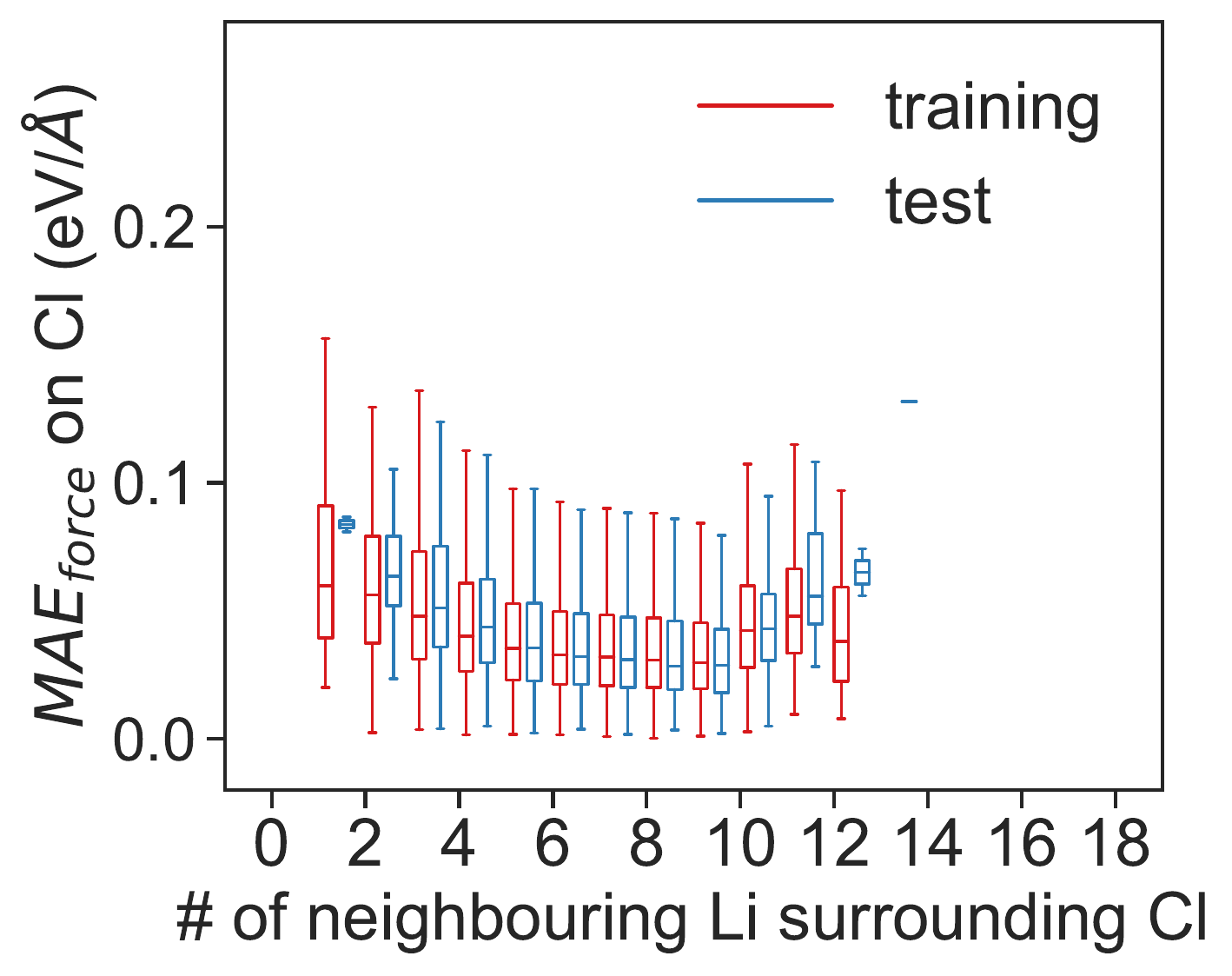}
    	\caption{\ce{Li3YCl6}}
    \end{subfigure}
    \begin{subfigure}[b]{0.32\textwidth}
        \centering
	    \includegraphics[height=3.2cm]{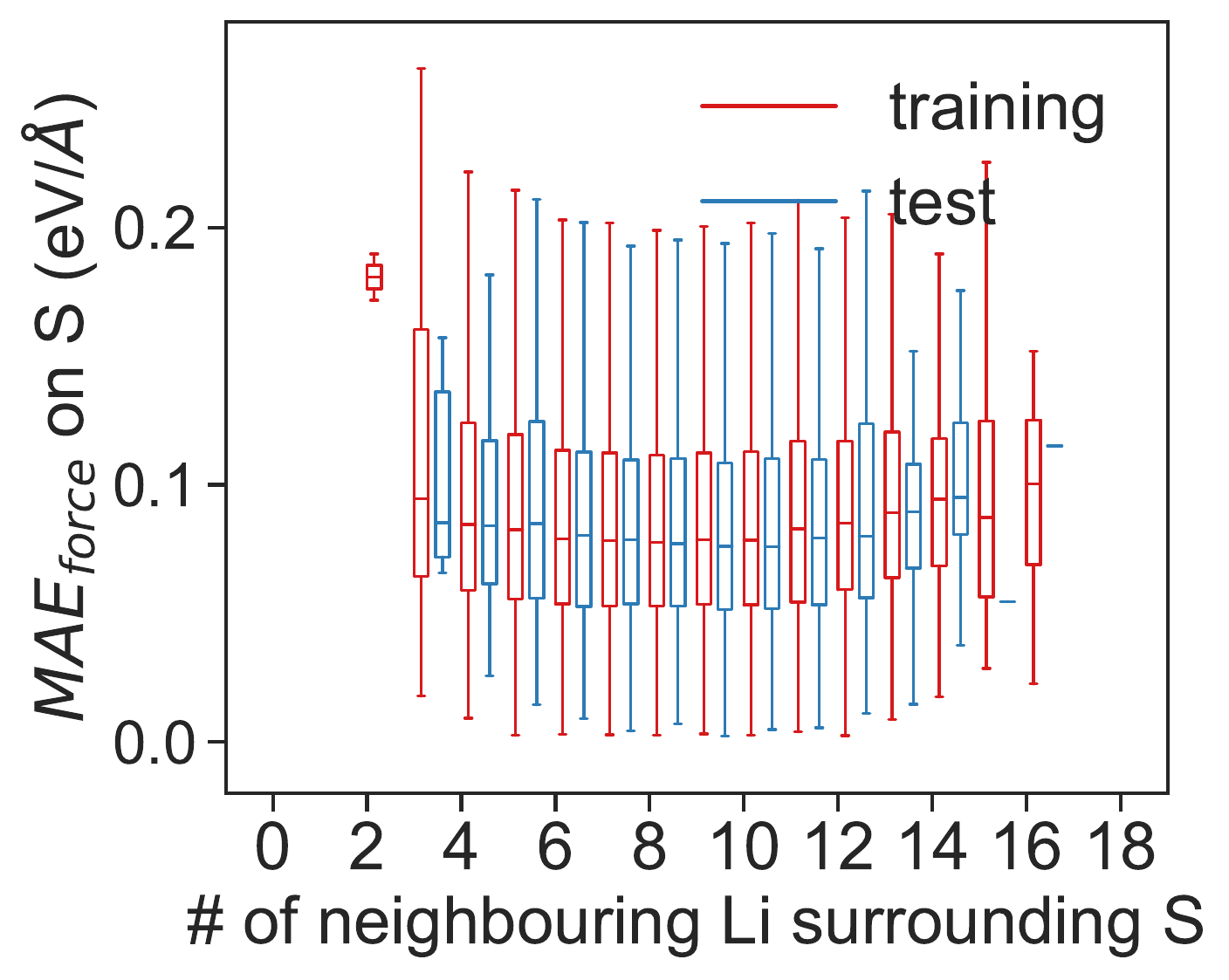}
    	\caption{\ce{Li7P3S11}}
    \end{subfigure}
           \begin{subfigure}[b]{0.33\textwidth}
        \centering
	    \includegraphics[height=3.2cm]{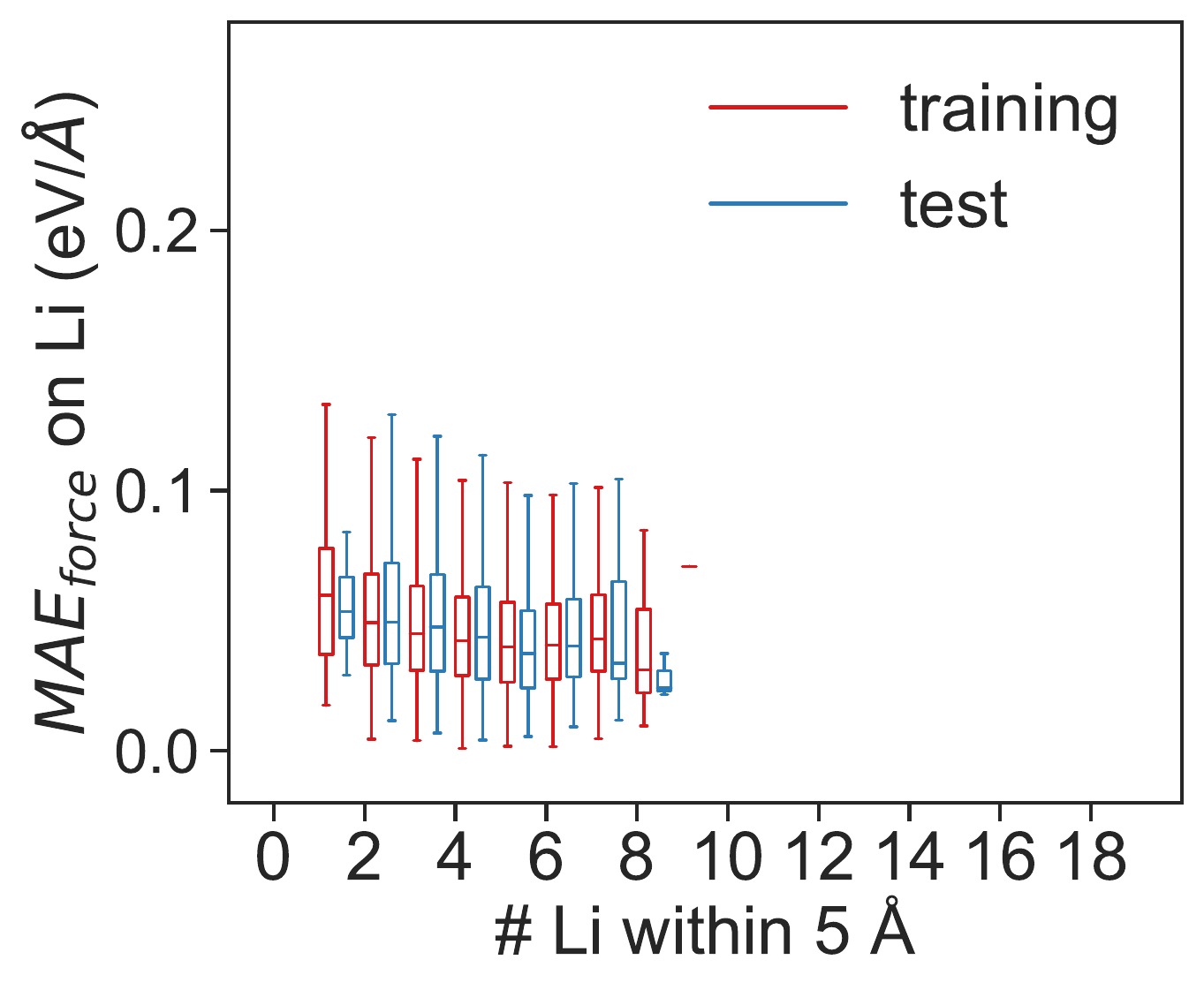}
    	\caption{\ce{LLTO}}
    \end{subfigure}
    \begin{subfigure}[b]{0.33\textwidth}
        \centering
	    \includegraphics[height=3.2cm]{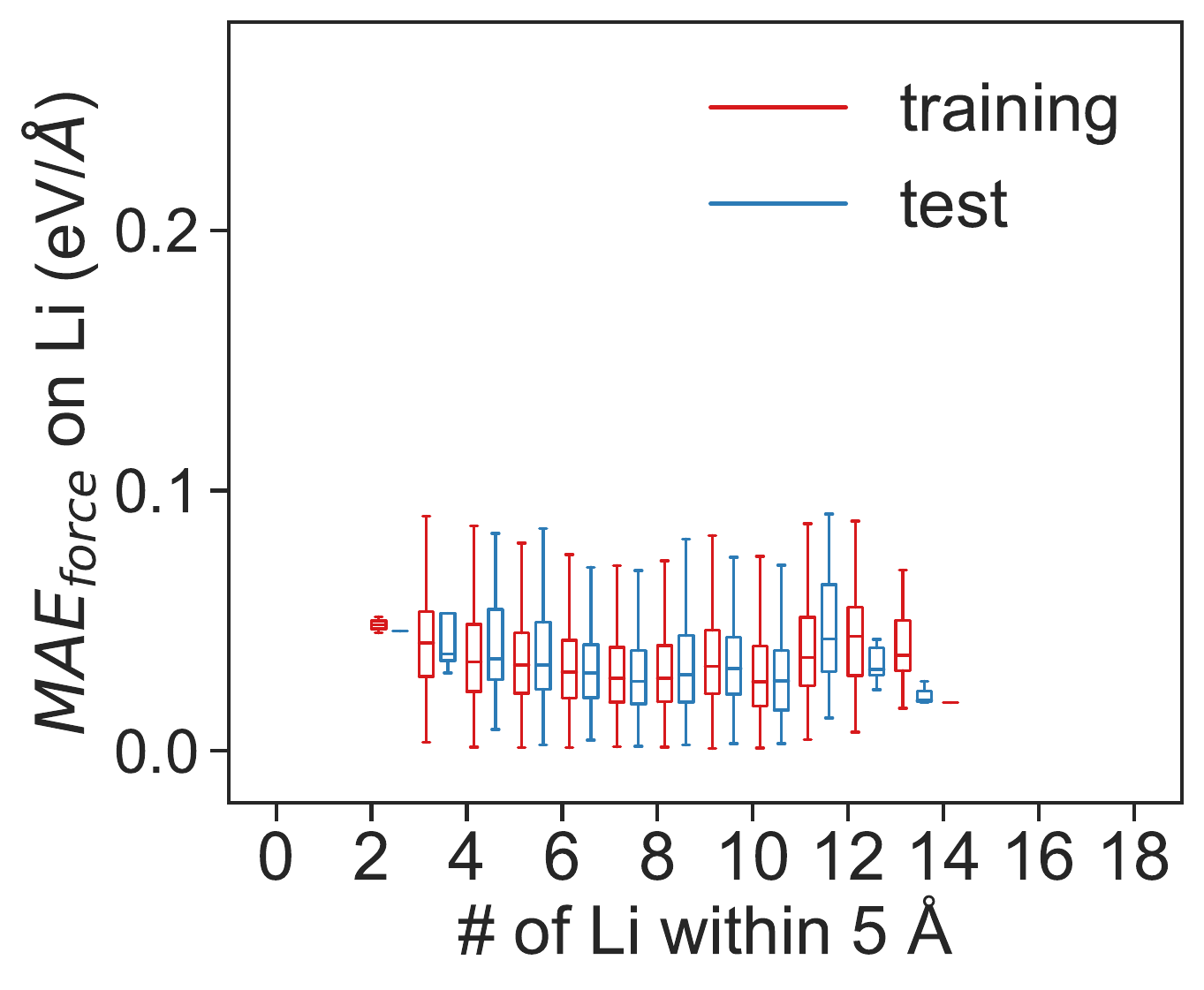}
    	\caption{\ce{Li3YCl6}}
    \end{subfigure}
    \begin{subfigure}[b]{0.3\textwidth}
        \centering
	    \includegraphics[height=3.2cm]{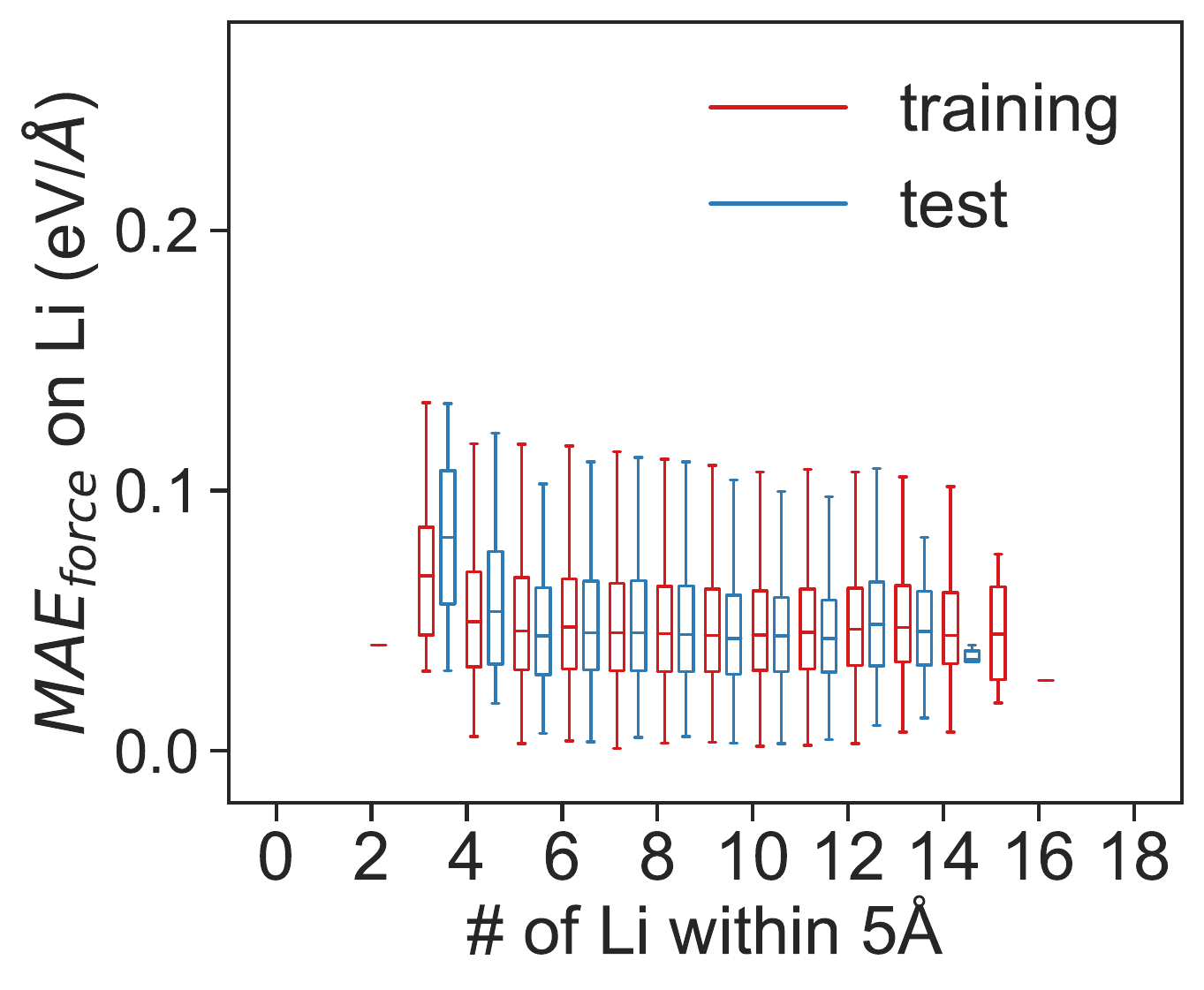}
    	\caption{\ce{Li7P3S11}}
    \end{subfigure}
    \caption{(a)-(c) The distribution of MAE$_{\rm{force}}$ on anions and (d)-(f) the distribution of MAE$_{\rm{force}}$ on Li ions with different local environments sampled by the training and test structures.}
\end{figure}
\begin{figure}[htb!]
\renewcommand{\thefigure}{S\arabic{figure}}
    \begin{subfigure}[b]{0.49\textwidth}
        \centering
	    \includegraphics[height=5cm]{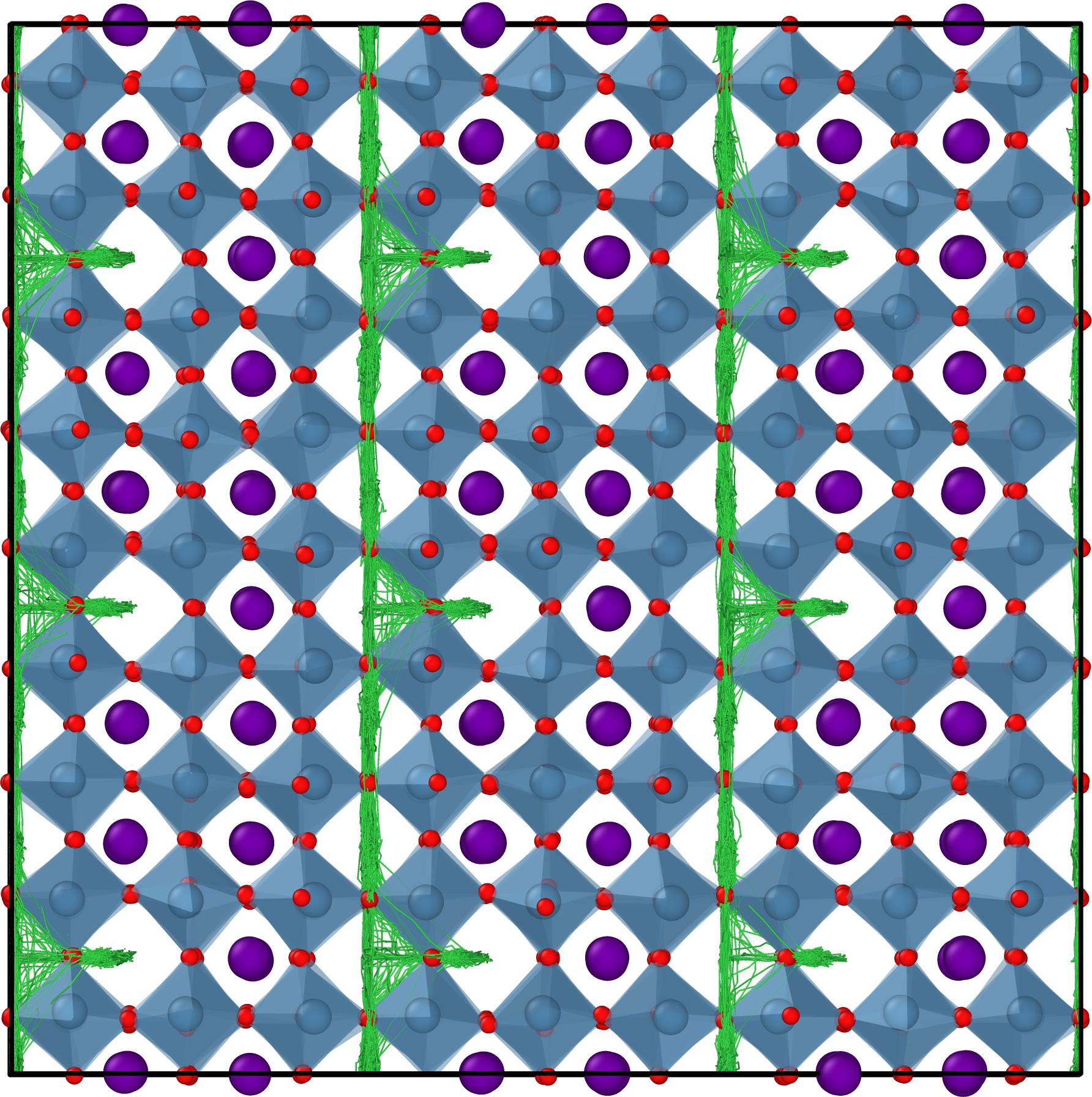}
    	\caption{LLTO 500K a-b plane}
    \end{subfigure}
    \begin{subfigure}[b]{0.49\textwidth}
        \centering
	    \includegraphics[height=5cm]{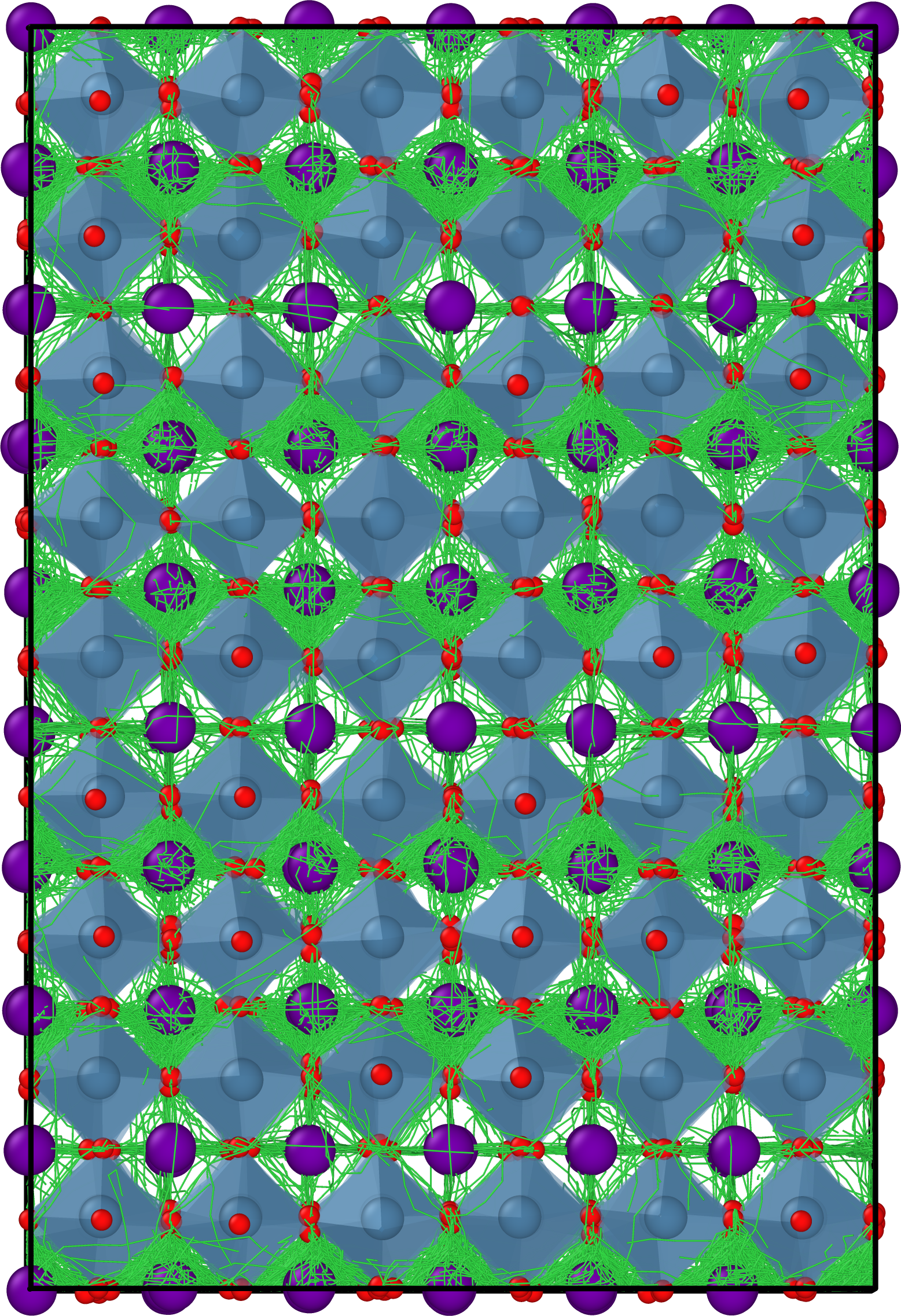}
    	\caption{LLTO 500K b-c plane}
    \end{subfigure}
    \begin{subfigure}[b]{0.49\textwidth}
        \centering
	    \includegraphics[height=5cm]{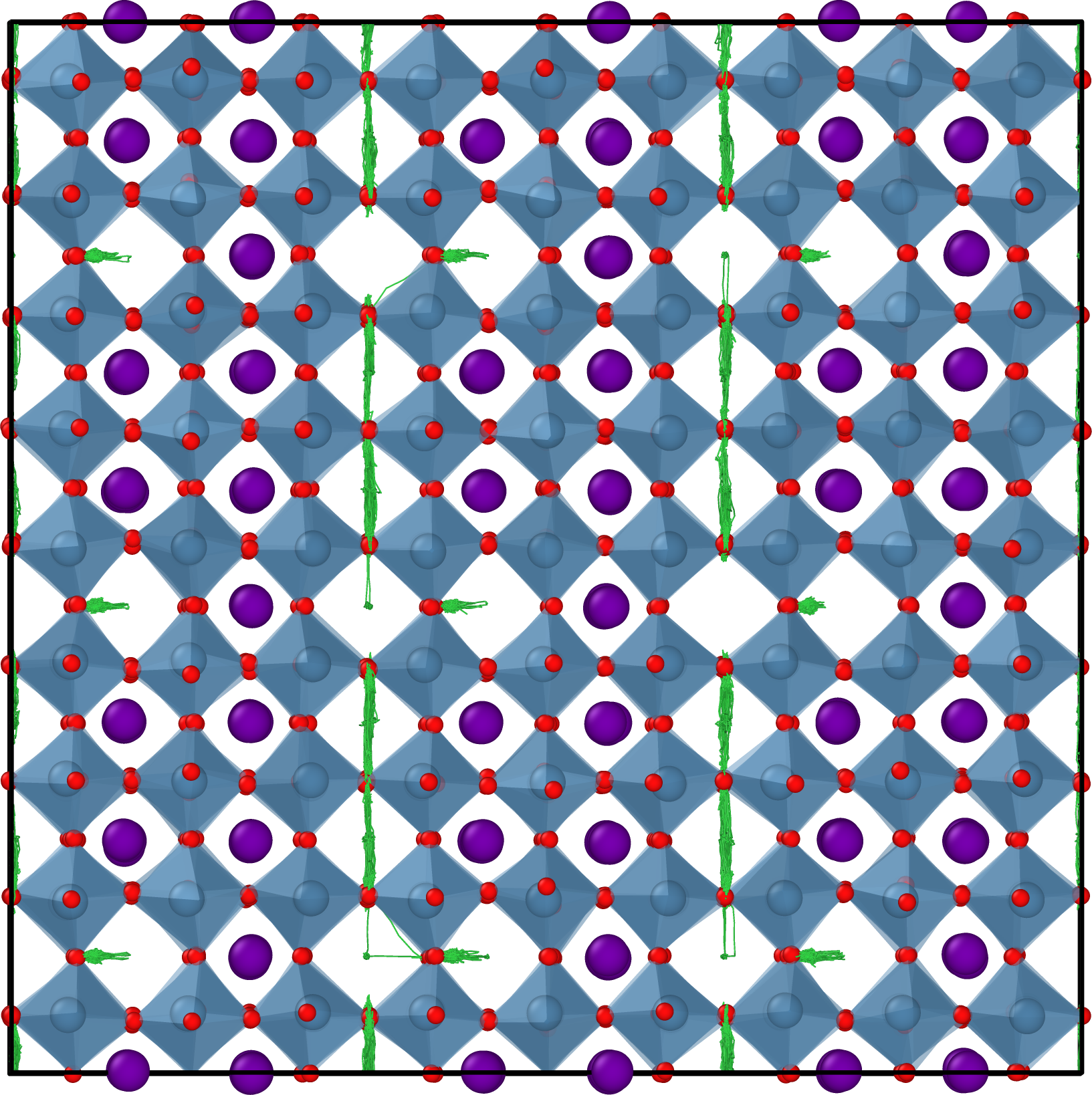}
    	\caption{LLTO 300K a-b plane}
    \end{subfigure}
    \begin{subfigure}[b]{0.49\textwidth}
        \centering
	    \includegraphics[height=5cm]{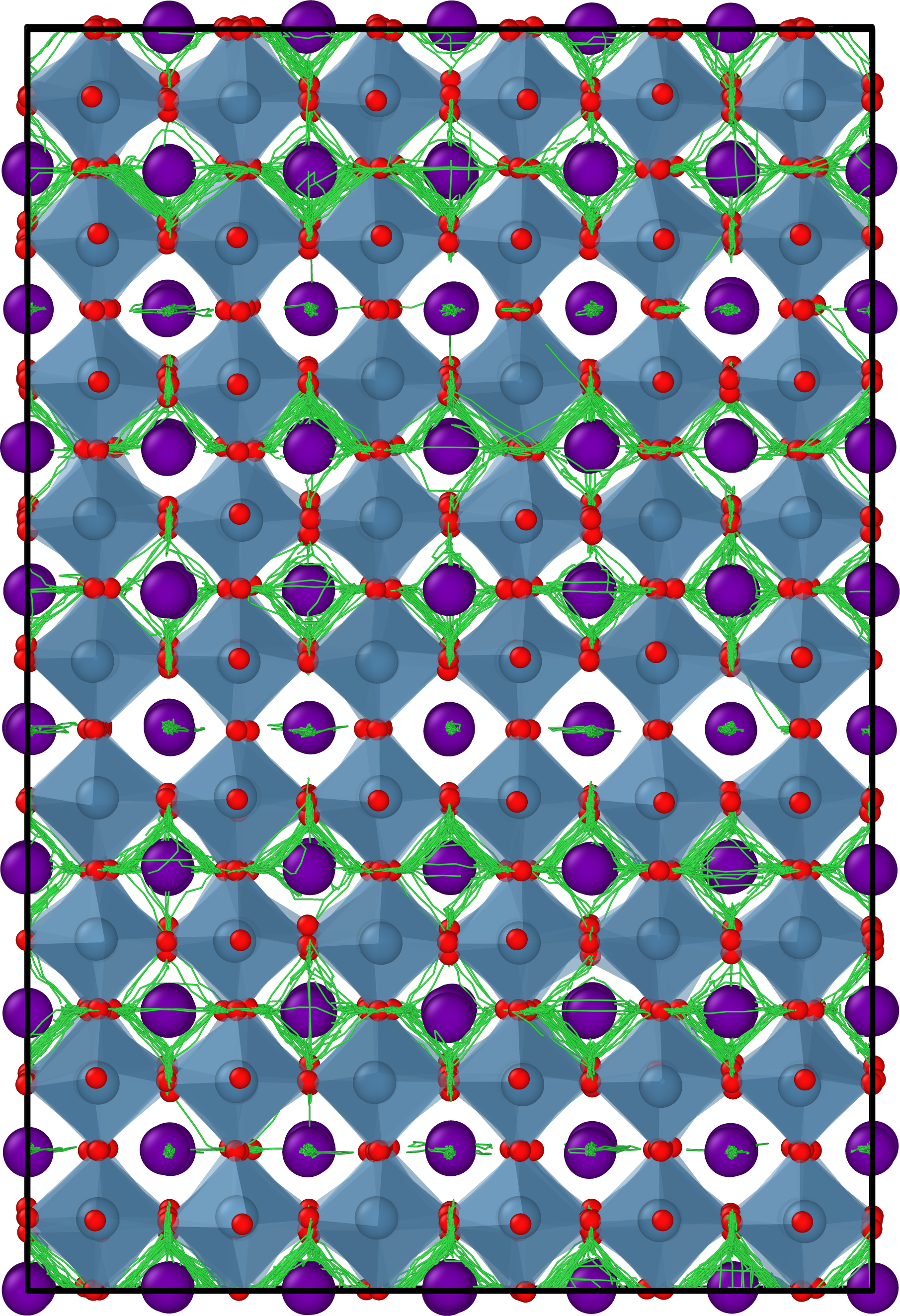}
    	\caption{LLTO 300K b-c plane}
    \end{subfigure}
    \caption{Li trajectories (colored as green) from MTP$_{\mathrm{PBE,optB88}}$ MD simulations of the LLTO at 300K and 500K, projected in the crystallographic a-b and b-c planes.}
\end{figure}

\begin{figure}[htb!]
\renewcommand{\thefigure}{S\arabic{figure}}
    \begin{subfigure}[b]{0.52\textwidth}
        \centering
	    \includegraphics[height=3.5cm]{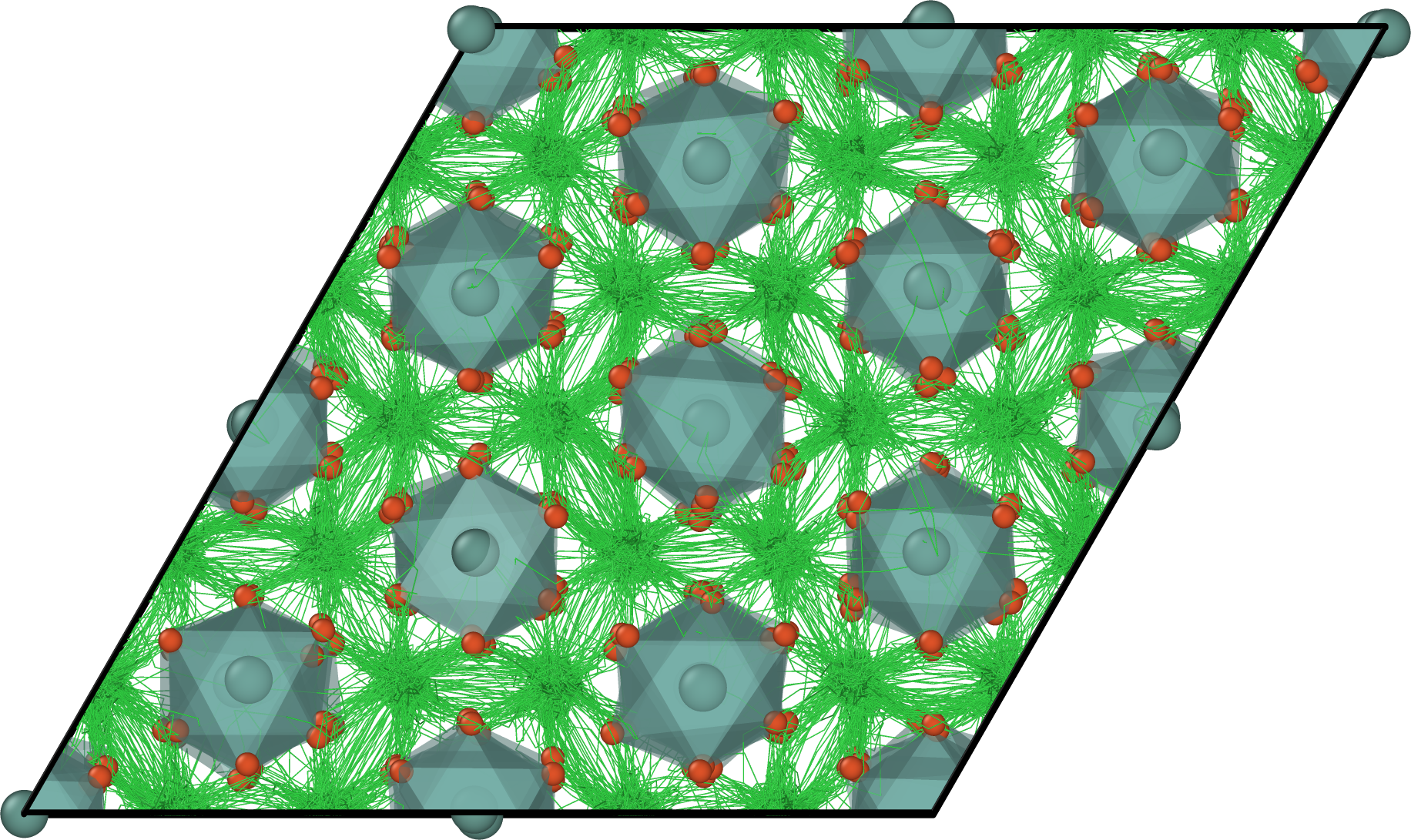}
    	\caption{\ce{Li3YCl6} 500K a-b plane}
    \end{subfigure}
    \begin{subfigure}[b]{0.47\textwidth}
        \centering
	    \includegraphics[height=3.5cm]{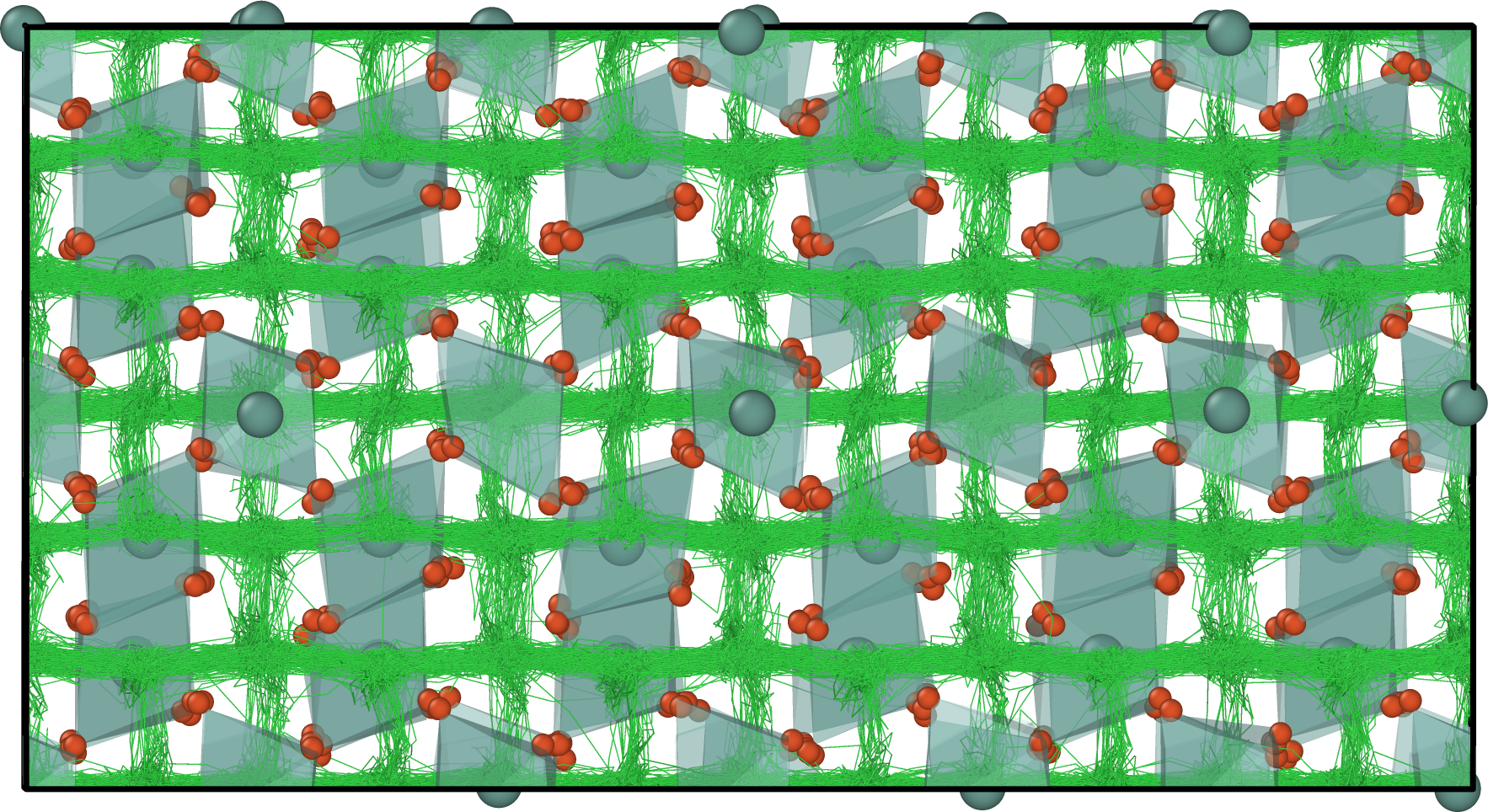}
    	\caption{\ce{Li3YCl6} 500K a-c plane}
    \end{subfigure}
    \begin{subfigure}[b]{0.52\textwidth}
        \centering
	    \includegraphics[height=3.5cm]{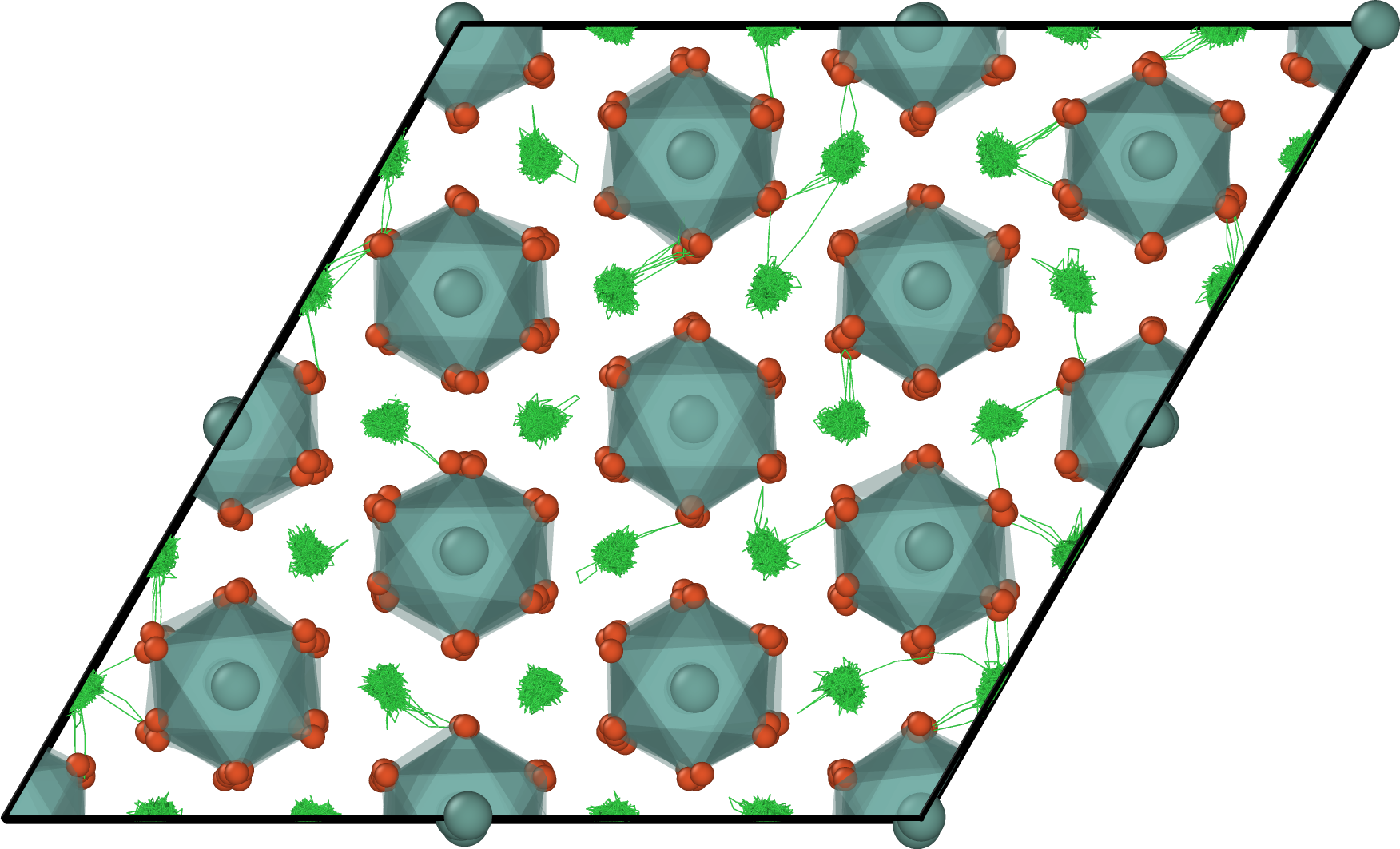}
    	\caption{\ce{Li3YCl6} 300K a-b plane}
    \end{subfigure}
    \begin{subfigure}[b]{0.47\textwidth}
        \centering
	    \includegraphics[height=3.5cm]{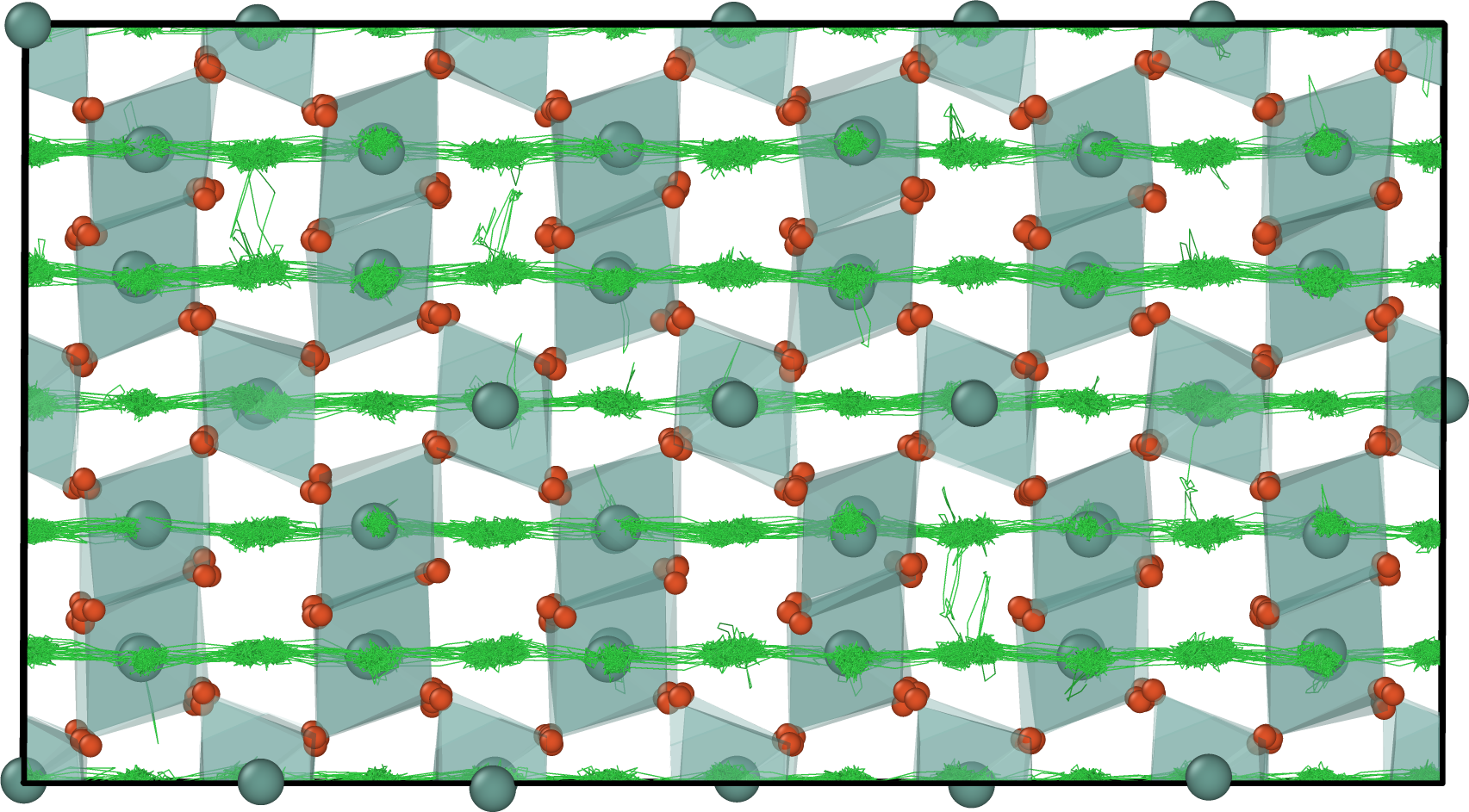}
    	\caption{\ce{Li3YCl6} 300K a-c plane}
    \end{subfigure}
    \caption{Li trajectories (colored as green) from MTP$_{\mathrm{PBE,optB88}}$ MD simulations of the \ce{Li3YCl6} at 300K and 500K, projected in the crystallographic a-b and a-c planes.}
\end{figure}

\begin{figure}[htb!]
\renewcommand{\thefigure}{S\arabic{figure}}
    \begin{subfigure}[b]{0.65\textwidth}
        \centering
	    \includegraphics[height=3.5cm]{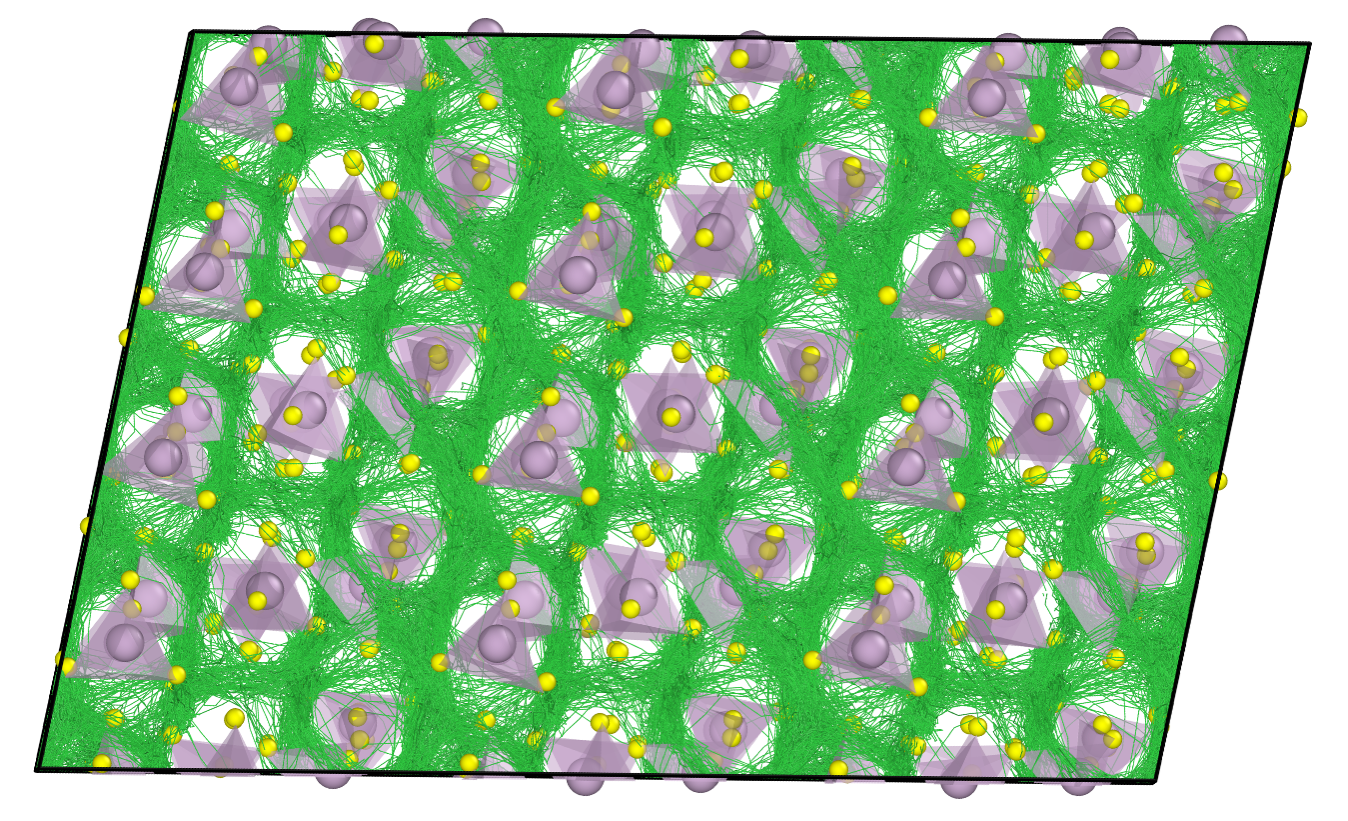}
    	\caption{\ce{Li7P3S11} 500K a-b plane}
    \end{subfigure}
    \begin{subfigure}[b]{0.34\textwidth}
        \centering
	    \includegraphics[height=3.5cm]{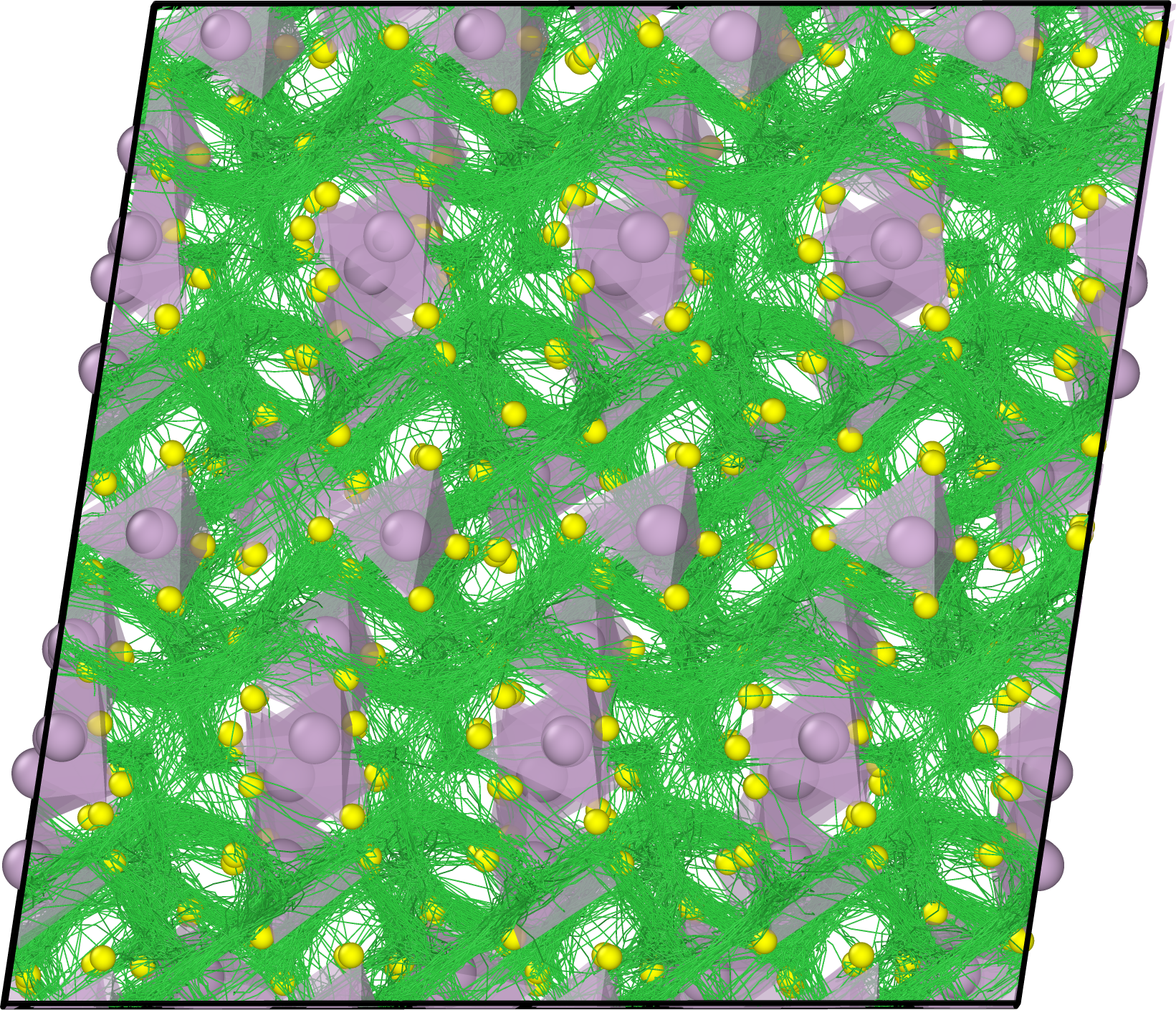}
    	\caption{\ce{Li7P3S11} 500K b-c plane}
    \end{subfigure}
    \begin{subfigure}[b]{0.6\textwidth}
        \centering
	    \includegraphics[height=3.5cm]{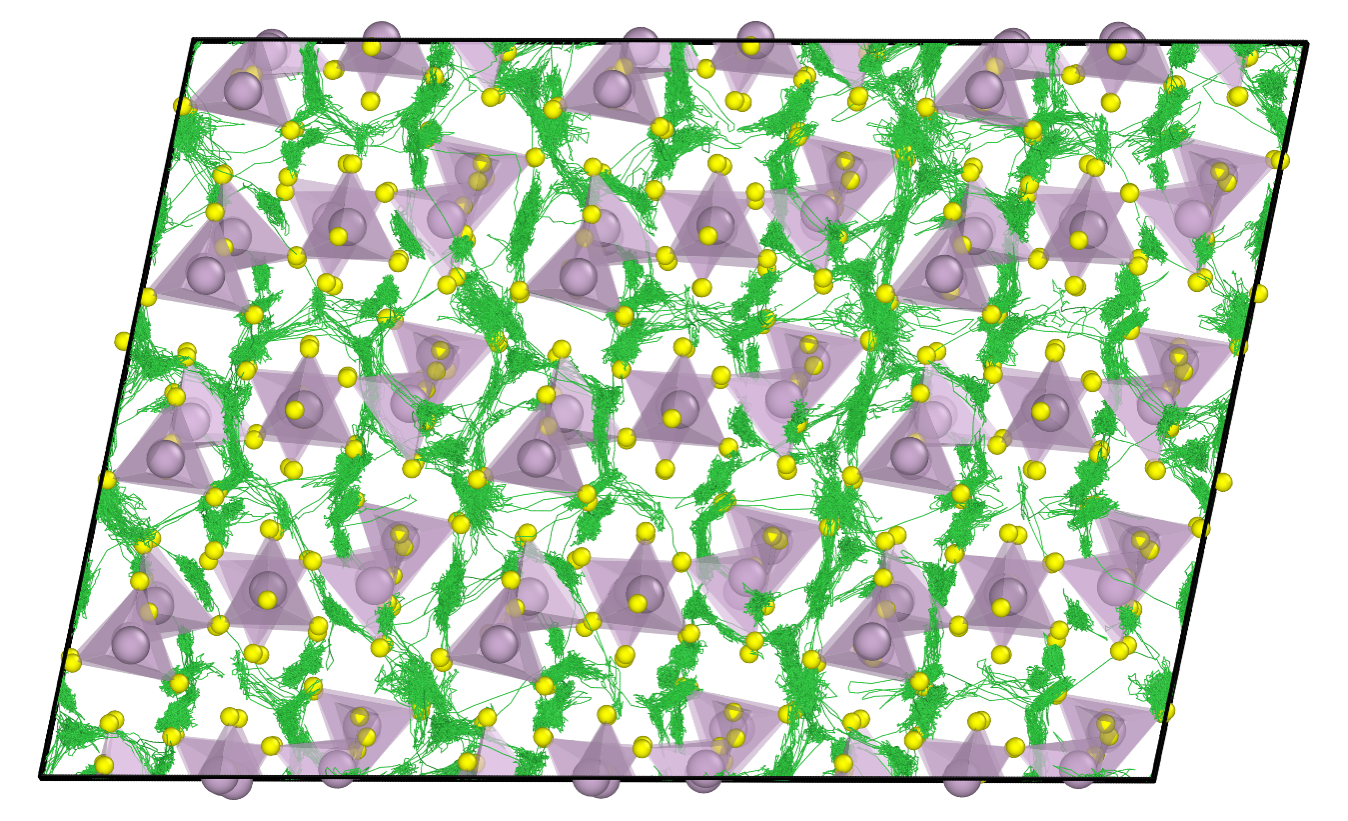}
    	\caption{\ce{Li7P3S11} 300K a-b plane}
    \end{subfigure}
    \begin{subfigure}[b]{0.34\textwidth}
        \centering
	    \includegraphics[height=3.5cm]{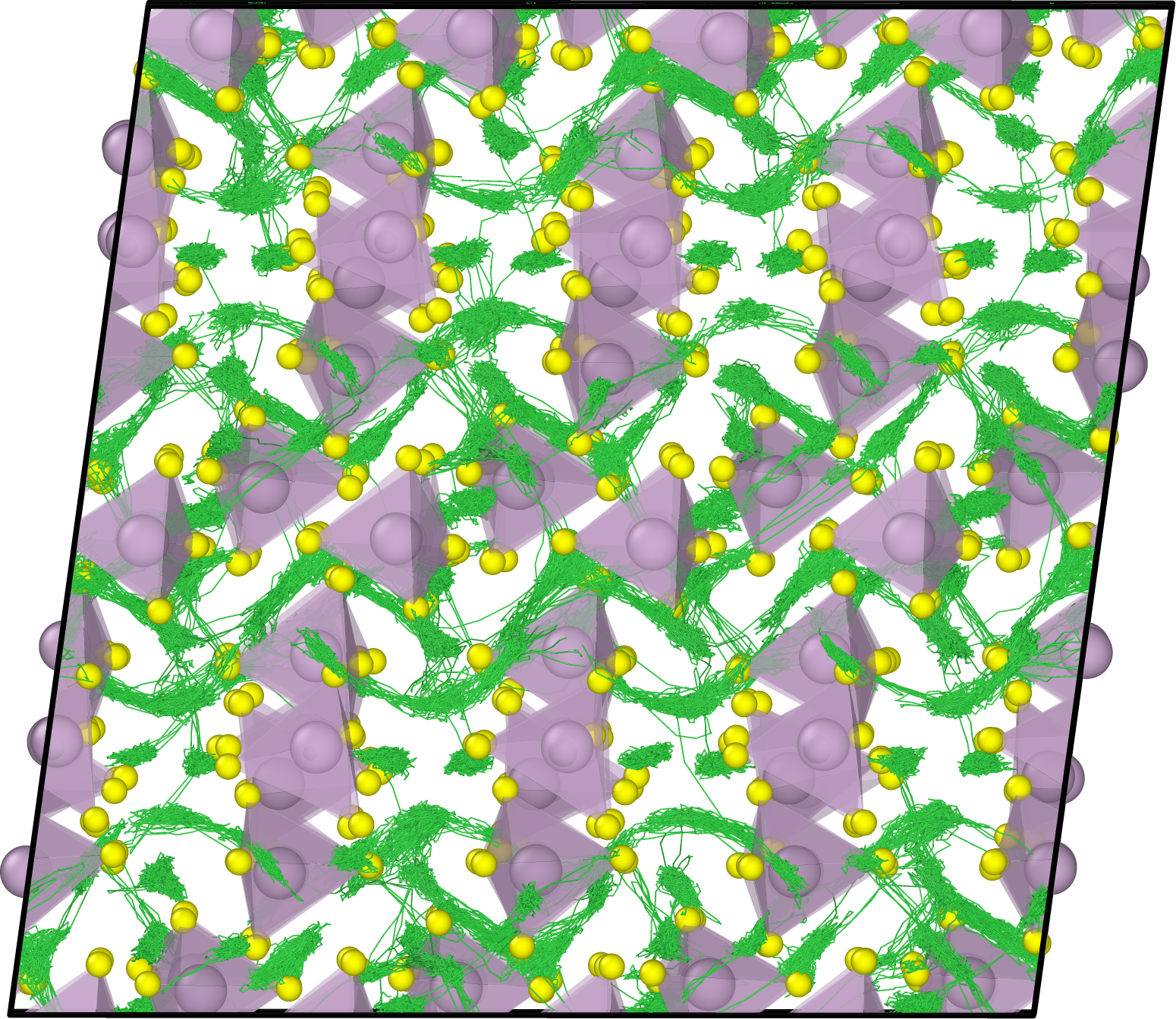}
    	\caption{\ce{Li7P3S11} 300K b-c plane}
    \end{subfigure}
    \caption{Li trajectories (colored as green) from MTP$_{\mathrm{PBE,optB88}}$ MD simulations of the \ce{Li7P3S11} at 300K and 500K, projected in the crystallographic a-b and b-c planes.}
\end{figure}

\begin{table}[hbt!]
\renewcommand{\thetable}{S\arabic{table}}
  \centering
  \caption{The computational cost of MTP MD NPT and AIMD NPT for different systems. The unit of computational cost is CPU-seconds per MD step. MTPs used in these comparisons are the same MTP$_{\mathrm{PBE,optB88}}$ selected for each LSC.}
   \begin{tabular}{cccc}
   \hline\hline
    LSC  & LLTO  & \ce{Li3YCl6} & \ce{Li7P3S11} \\
    No. of atoms & 132   & 60    & 84 \\\hline
    AIMD NPT & 1200  & 86 & 225 \\
    MTP MD NPT & 0.034 & 0.015 & 0.024 \\
    \hline\hline
    \end{tabular}%
  \label{table:MD_speed_MTP_AIMD}%
\end{table}%

\newpage

\FloatBarrier
\bibliography{Ref_MTP_Electrolytes_LLTO_LYC_Li7P3S11}